\documentclass[%
twocolumn,
superscriptaddress,
longbibliography,
amsmath,amssymb,
aps,
prx,
floatfix
]{revtex4-2}

\usepackage{graphicx}
\usepackage{float}
\usepackage{xcolor}
\usepackage{amssymb}
\usepackage{dsfont}

\makeatletter
\renewcommand{\l@subsubsection}[2]{}
\makeatother

\usepackage{hyperref}

\definecolor{darkblue}{rgb}{0,0,0.5}
\hypersetup{
    colorlinks=true,
    citecolor=darkblue,
    filecolor=black,
    linkcolor=black,
    urlcolor=black
}

\makeatletter
\DeclareFontFamily{OMX}{MnSymbolE}{}
\DeclareFontShape{OMX}{MnSymbolE}{m}{n}{
    <-6>  MnSymbolE5
   <6-7>  MnSymbolE6
   <7-8>  MnSymbolE7
   <8-9>  MnSymbolE8
   <9-10> MnSymbolE9
  <10-12> MnSymbolE10
  <12->   MnSymbolE12
}{}
\DeclareSymbolFont{mnlargesymbols}{OMX}{MnSymbolE}{m}{n}
\SetSymbolFont{mnlargesymbols}{bold}{OMX}{MnSymbolE}{b}{n}
\DeclareMathDelimiter{\llangle}{\mathopen}{mnlargesymbols}{'164}{mnlargesymbols}{'164}
\DeclareMathDelimiter{\rrangle}{\mathclose}{mnlargesymbols}{'171}{mnlargesymbols}{'171}
\makeatother

\newcommand{\bs}[1]{\boldsymbol{#1}}

\newcommand{\circV}{V}

\usepackage{xstring}
\makeatletter
\def\zpar@stop#1{%
  \IfEndWith{#1}{.}{}{%
  \IfEndWith{#1}{?}{}{%
  \IfEndWith{#1}{!}{}{%
  \IfEndWith{#1}{:}{}{%
  \IfEndWith{#1}{;}{}{.}}}}}}

\def\zpar#1{%
  \par\addvspace{0.7\baselineskip}%
  \noindent\textit{#1\zpar@stop{#1}}\hspace{0.6em}\ignorespaces}
\makeatother

\usepackage{supertabular}  %
\usepackage{tikz}
\usetikzlibrary{quantikz2,calc,positioning,arrows.meta}
\tikzset{
  gatenoisy/.style={fill=red!10},     %
  gateaccent1/.style={fill=blue!10},
  gateaccent2/.style={fill=green!10},
}

\IfFileExists{localdev.tex}{\input{localdev.tex}}{}

\begin{document}

\title{Quantum error mitigation from information dynamics}

\author{Philippe Suchsland}
\thanks{These authors contributed equally to this work.}
\affiliation{Google Quantum AI, Santa Barbara, CA, USA}
\author{Thomas Schuster}
\thanks{These authors contributed equally to this work.}
\affiliation{Google Quantum AI, Santa Barbara, CA, USA}
\author{Manuel S. Rudolph}
\affiliation{Google Quantum AI, Santa Barbara, CA, USA}
\affiliation{Institute of Physics, \'Ecole Polytechnique F\'ed\'erale de Lausanne (EPFL), CH-1015 Lausanne, Switzerland}
\author{Nicholas Noll}
\affiliation{Google Quantum AI, Santa Barbara, CA, USA}
\author{Thomas E. O'Brien}
\affiliation{Google Quantum AI, Santa Barbara, CA, USA}
\author{Zlatko K. Minev}
\affiliation{Google Quantum AI, Santa Barbara, CA, USA}

\begin{abstract}
Overcoming experimental errors is a central challenge in quantum science.
This challenge is particularly acute in large and complex quantum circuits, where quantum information spreads into exponentially many paths, foiling any direct attempt to understand the impact of errors.
In this work, we present a novel approach to understanding and mitigating experimental errors, based on a  surprisingly compact representation of a quantum circuit's information dynamics, the reactivity function.
The shape of a circuit’s reactivity function governs its noise response, and its amenability to different error mitigation methods.
From this viewpoint, we introduce two such complementary methods.
The first, Pauli-path zero-noise extrapolation, fits a parameterized model to a circuit's reactivity function and uses it to accurately extrapolate to zero noise.
The second, tunable error cancellation,
filters a circuit's reactivity function to cancel noise on all information up to a threshold locality.
Crucially, both methods feature a tuning parameter that allows one to systematically improve their accuracy until  convergence.
Across three circuit families from published experiments, our methods overcome conventional zero-noise extrapolation biases as large as 30\%; at $\sim 10^{-2}$ root-mean-square error, they reduce sampling cost relative to probabilistic error cancellation by up to a factor of $10^3$.
\end{abstract}

\maketitle

\makeatletter
\addtocontents{toc}{%
  \protect\begingroup
  \protect\let\protect\l@section\protect\@gobbletwo
  \protect\let\protect\l@subsection\protect\@gobbletwo
}
\makeatother

\begin{figure*}[t]
    \includegraphics[scale=1]{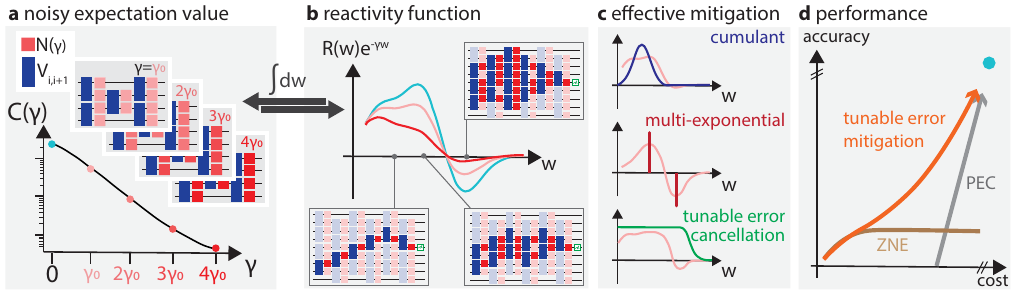}
    \caption{Schematic illustration of our results, which relate noisy quantum circuit expectation values, the reactivity function, and quantum error mitigation.
    {$\textbf{a}$} The quantum circuit expectation value, $C(\gamma)$, as a function of the noise rate $\gamma$.
    We assume local noise channels $\mathcal{N}(\gamma)$ (red) after every two qubit gate (dark blue).
    The goal of error mitigation is to infer the noiseless expectation value, $C(0)$, from noisy quantum experiments; here $\gamma_0$ denotes the native device noise rate.
    {$\textbf{b}$} The behavior of $C(\gamma)$ can be understood by decomposing the expectation value as a sum of paths of quantum information (insets).
    The reactivity function, $R(w)$, quantifies the total contribution from all paths of weight $w$ under the noise channel.
    The noisy expectation value is related by a Laplace transform, $C(\gamma) = \sum_w e^{-\gamma w} R(w)$; contributions from large $w$ decay quickly with noise, while those from small $w$ decay slowly.
    {$\textbf{c}$} Our quantum error mitigation methods build on this connection, by either fitting effective descriptions to the reactivity function (blue, red) or filtering it to cancel noise below a threshold weight (green).
    {$\textbf{d}$} Our methods are each tunable, in the sense that their accuracy can be continuously systematically improved at the expense of higher sampling cost.
    They bridge non-tunable conventional methods, such as zero-noise extrapolation (ZNE) and probabilistic error cancellation (PEC), and can yield substantial accuracy improvements and cost savings over them.
    }
    \label{fig:overview}
\end{figure*}

Extracting reliable observations from quantum processors subject to  noise remains a central challenge in quantum computing and simulation.
A seminal development in this direction was quantum error mitigation~\cite{temme2017error,li2017efficient,cai2023quantum},
a general class of methods for estimating noiseless expectation values from noisy experiments.
As quantum processors and noise-characterization methods have advanced~\cite{wallman2016noise,harper2020efficient,van2022model,van2023probabilistic,seif2024entanglement}, error mitigation has become indispensable to quantum computing and simulation experiments, enabling investigations and discoveries at scales otherwise out of reach~\cite{kim2023evidence,kim2023scalable,abanin2025constructive,zhang2025quantumcomputationmoleculargeometry,haghshenas2026digitalquantummagnetismtrappedion,qedma2025qesem,leviatan2026resolving,martiel2026sampling,barron2026observable,granet2025dilution,obrien2023purification,shtanko2025uncovering,minev2025fibonacci}.
It is also increasingly expected to remain important during the early fault-tolerant era, by suppressing residual logical errors to enable lower code distances and fewer physical qubits~\cite{piveteau2021error,lostaglio2021error,suzuki2022universal,wahl2023zero,zimboras2025myths,aharonov2025importance,dutkiewicz2025error,huggins2025flasq,zhang2025logical,eisert2025mind}.

Nevertheless, a tension has remained throughout this success.
Any mitigation method that succeeds for arbitrary quantum circuits requires a sampling cost scaling exponentially in the number of noise events per circuit~\cite{takagi2022fundamental,tsubouchi2023universal,takagi2023universal,quek2022exponentially}; a canonical example is probabilistic error cancellation (PEC)~\cite{temme2017error,endo2017practical,van2023probabilistic}.
This has motivated more efficient approximate methods, such as zero-noise extrapolation (ZNE)~\cite{temme2017error,li2017efficient,kandala2019error,giurgica2020digital,kim2023evidence}, which can be highly accurate for some circuits, yet suffer large biases for others~\cite{cai2021multi,anand2023classical}.
Unfortunately, despite an abundance of proposals and studies~\cite{cai2023quantum,huggins2021virtual,koczor2021exponential,o2021error,bonet2018low,krebsbach2022optimization,czarnik2021error,liao2024machine,filippov2023scalable}, two barriers have hindered the development and application of such methods so far.

First, we often lack a physical understanding of which properties of a circuit and observable govern the success of a mitigation method.
This leaves little intuition for when existing methods should succeed, or how better ones should be designed.
Second, approximate mitigation methods often lack a simple \emph{tuning} parameter that allows one to systematically improve their accuracy until convergence.
This contrasts with successful approximate methods elsewhere in science, such as tensor networks, where the bond dimension provides precisely such control~\cite{orus2019tensor}.
This limits the flexibility of approximate mitigation methods, and the ability to build confidence in their output and residual biases in practice.

In this work, we establish a physical connection between quantum error mitigation and a novel measure of quantum information dynamics---the reactivity function~\cite{schuster2026probing}---and build on this connection to develop several advantageous tunable mitigation methods.

As foreshadowed above, our first contribution is conceptual.
We show that the reactivity function
provides a unifying approach for understanding the success and failure of error mitigation methods.
The reactivity function sorts the exponentially many paths of quantum information that contribute to any quantum circuit and observable by their \emph{weight} in space and time~\cite{schuster2026probing}.
Different quantum circuits and observables feature different reactivity functions; these determine the circuit's response to noise, and which mitigation methods can be  successfully applied.
Crucially, we show that reactivity functions possess surprisingly compact \emph{effective} descriptions, which are indistinguishable from their exact values in any finite-resolution experiment.
We also extend this  framework to general noise models of experimental relevance, including coherent and non-unital noise.

Our second contribution builds on these observations to develop two tunable error mitigation methods.
Our methods exploit common, simplifying features of circuit reactivity functions to enable efficient and convergent mitigation.
Our first method, \emph{Pauli-path zero-noise extrapolation} (PP-ZNE)~\cite{zhang2025quantumcomputationmoleculargeometry}, fits a systematically improvable ansatz for the reactivity function to noisy experimental data, and uses it to reliably extrapolate to zero noise.
Our second method, \emph{tunable error cancellation} (TEC), uses signed linear combinations of experiments to \emph{filter}  reactivity functions, and precisely cancel noise on all information below a threshold weight. %
From an experimental perspective, both of our methods can be implemented using either amplified noise or discrete noise event insertions; we show that the latter, when possible, is strictly more powerful.
The tunability of our methods also yields unexpected unifications: Richardson ZNE and PEC, conventionally viewed as inherently disparate mitigation methods, are merely two limits of TEC.

We substantiate these results by performing extensive numerical studies of the reactivity function and our error mitigation methods across three circuit families drawn from published experiments: out-of-time-order correlators in nuclear spin dynamics~\cite{zhang2025quantumcomputationmoleculargeometry}, Stark many-body localization~\cite{Morong_2021}, and the two-dimensional transverse-field Ising model~\cite{haghshenas2026digitalquantummagnetismtrappedion}.
We find that our methods can yield multiple order-of-magnitude advantages over conventional methods in each setting.
As part of these studies, we develop a substantial toolkit towards the practical application of each of our methods, incorporating realistic experimental concerns.
We also use these studies to build physical intuition for where each method is best applied, based on reactivity function dynamics.

Our manuscript is organized as follows.
In Section~\ref{sec:theory}, we introduce the reactivity function and its connection to quantum error mitigation. %
In Section~\ref{sec:PP-ZNE}, we present our first mitigation method, Pauli-path zero-noise extrapolation.
In Section~\ref{sec:TEC}, we present our second mitigation method, tunable error cancellation.
In Section~\ref{sec:num_reactivity}, we present our numerical studies of reactivity functions in quantum many-body systems.
In Section~\ref{sec:num-mitigation}, we present our studies of the performance and comparative advantage of each of our methods, as well as  details regarding their practical implementation.
We conclude with an outlook in Section~\ref{sec:outlook}.
Further details, supporting studies, and derivations are provided in the Appendix.

\section{Quantum error mitigation and reactivity functions}\label{sec:theory}

In this section, we introduce the connection between quantum error mitigation and the reactivity function.
As aforementioned, this forms the backbone of our mitigation methods presented in the subsequent sections.
Our results also include novel findings on the reactivity function itself, which may be of interest more broadly. %

We begin by providing a brief review of quantum error mitigation and a pedagogical introduction to the reactivity function.
We then present a collection of our own results, on extending the reactivity to general noise models, effective reactivity functions, and the advantage of discrete noise insertions in quantum error mitigation.

\subsection{Quantum error mitigation}

The aim of quantum error mitigation is to estimate expectation values of noiseless quantum circuits using data from a noisy quantum device.
That is, we aim to estimate
\begin{align}
    C_\mathrm{exact} = \mathrm{Tr}[O U\rho U^\dagger],
\end{align}
where $\rho$ is an initial state, $U = \prod_{t=1}^T U_t$ is a unitary circuit of depth (i.e.~time) $T$, and $O$ is an observable.
In the absence of any error mitigation, we assume that the device instead measures a noisy expectation value,
\begin{align}
    C_\mathrm{noisy} = \mathrm{Tr}[O \mathcal{C}(\rho) ],
\end{align}
where $\mathcal{C} = \prod_{t=1}^T (\mathcal{N}_t \circ \mathcal{U}_t)$ is a noisy quantum circuit,  $\mathcal{U}_t(\cdot) \equiv U_t (\cdot) U_t^\dagger$, and $\mathcal{N}_t$ is a noise channel.
We let $N$ denote the number of qubits, and $V$ the circuit volume.

In order to perform error mitigation, one typically assumes one of two experimental capabilities~\cite{cai2023quantum}.
The first, and more common, is the ability to uniformly amplify the noise rate of the channels $\mathcal{N}_t$.
That is, we assume  each noise channel can be written in the form $\mathcal{N}_t(\gamma) = e^{\gamma \mathcal{L}_t}$, such that $\mathcal{N}_t(\gamma_0) \equiv \mathcal{N}_t$ is the device noise channel, with rate $\gamma_0$, and $\mathcal{N}_t(0) \equiv \mathcal{I}$ yields the ideal circuit, with zero noise (here, $\mathcal{I}$ is the identity channel). %
In practice, noise amplification can be achieved via randomized circuit compilation~\cite{wallman2016noise}, efficient Pauli-Lindblad device modeling~\cite{van2023probabilistic}, or probabilistic error amplification (PEA)~\cite{kim2023evidence}.
From this, one can access the noisy expectation value,
\begin{equation}\label{eq:noisy_signal}
    C(\gamma) = \mathrm{Tr}[O \mathcal{C}_\gamma(\rho)],
\end{equation}
where $\mathcal{C}_\gamma \equiv \prod_{t=1}^T (\mathcal{N}_t(\gamma) \circ \mathcal{U}_t)$, for $\gamma \geq \gamma_0$.
The goal of mitigation is then to recover $C_\mathrm{exact} \equiv C(0)$ from the noisy expectation values $C(\gamma)$ [Fig.~\ref{fig:overview}(a)].

A second, more powerful, capability is to insert discrete noise events drawn from the channel generators $\mathcal{L}_t$ into the experiment.
This capability is central to PEC and PEA.
Here, one typically assumes that the dominant noise is Pauli noise, i.e.~$\mathcal{L}_t(\rho) = - \sum_E q_t(E) (\rho - E \rho E^\dagger)$, where $E$ are Pauli operators and $q_t(E) \geq 0$ are rates.
If the rates $q_t(E)$ are known, then one can insert an individual noise event $E$ at time $t$ into the circuit, drawn from the probability distribution $p(E,t) \equiv q_t(E)/(\sum_{E,t} q_t(E))$.
This capability is more powerful than noise amplification, since one can implement amplification by drawing a mixture of noise events determined by $\gamma-\gamma_0$ (as in PEA).
In practice, discrete noise insertion requires twirling to ensure Pauli noise and precise device characterization to learn the rates $q_t(E)$~\cite{kim2023evidence}.

\subsection{Reactivity functions}\label{sec:reactivity}

We can now describe the reactivity function.
The reactivity function was introduced in Ref.~\cite{schuster2026probing}, building on ideas from earlier works~\cite{rakovszky2022dissipation,von2022operator,kechedzhi2023effective,gao2018efficient,aharonov2023polynomial}, in the context of classical algorithms for quantum circuits and many-body systems.
Our work identifies a powerful and distinct application of the reactivity function, to quantum error mitigation.

The reactivity function characterizes the `reaction' of a quantum experiment to noise~\cite{schuster2026probing}, and we will introduce it pedagogically with this motivation in mind.
To ease notation, we use the double bracket notation,
\begin{equation}
    |\rho\rrangle\equiv \rho,
    \hspace{0.4cm}
    \llangle O| \equiv \frac{1}{2^N}\mathrm{Tr}[O^\dagger (\cdot)],
    \hspace{0.4cm}
    \mathcal{U}|\rho\rrangle\equiv|U\rho U^{\dagger}\rrangle,
\end{equation}
with inner product $\llangle P|P'\rrangle = \tfrac{1}{2^N}\mathrm{Tr}[PP']$~\footnote{The factor $2^N$ in this equation makes the Pauli operators an orthonormal basis for this vector space, which is balanced by the factor of $2^N$ when considering expectation values.}.
This leads to the resolution of the identity,
\begin{equation}
     \mathcal{I}=\sum_{P}|P\rrangle \llangle P| = \sum_{P}\tfrac{1}{2^N}P\,\mathrm{Tr}[P(\cdot)].
\end{equation}
From this, we have $C(\gamma) = 2^N \llangle O |\prod_t (\mathcal{N}_t(\gamma) \circ \mathcal{U}_t) |\rho \rrangle$.

We begin with local depolarizing noise as in Ref.~\cite{schuster2026probing}, and extend to more general noise models after.
Local depolarizing noise is the simplest example of Pauli noise, setting $q_t(E) = 1/4$ for all single-qubit Pauli operators $E$.
In the double bracket notation, it has the simple action
\begin{equation}\label{eq:noise_channel_diag}
    \mathcal{N}_\mathrm{d}(\gamma) = \sum_{P} e^{-\gamma w_{P}}|P\rrangle\llangle P|.
\end{equation}
That is, it damps the amplitude of each Pauli operator $P$ exponentially in its \emph{weight}, $w_P$, which is equal to its number of non-identity components.
The Pauli operators are therefore the eigenbasis of the noise channel.

To understand the reaction of a quantum circuit to noise, one can insert a resolution of the identity in the Pauli basis at each circuit layer~\cite{gao2018efficient,aharonov2023polynomial}, which yields
\begin{align}
    C(\gamma) = & 2^N \llangle O|\prod_t \Big( \sum_{P_t}e^{-\gamma w_{P_t}}|P_t\rrangle\llangle P_t| \Big) \mathcal{U}_t|\rho \rrangle.
\end{align}
We can further simplify notation by abbreviating each sequence of Pauli strings, $\vec{P} \equiv (P_1,P_2,\ldots,P_T)$, as a \emph{Pauli path}, $\vec{P}$.
Each path's contribution to the ideal expectation value is given by $A_{\vec{P}}= 2^N \llangle O|P_T \rrangle (\prod_{t=2}^T \llangle P_t| \mathcal{U}_t |P_{t-1}\rrangle) \llangle P_1|\mathcal{U}_1|\rho\rrangle$.
In the presence of local depolarizing noise, each path is damped by an amount $e^{-\gamma w_{\vec P}}$, where $w_{\vec{P}}=\sum_t w_{P_t}$ is the \emph{path weight}.
This yields the decomposition,
\begin{align}
    C(\gamma) = & \sum_{\vec{P}}e^{-\gamma w_{\vec{P}}} A_{\vec{P}}, \label{eq:pauli_paths}
\end{align}
which is known as the Pauli path decomposition~\cite{gao2018efficient,aharonov2023polynomial}.

The Pauli path decomposition can be viewed as a Feynman path integral in operator space~\cite{aharonov2023polynomial}.
Physically, each path corresponds to a path of \emph{information} in space and time.
Quantum circuits that are dominated by large-weight paths are influenced by information stored highly non-locally at many time steps.
This leads to both a high susceptibility to decoherence, and a difficulty of simulation with many classical algorithms~\cite{aharonov2023polynomial,beguvsic2023fast,fontana2023classical,rudolph2023classical,gonzalez2024pauli,angrisani2024classically,schuster2025polynomial,abanin2025constructive,rudolph2025pauli,beguvsic2025real}.
Meanwhile, quantum circuits that are dominated by small-weight paths are influenced primarily by local information.
Pauli path dynamics probe a new type of quantum information physics compared to conventional studies of information scrambling~\cite{cotler2023information}\footnote{Fundamentally, this arises because the Pauli path coefficients, $A_{\vec P}$, weight each path of information according to its contribution to a specific expectation value of interest. Pauli path dynamics can also be efficiently measured in time-forward quantum experiments~\cite{schuster2026probing}. This contrasts with conventional scrambling diagnostics, which can only be efficiently measured in time-reversal experiments~\cite{cotler2023information,schuster2026probing}.}.
Their study has been initiated in several works~\cite{rakovszky2022dissipation,von2022operator,yoshimura2025operator,nahum2022real,schuster2026probing,abanin2025constructive}, but is still nascent compared to more conventional fields.

While the Pauli path expansion, Eq.~\eqref{eq:pauli_paths}, is the most fundamental object in Pauli path dynamics, it is also unwieldy, as it contains an exponential number of terms, $4^{NT}$.
To simplify this, one can group together all paths of the same weight $w$, and write
\begin{equation} \label{eq:laplace}
    C(\gamma) = \sum_w e^{-\gamma w} R(w),
\end{equation}
where $R(w) \equiv \sum_{\vec P} \delta_{w_{\vec P},w} A_{\vec P}$ is the reactivity function~\cite{schuster2026probing}.
The reactivity function quantifies the total contribution from all Pauli paths of weight $w$ to the ideal expectation value.
It can take negative values and is in general not normalized~\footnote{In particular, one can bound the sum of squared path amplitudes, $\sum_{\vec{P}}A_{\vec{P}}^2\leq 1$. However, the grouping of paths with the same $w$ in the reactivity function, implies that $\sum_w R(w)^2$ can be exponentially large.}.
From the above, we see that the reactivity function is the inverse Laplace transform of the noisy expectation value $C(\gamma)$ [Fig.~\ref{fig:overview}(b)].
As such, it is sufficient to fully explain the behavior of the expectation value in the presence of noise.

The relation between the noisy expectation value and the reactivity function [Eq.~\eqref{eq:laplace}] forms the basis of our work (Fig.~\ref{fig:overview}).
It connects the central object in quantum error mitigation to a physical quantity that one can investigate and obtain insight about.
Similar decompositions have been performed before in quantum error mitigation~\cite{cai2021multi,endo2017practical,kechedzhi2023effective,zhang2025quantumcomputationmoleculargeometry}; a goal of our work is to emphasize the importance of this decomposition, and to centralize the physical behavior of the reactivity function in mitigation studies.
As we will see, this perspective leads to new insights on existing mitigation methods, as well as new methods entirely.
For example, one can improve zero-noise extrapolation using  ans\"atze for the reactivity function that would not be obvious from the perspective of the noisy expectation value (Section~\ref{sec:PP-ZNE}), and lessen the cost of  error cancellation for quantum circuits where the path weight has bounded range, $w \leq w_*$ (Section~\ref{sec:TEC}).
This connection also opens the door to a fruitful interplay between error mitigation and physical investigations of quantum information dynamics in future work.

\subsection{Extension to general noise models}\label{sec:general-noise}

While our discussion of the reactivity function has thus far been restricted to local depolarizing noise, the identification of the Pauli basis as the eigenbasis of the local depolarizing channel enables a natural extension to more general noise models.
We restrict to noise channels with self-adjoint Lindblad generators $\mathcal{L}_t$, which includes Pauli noise, for the present discussion, and present a more sophisticated framework that incorporates non-unital and coherent noise models in  Appendix~\ref{app:generalized_reactivity}.

To maintain familiar notation, we let $| Q_t \rrangle$ denote the orthonormal eigenstates of each $\mathcal{L}_t$, and $-v_{Q_t}$ their real negative eigenvalues.
Hence, each noise channel can be diagonalized as before, $\mathcal{N}_t(\gamma) = \sum_{Q_t}e^{-\gamma v_{Q_t}}|Q_t\rrangle\llangle Q_t|$.
Repeating our previous derivation with this choice of basis yields the path decomposition, $C(\gamma) = \sum_{\vec Q} e^{-\gamma v_{\vec Q}} A_{\vec Q}$, where each path $\vec{Q} = (Q_1,\ldots,Q_T)$ now corresponds to a sequence of eigenoperators, and $v_{\vec Q} = \sum_t v_{Q_t}$ to their summed eigenvalues.
This yields the decomposition,
\begin{equation}
    C(\gamma) = \sum_v e^{-\gamma v} R(v),
\end{equation}
with reactivity function $R(v) \equiv \sum_{\vec{Q}} \delta_{v_{\vec{Q}},v} A_{\vec Q}$.

Unlike before, the eigenoperators $Q_t$ of the noise channels will not typically be tensor product operators, and the path weights $v$ not integer-valued.
Nevertheless, we expect that the physics of the generalized reactivity function, $R(v)$, is likely similar to that of $R(w)$ for generic local noise models.
For the former, this follows from the locality of $\mathcal{L}_t$.
In particular, the lowest eigenoperator is $| 1 \rrangle$ with eigenvalue zero, and we generically expect the next lowest eigenoperators to be $\mathcal{O}(1)$-local with eigenvalues $-\mathcal{O}(1)$, and the higher eigenoperators to be $\mathcal{O}(v)$-local with eigenvalue $-v$.
Hence, whether an experiment receives contributions primarily from high weights or low weights will be mostly unchanged, even if the precise values of the reactivity function may be different.
For the latter, the non-integer nature of $v$ is not particularly important in light of our smoothing procedure in the following subsection.

\subsection{Effective reactivity functions}\label{sec:effective}

We now introduce a key observation of our work.
We show that the reactivity function, $R(w)$, can be replaced by an \emph{effective} reactivity function, $\tilde{R}(w)$, which approximately reproduces all noisy expectation values, $C(\gamma)$, while retaining substantially less information.

This fact may appear surprising at first.
The expectation value $C(\gamma)$, when viewed as an analytic function of $\gamma$, is the Laplace transform of $R(w)$, and hence contains full information about its functional form.
Nonetheless, in practice, one can only measure $C(\gamma)$ up to a finite precision due to sampling noise.
This, combined with the ill-conditioned nature of the inverse Laplace transform, implies that many features of $R(w)$ are effectively irrelevant from the perspective of $C(\gamma)$.
The effective reactivity seeks to isolate the  relevant features.

We do so via a smoothing procedure. %
Namely, we convolve $R(w)$ with a Gaussian distribution of width $\gamma_G^{-1}$,
\begin{align}
    \tilde{R}_{\gamma_G}(w) \equiv \int_0^\infty\!\mathrm{d}w' \, \frac{\gamma_G}{\sqrt{2\pi}}e^{-\gamma_G^2(w'-w)^2/2} R(w').
\end{align}
This erases information at scales $\delta w \lesssim 1/\gamma_G$ while retaining coarse-grained features at larger scales.
The effective reactivity leads to expectation values,
\begin{align}
    \tilde{C}_{\gamma_G}(\gamma) \equiv \sum_w e^{-\gamma w} \tilde{R}_{\gamma_G}(w) = e^{\gamma^2/(2\gamma_G^2)} \, C(\gamma).
\end{align}
Hence, if one is interested in the signal at only small noise rates $\gamma$ (as is the case in quantum error mitigation),
then one can set $\gamma_G \gtrsim \gamma$,
and the effective reactivity produces a signal that differs from the original signal by only a small relative amount, $\mathcal{O}(\gamma^2/\gamma_G^2)$.

This result is a remarkable simplification.
In practice, we will be mainly interested in error-mitigating quantum circuits whose $R(w)$ have support up to weights $w_*$ that are at most a few multiples of the inverse noise rate, $w_* \gamma \lesssim 5$.
This follows because the sampling overhead of mitigation  scales exponentially in $\gamma w_*$.
For such circuits, our smoothing procedure leaves only an extremely small number, $\gamma_G w_* = \mathcal{O}(1)$, of features remaining.
Hence, the noise response of quantum circuits that is relevant for error mitigation can be shockingly compactly described.

We also note that, by the transitive property, any ansatz  for the reactivity function, $\hat{R}(w)$, which produces the same effective reactivity as the true reactivity function, $R(w)$, after smoothing, is indistinguishable from $R(w)$ for all practical purposes.
This fact, combined with the above, explains the success of our first mitigation method, Pauli-path ZNE, which uses extremely coarse physical ans\"atze for the reactivity (Fig.~\ref{fig:overview_error_mit}).

\subsection{The advantage of discrete noise insertions}\label{sec:k_error}

We conclude this section by discussing the impact of discrete noise insertions within the reactivity framework.
We focus here on local depolarizing noise, for which each discrete noise insertion is drawn uniformly at random from the single-qubit Pauli operators at all circuit locations. %
An identical analysis applies to general Pauli noise models, for which each insertion is drawn from $p(E,t)$ (see Appendix~\ref{app:discrete_insertion}).

Consider a quantum circuit with device noise rate $\gamma_0$, and $k$ independent discrete noise events inserted.
Averaging over the $\circV$ choices for each noise event,
one finds the resulting expectation value (see Appendix~\ref{app:discrete_insertion})~\cite{schuster2026probing},
\begin{equation}
    C(\gamma_0,k)=\sum_w e^{-\gamma_0 w}\left(1-\frac{4w}{3\circV}\right)^kR(w).
    \label{eq:discrete_noise}
\end{equation}
Hence, the reaction to discrete noise insertions can also be  understood via the reactivity function. %

As aforementioned, the ability to insert discrete noise events is strictly more powerful than the ability to continuously amplify the noise rate.
Intuitively, the damping associated with discrete noise insertions, $(1-4w/3\circV)^k = e^{k\ln(1-4w/3\circV)}$, is better able to discern different weights $w$ than the damping associated with noise amplification, $e^{-\gamma w}$, due to the concavity of the logarithm~\footnote{As the simplest possible example, consider the difficulty of distinguishing a quantum circuit whose reactivity has support entirely at weight $n/2$, $R(w) = \delta_{w,n/2}$ from one with support at weight $n$, $R(w) = \delta_{w,n}$.
Amplified noise produces signals $e^{-\gamma n/2}$ and $e^{-\gamma n}$, respectively, whose difference is maximized at $e^{-\gamma_* n} = 1/4$ with value $1/2-1/4=1/4$.
Meanwhile, the insertion of a single discrete noise event (i.e.~$k=1$) produces signals $1/2$ and $0$, respectively, whose difference is $1/2 > 1/4$.
While this example produces only a twofold advantage for discrete noise events, one finds that in general this advantage can be exponentially large (see Section~\ref{sec:TEC} and our numerical studies for examples).}.
This is particularly true at large weights, when $w$ is an appreciable fraction of the circuit volume $\circV$.

To see that the inequality between the two abilities is strict, we simply note that noise amplification data can always be obtained from discrete noise insertion data by taking a probabilistic mixture. %
Indeed, this forms the basis of PEA.
To see this explicitly, define the mean number of noise insertions at a noise rate $\gamma$ as $K_\gamma=3\circV(\gamma-\gamma_0)/4$. We then have,
\begin{equation}\label{eq:poisson_main}
    e^{-(\gamma-\gamma_0)w}
    =
    \sum_{k=0}^{\infty}
    e^{-K_\gamma}
    \frac{(K_\gamma)^k}{k!}
    \left(1-\frac{4w}{3\circV}\right)^k.
\end{equation}
Hence, amplifying noise from $\gamma_0$ to $\gamma$ is equivalent to inserting a Poisson-distributed number of noise events, $p^\gamma_k = e^{-K_\gamma} (K_\gamma)^k / k!$.
A reverse translation is not possible without significant sampling overhead---an indication that there is significant information loss when converting from discrete noise insertions to noise amplification.

For these reasons, whenever one has the ability to perform discrete noise insertions, one should perform mitigation using them directly, rather than converting them to amplifications.
We demonstrate this advantage for our  mitigation methods in our numerical studies.

\section{Pauli-path zero noise extrapolation}\label{sec:PP-ZNE}

In this section, we introduce our first error mitigation method, \emph{Pauli-path zero-noise extrapolation}.
Our method encompasses a general approach to error mitigation, in which one begins with a prior ansatz for the functional form of the reactivity function of a circuit of interest, fits this form to noisy experimental data, and then uses this fit to extrapolate the ideal signal at zero noise rate.
Crucially, this method builds on our observation in the previous section, that \emph{effective} reactivity functions, containing only a small number of smoothed features, can yield equivalent descriptions of noisy quantum circuit expectation values to exact reactivity functions.
This enables functional ans\"atze with shockingly few fitting parameters to yield accurate predictions for the ideal signal, and underlies the efficiency of our method.

We implement Pauli-path ZNE by introducing two explicit physics-inspired ans\"atze for the reactivity function.
Our first ansatz corresponds to a cumulant expansion of the reactivity function, and corresponds to a polynomial fit to the logarithm of the noisy expectation value decay.
Our second ansatz corresponds to an expansion of the reactivity function as a small discrete number of delta functions, and corresponds to a multi-exponential fit to the noisy expectation value decay.
A key feature of both ans\"atze is that they feature a systematic tuning knob: the degree of the polynomial fit for the cumulant expansion, and the number of discrete delta functions for the multi-exponential fit.
This increases the expressibility of each ansatz, and allows one to systematically check their accuracy by observing convergence as the tuning knob is increased.
We find in practice that both methods can yield accurate convergence and large efficiency gains with as few as 2 to 8 fitting parameters.

We remark that, in some sense, the functional fits that we perform could have been performed without any knowledge of the reactivity function or information dynamics.
Indeed, the multi-exponential fit we consider was previously studied in Ref.~\cite{cai2021multi}.
In this case, the fits would be physically unmotivated, yet could nonetheless be applied and observed to converge on data.
From this perspective, one main contribution of ours is in physical interpretation.
This informs the conception of new fitting functions (such as our cumulant expansion) and of the physical settings to which each should be applied.
The connection to the reactivity also yields distinct benefits: for example, one can now apply the same functional fits to experimental data from discrete noise insertions, which can yield substantial efficiency gains.

\begin{figure}
    \centering
    \includegraphics[scale=1]{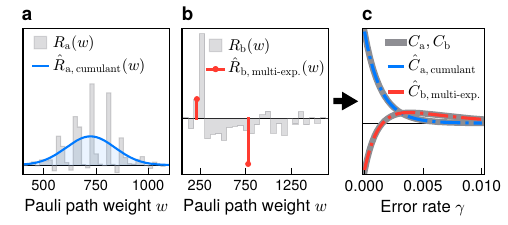}
    \caption{
    Reactivity $R(w)$ examples, different effective descriptions $\hat{R}(w)$ and the resulting noisy signal decay curve $C(\gamma)$, Eq.~\eqref{eq:laplace}. \textbf{a} Example 1: $R(w)$ is predominantly positive yielding a good agreement between $C(\gamma)$ (c, gray) and the fit using the cumulant expansion (c, blue). \textbf{b} Example 2: $R(w)$ splits into a negative and a positive part so that the multi-exponential ansatz fit with two exponentials (c red) yields good agreement with $C(\gamma)$ (c gray). \textbf{c} Comparison between noisy signal $C(\gamma)$ and fits corresponding to the effective reactivity descriptions shown in a and b. %
    }
    \label{fig:overview_error_mit}
\end{figure}

To introduce our method,  we first present a general perspective on error mitigation methods that fit a functional form to the noisy expectation value decay.
We then introduce our two physical ans\"atze and demonstrate Pauli-path ZNE.

\subsection{Error mitigation via reactivity function fitting}

We consider any quantum error mitigation protocol in which the estimate for the ideal expectation value is obtained from a functional fit to experimental data, using either amplified noise rates or discrete noise insertions.
We recall that any noisy quantum expectation value can be written as
\begin{equation}
    C(\gamma_0,r) = \sum_w e^{-\gamma_0 r w} R(w)
\end{equation}
for noise amplified by factor $r$,
and
\begin{equation}\label{eq:disc_noise_insertion}
    C(\gamma_0,k) = \sum_w e^{-\gamma_0 w} \left(1-\frac{4w}{3\circV}\right)^k R(w)
\end{equation}
for $k$ inserted discrete noise events. This relation between reactivity and noisy signal is used to fit reactivity ans\"atze for both techniques, the technically simpler noise amplification or the more efficient discrete noise insertion. %

As discussed in the previous section, there is significant information loss when converting from $R(w)$ to expectation values.
Hence, in many cases, $R(w)$ can be replaced with much simpler effective reactivity functions, $\tilde{R}(w)$.

The central choice of any fitting-based error mitigation method is the choice of function to fit.
From the above, we see that any fitting-based error mitigation method is, in essence, equivalent to an ansatz for the functional form of $R(w)$ (or $\tilde{R}(w)$).

The simplest example is exponential zero-noise extrapolation (ZNE).
Here, one assumes an expectation value follows an exponential decay, $C(\gamma) \sim \! e^{-\gamma W}$, with the noise rate $\gamma$, and seeks to estimate the decay rate $W$ from data on amplified noise rates.
This assumption is intuitive, as many simple systems feature exponential decays in the presence of noise.
However, we see that it implies a rather strong assumption on the reactivity function: it must be effectively described by a delta function, $\tilde{R}(w) \approx \delta_{w,W}$.
This can be true, for example in Clifford circuits, OTOCs in short-range geometrically-local systems at late times~\cite{mi2021information,schuster2023operator,abanin2025constructive}, or for systems with integrable or dual unitary structure. %
However, it does not appear to be true in general (see Fig.~\ref{fig:overview_error_mit}).
Hence, outside of specific cases, one should not expect a priori that exponential ZNE produces any reliable accuracy.
Other examples of reactivity functions, and their associated expectation value decay, are described in the following sections.

Before proceeding to our methods, we illustrate the difficulty with two examples.
The first comes from quantum simulations of molecular out-of-time-ordered correlators (OTOCs), used to infer molecular geometry from NMR experiments~\cite{zhang2025quantumcomputationmoleculargeometry}.
These circuits contain hundreds of entangling gates, making probabilistic error cancellation prohibitively expensive.
Both single-exponential ZNE and Loschmidt echo rescaling proved insufficient experimentally, yielding RMSEs of order $0.2$ against a target of $0.04$.
The reactivity reveals why: contributing Pauli paths span a broad range of accumulated noise weights, even after fine structure is smoothed away [Fig.~\ref{fig:nmr_otoc_numerical_examples}(c,d)].
Increasing the noise preferentially suppresses higher-weight contributions, changing the apparent decay rate and biasing a single-exponential fit.
Echo rescaling instead assumes that the reference echo and OTOC share the same relative noise response, which the butterfly perturbation changes by altering the contributing paths.
Neither mismatch can be removed by additional measurements.
Accounting for finite reactivity width through PP-ZNE enabled substantially more accurate mitigation, reaching RMSE as low as $0.018\pm0.010$ in a DMBP benchmark.
Below, we develop systematic extensions of this experimentally demonstrated approach.

The second example concerns models for which the signal originates from distinct processes. The information dynamics of each process can be different, yielding distinct contributions to the reactivity. An example is shown in Fig.~\ref{fig:overview_error_mit} b, with a large, positive contribution at $w=250$ and negative contributions for $250 \leq w \leq 800$. The observation of distinct contributions motivates an effective reactivity description as a sum of $\delta$-peaks, which yields the multi-exponential ansatz discussed in Sec.~\ref{sec:multi_exp_intro}. In fact, such an ansatz is more general as features on scales $\Delta w \ll 1/\gamma$ cannot be resolved so that a small number of $\delta$-peaks can effectively describe any reactivity function.

\subsection{Cumulant Expansion}\label{sec:cumulant}

We showed above that exponential zero-noise extrapolation is equivalent to fitting the reactivity with a single $\delta$ function and argued it to be insufficient for more generic cases. Hence, we systematically extend the $\delta$-function ansatz using the cumulant expansion of the reactivity $e^{-\gamma_0 w} R(w)$. It predicts a noisy signal of the form
\begin{align}
	\log\!\left(C(\gamma)\right)=\sum_{m= 0}^M \frac{(\gamma_0-\gamma)^m}{m!}\,\kappa_{m}, \label{eq:cumulant_exp}
\end{align}
where $\kappa_{m}$ are the cumulants of $e^{-\gamma_0 w} R(w)$. We fix $\gamma_0$ in advance to avoid singularities of $\log\!\left(C(\gamma)\right)$ for unphysical $\gamma<0$; any shift in $\gamma_0$ can be absorbed into the $\kappa$ values. The first three cumulants follow from Eqs.~\eqref{eq:laplace},~\eqref{eq:cumulant_exp}
\begin{align}
    \kappa_0  &= \log\left(\sum_w e^{-\gamma_0 w}  R(w)\right), \; \kappa_1 = \sum_w w e^{-\gamma_0 w}  R(w) e^{-\kappa_0} \nonumber \\
	\kappa_2 & = \sum_w w^2 e^{-\gamma_0 w}  R(w)e^{-\kappa_0}-\kappa_1^2.
\end{align}
We allow $\kappa_{m>0}$ to be real and $\kappa_{0}$ to be complex, which would not be allowed for either $\kappa_{0}$ or $\kappa_{2}$ if $R(w)$ were a real probability distribution.

The cumulant expansion is well suited for reactivities which effectively correspond to unnormalized, localized probability distributions. Such phenomenology has been observed in, e.g., simulated 1D Hamiltonian evolution with no conserved quantities: here the size distribution of an evolving Pauli operator is Gaussian~\cite{schuster2023operator}, which suggests similar results for reactivities. Moreover, if the reactivity is tightly peaked and $(\gamma-\gamma_0)^m\kappa_m/m! \ll 1$ for $m\geq 2$, $C(\gamma) \approx e^{\kappa_0 + (\gamma_0-\gamma) \kappa_1}$ is a sufficiently accurate model, and the cumulant expansion reduces to zero-noise extrapolation.
More generally, if the reactivity is peaked but with finite width, %
we expect Eq.~\eqref{eq:cumulant_exp} to converge well and the cumulants of the function to decay quickly; if $\kappa_m=0$ for $m\geq 3$, the functional form corresponds to a Gaussian.
In App.~\ref{app:cumulant_expansion_examples} we show the cumulant expansions for various common distributions and discuss when they converge.
The systematic bias of single-exponential extrapolation for a finite-width reactivity is analyzed in App.~\ref{app:cumulant_bias}.

To implement this method in practice, we obtain the cumulants $\hat{\kappa}_m$ by fitting either $C(\gamma)$ using Eq.~\eqref{eq:cumulant_exp} at noise rates accessible in experiment $\gamma_0 \leq \gamma \leq \gamma_\mathrm{max}$ or for discrete noise insertions we fit the prediction Eq.~\eqref{eq:disc_noise_insertion} with the ansatz $\hat{R}_{M=1}(w)= e^{\hat{\kappa}_0 + \gamma_0 \hat{\kappa}_1} \delta_{w,\hat{\kappa}_1}$ or $\hat{R}_{M=2}(w) = \exp( \sum_{m = 0}^2 \hat{c}_m w^m )$, see also App.~\ref{sec:app_pp_zne_otoc}. Note that the estimates $\hat{\kappa}_m$ may differ from $\kappa_m$ as they are obtained via fitting $\log(C(\gamma))$ for noise rates $\gamma_0 \leq \gamma \leq \gamma_\mathrm{max}$ (or a finite number of discrete noise insertion events), so that they also incorporate the impact of terms with $m>M$.
This can be understood as $\hat{\kappa}_m$ corresponding to an effective
reactivity $\hat{R}(w)$, with features of $R(w)$ that are unresolved over the
accessible noise interval $\gamma_0\leq\gamma\leq\gamma_{\max}$ removed.

In practice, the most suitable order $M$ for the error mitigation task at hand may not be known.
Instead, we can fit multiple orders to the taken data $C(\gamma_r)$ and combine the fitting results based on their uncertainty~\cite{giles2007a} or goodness-of-fit using the Akaike information criterion (AIC)~\cite{akaike_new_1974,burnham2002model,Symonds2010ABG} to obtain a refined estimate $\hat{C}(0)$.
A detailed discussion is provided in App.~\ref{sec:optimal_combination_estimates}.
We find in practice that for a small number of data points $n_r$, a telescopic sum approach works best, where we add the next order with weight
\begin{align}
    \hat{C}_{\leq M}(0) = \hat{C}_{\leq M-1}(0) + \mathrm{max}\left(0,1 - \sigma_{\hat{D}_M}^2/\hat{D}_M^2\right)\hat{D}_M, \label{eq:telesc_sum}
\end{align}
where $\hat{D}_M=\hat{C}_{M}(0)-\hat{C}_{M-1}(0)$ and $\sigma_{\hat{D}_M}^2$ is its estimated variance.
This weighting minimizes the MSE of $\hat{C}_{\leq M}(0)$ compared to its zero variance value, see App.~\ref{sec:optimization_weighting}.
For a larger number of datapoints $n_r$, we find the AIC to work better, which assigns Akaike weights $a_M$ to all models $M$
\begin{align}
    a_M \propto \exp\left(\chi^2-\frac{n_r d_M}{n_r-d_M-1}\right), \label{eq:akaike_weight}
\end{align}
where $\chi^2$ is the log-likelihood of the fitting result, $d_M$ the number of fitting parameters of the model and $a_M$ are normalized so that $\sum_M a_M=1$, as explained in detail in App.~\ref{sec:optimization_aic}.

\subsection{Multi-Exponential Model}\label{sec:multi_exp_intro}
An alternative model for the reactivity is to consider a tightly peaked function around a few $w_m$; in an extreme case this corresponds to a sequence of $M$ delta distributions of different weights
\begin{align}\label{eq:multi-exponent}
    \hat{R}(w) = \sum_{m = 1}^{M} \delta_{w,\hat{w}_m} \hat{R}_{m}.
\end{align}
This corresponds to fitting the noisy signal $C(\gamma)$ to a sum of exponentially decaying signals, as was studied in Ref.~\cite{cai2021multi}.
Our work provides extensions of and alternative theoretical justifications for such a model.

The accuracy of the above model relies on the number of exponentials $M$ needed to describe the reactivity; in principle as $M\rightarrow\infty$ this becomes exact (but with a large variance cost to fit the model).
However, this is hardly necessary, as one can only resolve two $\delta$ functions separated by $\Delta w$ when $\Delta w \gtrsim 1/\gamma$.
Hence, the number of needed exponentials scales with noise as $M\sim 1/\Delta w \sim \gamma$.
In App.~\ref{app:details_multi_exp} we formalize this, and show that under reasonable assumptions, the discretization error with an equidistant spacing $\Delta w$ scales to lowest order in $\gamma \Delta w$ as $|\hat{C}_M(\gamma)-C(\gamma)| \leq |\gamma \Delta w C(\gamma)| + \mathcal{O}(\gamma^2 \Delta w^2)$ for sufficiently large $M$.
In the case where the maximal distance $\max_{m,m'}|w_m-w_{m'}|\ll 1/\gamma$, a single exponential is sufficient, which again reduces to ZNE.

As was true for the cumulant case, it is non-trivial to fit Eq.~\eqref{eq:multi-exponent} to the experimentally obtained data $C(\gamma,k)$ through the Laplace or $k$-error transform.
One wishes to compare the predicted noisy signal
\begin{align}
    \hat{C}_M(\gamma,k) =  \sum_{m = 1}^{M} \hat{R}_m  e^{-\gamma \hat{w}_m} \left(1-\frac{4\hat{w}_m}{3\circV}\right)^k\label{eq:sum_exp_ansatz}
\end{align}
to $C(\gamma,k)$ and fit the parameters $\hat{w}_m$ and $\hat{R}_m$, however this is a non-linear fitting problem.
There are various approaches for fitting a multi-exponential function, such as the Prony, LASSO or the Matrix Pencil method.
However, in practice we find these methods unstable.
Instead, we take a two-step approach: we first discretize $w$ and then, for every possible combination of $\hat{w}_m$ up to order $M$, we fit the $\hat{R}_m$ using least-squares, and then select those $\hat{w}_m$ which yielded the best fit.
(See App.~\ref{app:details_multi_exp} for further details and comparison.)
We have found this approach to be the most stable without being subject to overfitting for the presented case studies; in other cases other fitting methods might be more suitable.

The cost of this error mitigation scheme is expected to scale with the fidelity squared. In App.~\ref{ssec:cost_estimate_sum_exp}, we discuss fitting two exponentials $M=2$ assuming two decay rates $w_2>w_1$ using a least squares optimization. In this scenario, it appears to be best to measure at two noise rates with $\gamma_\mathrm{max} -\gamma_0 \sim 1/w_2$  yielding a sampling overhead of
\begin{align}
    N_\mathrm{samples}\sim e^{2 \gamma_0 w_2}(1+\mathrm{const} \,\gamma_0 w_2).
\end{align}
The optimal choice $\gamma_\mathrm{max} \approx \gamma_0 + 1/w_2$ corresponding to an additional damping of the signal of $\mathcal{O}(1)$ provides guidance for the range of $\gamma$ used for error mitigation.

\section{Tunable error cancellation} \label{sec:TEC}

In this section, we introduce our second error mitigation method: \emph{tunable error cancellation} (TEC).
Compared to our first method, we expect tunable error cancellation to be more generally applicable, as its success is tied to less restrictive assumptions on the reactivity function of the system of interest.
On the other hand, for the same reason, it can be much less efficient than our first method in settings where physical priors about the reactivity function are strong.

The name of our method is inspired by probabilistic error cancellation (PEC), which uses large signed linear combinations of quantum experiments to precisely cancel the effects of noise~\cite{temme2017error}.
Our method uses smaller signed linear combinations to cancel the effects of noise on all information up to a threshold path weight $w_*$.
The motivation for this feature comes from a broad range of studies, which suggest that many quantum circuits and dynamics are dominated by information of weight much smaller than their full circuit or light-cone size~\cite{white2018quantum,rakovszky2022dissipation,rudolph2025pauli,angrisani2024classically,schuster2026probing,nahum2022real}.
Crucially, this remains true even if one considers approximate light-cone notions, such as the shaded light-cone~\cite{eddins2024lightcone} or active or effective circuit volume~\cite{aharonov2025importance,kechedzhi2023effective}.
Intuitively, every gate in the active volume may be acted upon by some important Pauli path(s).
However, each important Pauli path may act on only a small fraction of the total active gates.
Our method exploits this behavior to achieve more efficient mitigation~\footnote{At the same time, our method also benefits from being applied in a smaller active volume, and hence can naturally combine with existing volume-reduction approaches.}.

We remark that classical simulation algorithms can exploit similar simplifications~\cite{schuster2025polynomial}.
However, their runtime in general scales exponentially in the path weight or maximum Pauli weight, $e^{\mathcal{O}(w_*)}$, while the sampling overhead of our mitigation method scales exponentially in the path weight multiplied by the noise rate, $e^{\mathcal{O}(\gamma w_*)}$.
Hence, the overhead of  classical simulation compared to error mitigation grows exponentially in the inverse noise rate $1/\gamma$.

To introduce our method,  we first present a general framework for analyzing  error mitigation methods in terms of \emph{filter functions}~\cite{schuster2026probing}.
The filter function, $h(w)$, characterizes how much a Pauli path of weight $w$ is amplified by a given  method.
We then show how two paradigmatic methods---PEC and Richardson zero-noise extrapolation (RZNE)---are naturally viewed within the filter function framework.
Finally, we introduce our method, TEC, and show that it continuously interpolates between RZNE and PEC as the threshold weight $w_*$ is increased.

\begin{figure}[t]
  \centering
  \includegraphics[width=\columnwidth]{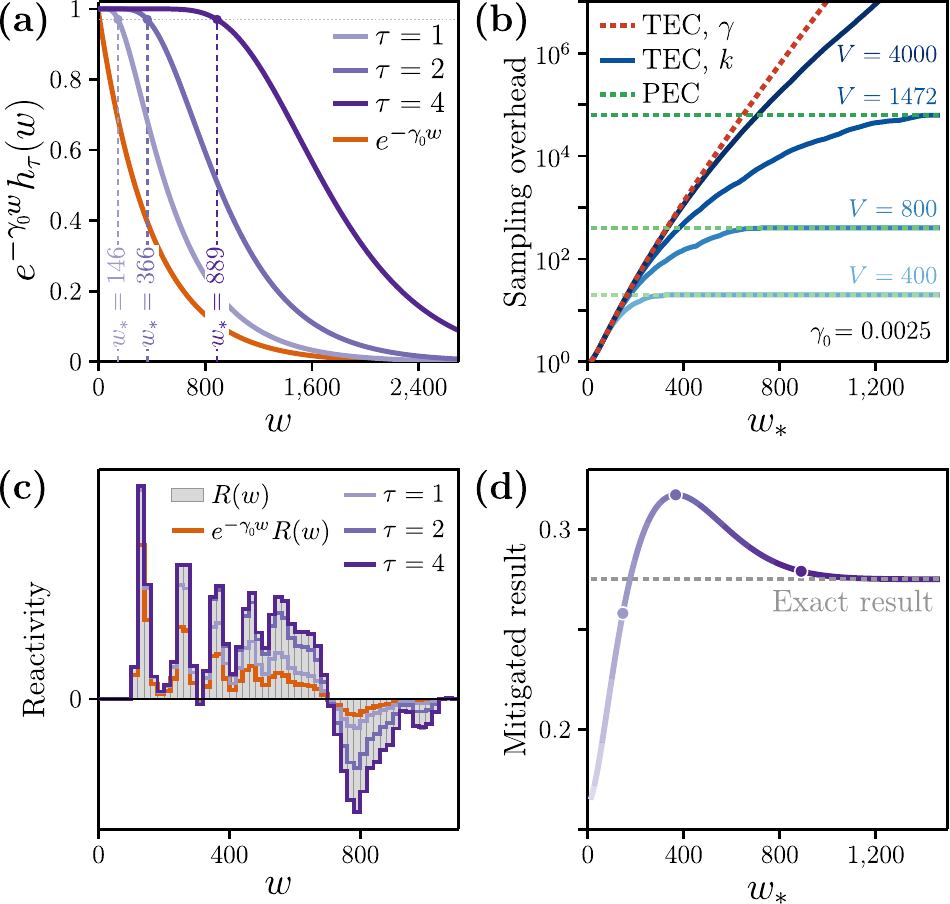}
  \caption{Illustration of tunable error cancellation (TEC).
  \textbf{(a)} The relative damping of a Pauli path of weight $w$ compared to its noiseless value, \(e^{-\gamma_0 w}h_{\tau}(w)\), for three values of the tuning parameter \(\tau\) at
  \(\gamma_0=0.0025\). Dashed lines
  indicate the threshold weight \(w_*\) such that
  \(|1-e^{-\gamma_0 w_*}h_{\tau}(w_*)|\leq 0.03\).
  Orange lines show relative damping, $e^{-\gamma_0 w}$, before mitigation.
  \textbf{(b)} Sampling overhead \(X(w_*)\) of
  TEC with noise amplification (red),
  TEC with discrete noise events (blue),
  and PEC (green).
  The first is independent of the circuit size $V$.
  \textbf{(c)} Reactivity function of the TFIM ($4\times 4$, $t = 23$, $V = 1472$), for the exact circuit (gray histogram), the noisy circuit (orange), and after TEC (purple).
  \textbf{(d)} The mitigated expectation value (purple), $\sum_w e^{-\gamma_0 w} h_\tau(w) R(w)$, converges to the noiseless expectation value (gray), $C(0) = \sum_w R(w)$, as $w_*$ is increased.}
  \label{fig:TEC}
\end{figure}

\subsection{Error mitigation as information filtering} \label{sec:info-filter}

Our framework applies to any quantum error mitigation protocol in which the estimate for the ideal expectation value is obtained as a linear combination of noisy expectation values.
This linear combination can consist of experiments with noise amplification by a rate $r$,
\begin{equation}\label{eq:linear_combination}
    \hat C(0)=\sum_{r\in\mathcal R}h_r C(\gamma_0,r),
\end{equation}
or of experiments with $k$ discrete noise insertions,
\begin{equation}\label{eq:linear_combination_k}
    \hat C(0)=\sum_{k\in\mathcal K}h_k C(\gamma_0,k).
\end{equation}
Here, $\mathcal{R}$, $\mathcal{K}$ are the amplifications or noise event numbers considered.
In this framework, the central question of  mitigation  is how to choose the coefficients $h_r$ or $h_k$.

Using the reactivity function, both estimators above can be written in the extremely simple form
\begin{equation}\label{eq:filter_main}
    \hat C(0)
    =
    \sum_w e^{-\gamma_0w}h(w)R(w),
\end{equation}
where we define the \emph{filter functions}~\cite{schuster2026probing},
\begin{align}\label{eq:filter_function_main}
    h(w)
    & =
    \sum_{r\in\mathcal R}h_r e^{-\gamma_0(r-1)w}, \\
    h(w)
    & =
    \sum_{k\in\mathcal K}h_k
    \left(1-\frac{4w}{3\circV}\right)^k.
\end{align}
We see that the total effect of  mitigation is to rescale the path contributions of the noisy experiment by the filter function $h(w)$.
Equivalently, compared to the ideal  experiment, the path contributions are rescaled by $e^{-\gamma_0 w}h(w)$.
The bias of the mitigation is therefore
\begin{equation}\label{eq:filter_bias_main}
    C(0)-\hat C(0)
    =
    \sum_w
    \left[1-e^{-\gamma_0w}h(w)\right]R(w).
\end{equation}
Hence, in order to construct as accurate an estimate as possible, one should ensure that $h(w)\approx e^{\gamma_0 w}$ on all values of $w$ where $R(w)$ is non-negligible.

A second consideration is the sample complexity of resolving the estimate $\hat{C}(0)$ to a small error $\varepsilon$.
Let us write the sample complexity as $X/\varepsilon^2$, where $X$ is the sampling overhead due to error mitigation.
The optimal overhead is easily found to be $X=(\sum_r|h_r|)^2$, obtained by allocating a fraction $|h_r| / \sum_r |h_r|$ of experimental shots to the measurement of each  $r$.
An identical analogous statement holds for discrete noise insertions.
In practice, the two goals of achieving small bias and moderate sampling overhead are in tension with one another.
Accurately reconstructing $e^{\gamma_0 w}$ generally requires large, oscillating coefficients, while constructing a sample-efficient estimator requires coefficients of moderate size.
The aim of TEC will be to achieve a minimal trade-off between these considerations, by seeking to invert noise only over the path weights that contribute non-negligibly to a given circuit.

Before proceeding, we remark that any filter function constructed with noise amplification, $h(w)=\sum_r h_r e^{-\gamma_0 (r-1)w}$, can also be constructed with discrete noise insertions, $h(w)=\sum_k\widetilde h_k(1-4w/3\circV)^k$, with coefficients $\widetilde h_k=\sum_r h_r p^{\gamma_r}_k$.
Moreover, because the probabilities $p^\gamma_k$ are positive and normalized, this translation can only decrease the sampling overhead, $\sum_k|\widetilde h_k|\leq \sum_r |h_r| \sum_k p^{\gamma_r}_k = \sum_r|h_r|$.
This decrease can be significant, as signed oscillations in nearby values of $h_r$ can cancel to yield much smaller values of $\tilde{h}_k$.

\subsection{Probabilistic error cancellation}

Let us now show how existing  error mitigation methods appear within our framework.
We begin with probabilistic error cancellation (PEC)~\cite{temme2017error,van2023probabilistic}.
Conventionally, PEC is viewed as a product of individual noise cancellations performed on every circuit gate.
Here, we will see that it can also be viewed as the ideal limit of our framework, in which the filter function precisely inverts the effects of physical noise at every weight $w$, $h_{\mathrm{PEC}}(w)=e^{\gamma_0 w}$.

To see this, we define $K\equiv3\gamma_0 \circV/4$, and simply rewrite the exponential  function and Taylor expand,
\begin{align}
    e^{\gamma_0 w}
    &=
    e^{K}
    e^{-K(1-\frac{4w}{3\circV})} =
    e^{K}
    \sum_{k=0}^{\infty}
    \frac{(-K)^k}{k!}
    \left(1-\frac{4w}{3\circV}\right)^k.
    \label{eq:pec_filter_main}
\end{align}
Thus, PEC corresponds precisely to the filter coefficients $h_k^{\mathrm{PEC}}=e^{K}(-K)^k/k!$.
The sampling overhead is
\begin{equation}\label{eq:pec_cost_main}
     X_{\mathrm{PEC}}
    =
    \Big(\sum_k|h_k^{\mathrm{PEC}}|\Big)^2
    =
    e^{4K}
    =
    e^{3\gamma_0 \circV},
\end{equation}
which is the standard PEC result~\cite{temme2017error,van2023probabilistic}.
We extend this derivation to general Pauli noise models in the appendix.

PEC  expends substantial sample resources to invert noise across the entire domain of Pauli path weights.
This makes the estimator unbiased for an arbitrary reactivity function, but requires paying substantial cost to mitigate the largest possible weights regardless of whether a particular experiment has significant reactivity there.

\subsection{Richardson zero-noise extrapolation}

We next turn to Richardson zero-noise extrapolation (RZNE)~\cite{temme2017error}.
Conventionally, RZNE is viewed as a procedure in which experimental data at large noise rates are extrapolated to recover the ideal result at zero noise.
Here, we show that, within our framework, RZNE from $m+1$ noise rates corresponds to a filter function, $h_{\text{RZNE}}(w)$, that  inverts the effects of noise  up to $m$-th order in $\gamma_0 w$.

Let $\mathcal R=\{r_0,\ldots,r_m\}$ denote a set of $m+1$ noise-amplification ratios, with $r_0=1$ corresponding to no amplification.
RZNE selects the coefficients $h_j$ such that~\cite{temme2017error}
\begin{equation}\label{eq:richardson_conditions_main}
    \sum_{j=0}^{m}h_j=1,
    \qquad
    \sum_{j=0}^{m}h_j r_j^\ell=0,
    \qquad
    \ell=1,\ldots,m,
\end{equation}
which can be solved to yield the coefficients $h_i=\prod_{j\neq i}r_j/(r_j-r_i)$.
The bias of RZNE is~\cite{temme2017error}
\begin{equation}\label{eq:richardson_exact_error_main}
    C(0)-\hat C(0)
    =
    \frac{(-\gamma_0)^{m+1} \prod_{j=0}^{m}r_j}{(m+1)!}
    E^{(m+1)},
\end{equation}
where $E^{(m+1)} \equiv C^{(m+1)}(\xi\gamma_0)$ is the $(m+1)$-th derivative of the expectation value with respect to the noise rate, at some $\xi\in[0,r_m]$.
For a fixed sampling overhead, the bias is minimized by choosing $r_j$ according to the tilted-Chebyshev ratios~\cite{krebsbach2022optimization}, $r_j
    =
    1+
    \Delta \sin^2\big(\frac{j\pi}{2(m+1)}\big)/\sin^2\big(\frac{\pi}{2(m+1)}\big)$,
where $\Delta$ controls the spacing between adjacent $r_j$.

In the filter function framework, the conditions,  Eq.~\eqref{eq:richardson_conditions_main}, correspond to setting the first $m$ derivatives of the filter function, $h_{\text{RZNE}}(w)=\sum_j h_j e^{-\gamma_0(r_j-1)w}$, equal to those of $e^{\gamma_0 w}$.
This ensures that $e^{-\gamma_0 w} h_{\text{RZNE}}(w) \approx 1$ up to  $(m+1)$-th order in $\gamma_0 w$.
The leading-order correction is given by,
\begin{equation}\label{eq:richardson_error_main}
    e^{-\gamma_0 w}h_{\text{RZNE}}(w)
    \approx
    1
    -
    \frac{\prod_j r_j}{(m+1)!}
    (\gamma_0 w)^{m+1},
\end{equation}
which is exactly the prefactor of the bias, Eq.~\eqref{eq:richardson_exact_error_main}.
Thus, RZNE recovers low-weight information with high accuracy, and becomes inaccurate only once the weight becomes sufficiently large, $\gamma_0 w_* \gtrsim (m+1) / (\prod_j r_j)^{\frac{1}{m+1}}$.
The tilted-Chebyshev ratios maximize the range of accurately mitigated weights for a fixed sampling overhead.

Interestingly, Richardson extrapolation already features a systematic tuning parameter, in the  noise-rate spacing $\Delta$.
Smaller $\Delta$ yields accuracy up to larger weights, while also requiring larger sampling overhead.
In the following, we build on this observation to construct our method of tunable error cancellation.

\subsection{Tunable error cancellation}

We now present our method of tunable error cancellation (TEC).
We begin with an analytic approach, and present a more optimized numerical approach after.
Our analytic approach corresponds to a particular choice of filter function, $h_\tau(w)$, parameterized by a tuning parameter $\tau$.
The choice of $\tau$ determines the threshold weight $w_*(\tau)$, below which $e^{-\gamma_0 w}h(w)\approx1$.
Our method can be implemented using either amplified noise rates or discrete noise insertions.
The two choices are equivalent for small  weights, $w_* \ll \circV$, while there are substantial efficiency gains using discrete noise insertions once $w_* = \mathcal{O}(\circV)$~\footnote{We expect this to be a general feature, beyond our specific method. To mitigate information at a small weight $w \ll \circV$, one will predominantly use a number of inserted noise events $k \approx \circV/w$, so that the impact of the inserted noise on the weight-$w$ information is appreciable. However, for a small weight Pauli path, the effect of such a large number of noise insertions is indistinguishable from amplified noise of rate $\gamma \approx k/\circV$. Hence, the two abilities should behave similarly when $w \ll \circV$.}.

We begin with amplified noise rates.
Here, TEC is  obtained as an unconventional limit, $m \rightarrow \infty$, of RZNE using the tilted-Chebyshev spacing. %
We find that, for this spacing, the $m\rightarrow \infty$ limit is convergent and advantageous over finite $m$.
This limit yields amplifications, $r_j
    =
    1+(\pi j/2\tau)^2$,
where $\tau \equiv \pi /2\sqrt{\Delta}$,
and coefficients,
\begin{equation}\label{eq:tec_main}
    h_0= \sinh(2\tau)/2\tau, \quad
    h_j=(-1)^j 2h_0/r_j,
\end{equation}
which alternate in sign and decay asymptotically as $|h_j|=\mathcal{O}(j^{-2})$.
This yields the filter function, $h_\tau(w) = \sum_j h_j e^{-\gamma_0(r_j-1)w}$,
which is plotted in Fig.~\ref{fig:TEC}(a).

To gain intuition for $h_\tau(w)$,  consider the product, $e^{-\gamma_0 w}h_\tau(w)$, which measures the remaining damping of a  path of weight $w$ after mitigation.
At small weights, $\gamma_0 w \ll \tau$, the error is non-perturbatively suppressed, $1-e^{-\gamma_0 w}h_\tau(w) \sim e^{
-\tau^2/\gamma_0 w}$,
since all derivatives vanish due to the $m\rightarrow\infty$ limit of RZNE.
As the weight increases, $\gamma_0 w\gtrsim\tau$, the filter function crosses over from this near-perfect cancellation and rapidly falls to zero [Fig.~\ref{fig:TEC}(a)]. %
A detailed analysis (see Appendix) finds that to cancel noise to accuracy $\varepsilon$ up to weight $w_*$ requires
\begin{equation}\label{eq:tau_main}
\tau
=
\gamma_0 w_*
+
\mathcal O\left(
\sqrt{\gamma_0 w_*\log(1/\varepsilon)}
+
\log(1/\varepsilon)
\right).
\end{equation}
In practice, we expect that the chosen value of $\tau$ will not be set a priori, but rather determined empirically, by gradually increasing $\tau$  until observing convergence.

The sampling overhead of TEC using amplified noise admits a simple expression, $X(\tau)=\cosh^2(2\tau)$.
Thus, at a constant fixed accuracy, we have
\begin{equation}\label{eq:tec_cost_main}
    X(w_*)
    =
    e^{4\gamma_0 w_*
    +
    \mathcal{O}(\sqrt{\gamma_0 w_*})}.
\end{equation}
Hence, while PEC pays an exponential cost in the entire active volume, $e^{3\gamma_0 \circV}$, TEC pays an exponential cost only in the threshold weight that one chooses to mitigate up to.
See Fig.~\ref{fig:TEC}(b) for an explicit comparison.

Our extension to discrete noise insertions is now simple.
We apply the mapping in Eq.~\eqref{eq:poisson_main}, which yields coefficients $\tilde{h}_k = \sum_r p^{\gamma_r}_k h_r$ and leaves $h_\tau(w)$ unchanged.
When $w_* \ll \circV$, the spacing of the amplified noise rates is sufficiently large that no cancellations occur in $\tilde{h}_k$, and the sampling overhead, $X(\tau) = ( \sum_k |\tilde{h}_k|)^2$, is unchanged.
On the other hand, when $w_* = \mathcal{O}(\circV)$, massive cancellations between adjacent $h_r$ occur, and lead to a crossover between   Eq.~\eqref{eq:tec_cost_main} and the PEC overhead, $e^{3\gamma_0 \circV}$ [Fig.~\ref{fig:TEC}(b)]. %
Remarkably, this correspondence extends beyond the sampling cost.
In the limit $w_* \gtrsim \circV$, the TEC  coefficients approach $\widetilde h_k(\tau)\rightarrow e^{K}(-K)^k/k!$, identical to PEC. Hence, PEC can be viewed as the most accurate and costly limit of TEC using discrete noise insertions.

\subsection{Numerical optimization and parameter selection} \label{sec:TEC-numerical}

We make two final remarks, regarding the practical implementation of TEC.
First, in practice, we find that one can often moderately further reduce the sampling overhead of TEC by optimizing the filter functions numerically.
Intuitively, the large size of the coefficients $h_\gamma, h_k$ allows small optimizations to create significant  savings.
Our optimization procedure is straightforward: we search for the set of linear coefficients $\{ h_k \}$ (or $\{ h_\gamma \}$) that minimize the sampling overhead $X = ( \sum_k |h_k|)^2$ while obeying the constraint $|1-e^{-\gamma w} h(w)| \leq \alpha (w/w_*)$ for all $w \leq w_*$; here, $\alpha$ is taken to be a small non-zero value. 
This is a convex optimization task and can be easily solved numerically.
An example study is shown in Fig.~\ref{fig:error_mit_results_2d_tfi} and further details and analysis are provided in the Appendix.

Second, we address the choice of the tuning parameter $\tau$ (or equivalently, the threshold weight $w_*$).
If one had perfect knowledge of the reactivity function of the experiment of interest, one would  set $w_*$ to be the maximum weight on which the reactivity  had significant support.
This would ensure accurate mitigation with minimum shot overhead.
In practice, this requisite $w_*$ will be unknown, and its value must be assessed from the convergence of the mitigation output as $w_*$ is increased.
In the Appendix, we describe an adaptive algorithm to implement this assessment and select a favorable $w_*$.
Our algorithm makes use of existing shots accrued at small values of $w_*$ to estimate the performance of mitigation at larger $w_*$, and allocates future experimental shots accordingly.
This enables it to avoid local plateaus in the bias as a function of $w_*$, and to make more efficient reuse of shots compared to more naive methods.

\subsection{Pauli-path zero-noise extrapolation as information filtering}\label{sec:fitting_as_filtering}

We conclude this section by noting that our first mitigation method, PP-ZNE, can also be viewed in the filter function framework.
In the limit of many experimental shots, the PP-ZNE prediction, $\hat{C}(0)$, can be linearly expanded in the fitting parameters.
When this expansion holds, the least-squares fit performed in PP-ZNE implies that $\hat{C}(0)$ is also linear in the noisy expectation values, $C(\gamma)$ [Eq.~\eqref{eq:linear_combination}] or $C(k)$ [Eq.~\eqref{eq:linear_combination_k}].
The coefficients of this linear expansion define a filter function, $h(w)$.
These coefficients depend on the sensitivity of the fitting function, $C_{\kappa}(\gamma)$, to the fitting parameters $\kappa$, as well as the uncertainty in the experimental data points; we derive the explicit relationship in Appendix~\ref{app:ppzne_as_filtering}.

This connection leads to two useful insights.
First, one can use this expansion to optimize the allocation of experimental shots to different values of noise amplification or discrete event insertions, as in Section~\ref{sec:info-filter}.
Second, this connection highlights a conceptual difference between PP-ZNE and TEC.
In TEC, the coefficients of the filter function are independent of the experiment to which it was applied (aside from a single choice, of the threshold weight $w_*$).
In PP-ZNE, this is not the case, as the coefficients depend on the values of the fitting parameters that are linearized around, which are obtained by fitting to data.
Hence, PP-ZNE can be viewed as an adaptive procedure, in which one first determines a filter function by (coarsely) fitting the model parameters, and then implements error mitigation via filter functions using this choice.
We give the exact expression for this in Appendix~\ref{app:ppzne_as_filtering}; investigating the resulting filter functions in more detail is an interesting direction for future work.

\section{Numerical studies of the reactivity}\label{sec:num_reactivity}

In this section, we study the reactivity function, and its connections to quantum error mitigation, in three exemplary models and observables that have been studied in the field of error mitigation before.
The first are out-of-time-ordered correlation functions (OTOCs) in nuclear spin dynamics in the molecules toluene and DMBP~\cite{zhang2025quantumcomputationmoleculargeometry}. The second are charge expectation values in the 2D XY model in a linear potential~\cite{Morong_2021}. The third are spin-spin correlation functions of the Ising order parameter in the 2D transverse-field-Ising model (TFIM)~\cite{haghshenas2026digitalquantummagnetismtrappedion}. We will find that different models and observables yield qualitatively different reactivity functions, $R(w)$.
These are in turn associated with different noise responses, $C(\gamma)$, to which error mitigation is applied.

\begin{figure}
    \centering
    \includegraphics[width=\linewidth]{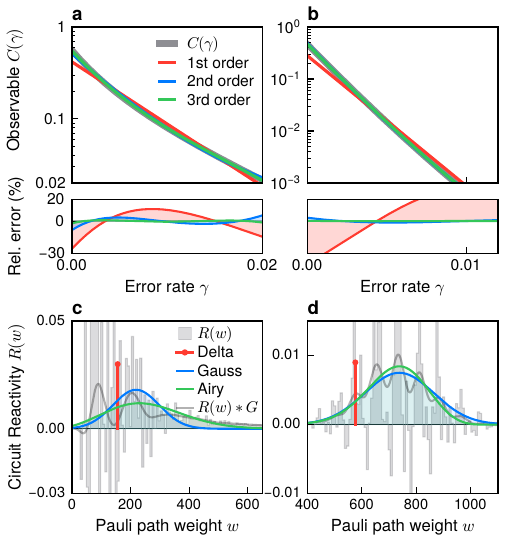}
    \caption{{Noise response and reactivity of the NMR OTOC on toluene (left) and DMBP (right) molecule systems.}
    (\textbf{a, b})~Decay of the expectation value $C(\gamma)$ under single qubit depolarizing noise at rate $\gamma$ (thick gray curve), compared with 1st-order (red), 2nd-order (blue), and 3rd-order (green) cumulant expansion fits.
    The panels directly below show the relative fit residuals $(C_{\text{fit}} - C_{\text{exact}}) / C_{\text{exact}} \times 100\%$, with shaded fills to the zero baseline. The 1st-order (exponential) fit exhibits systematic deviations up to ${\sim}30\%$, while the 2nd- and 3rd-order fits remain within a few percent across the full noise range.
    (\textbf{c, d})~Circuit reactivity spectra $R(w)$ as a function of total Pauli path weight~$w$. The gray filled staircase shows the binned exact reactivity obtained from the Pauli path decomposition. The dark curve shows $R(w)$ convolved with a Gaussian kernel ($R(w) * G$, $\gamma_G = 0.04$), which smooths out fine-grained fluctuations that are invisible at experimentally relevant noise rates. Overlaid are the effective reactivity models corresponding to each cumulant order: a delta function (red), a Gaussian (blue), and an Airy function (green). The convolved reactivity closely tracks the Gaussian and Airy models, confirming that low-order cumulant fits capture the coarse spectral shape governing $C(\gamma)$. The results are obtained via a density matrix simulation as explained in App.~\ref{app:density_matrix_fourier_trafo}.}
    \label{fig:nmr_otoc_numerical_examples}
\end{figure}

\subsection{OTOCs in nuclear spin dynamics}

We continue with discussing noisy OTOC measurements in the recently studied toluene and DMBP molecule, which serve as an example for a system whose reactivity is mostly centered. The observable is the expectation value of $Z_{q_1}$ on qubit $q_1$ in the state $B_{q_2}(t)|0\rangle$, where $|0\rangle$ is the $0$-computational basis state and $B_{q_2}(t)$ the Heisenberg-time evolved butterfly operator $B_{q_2}$ acting on qubit $q_2$, see Ref.~\cite{zhang2025quantumcomputationmoleculargeometry} for the details.
We simulate the impact of noise by inserting single qubit depolarizing channels $\mathcal{N}_\gamma[\rho]$ %
after every two qubit gate at rates $\gamma$ comparable to current quantum devices ($\gamma_0 \approx 0.0025$). The resulting decay of the signal is shown for two examples in Fig.~\ref{fig:nmr_otoc_numerical_examples}(a, b). For noise rates up to several multiples of current hardware noise, the signal decay exhibits clear curvature in log-space, indicating that the reactivity has finite width. Hence, a 1st-order (purely exponential) fit is insufficient for error mitigation: the red curve in Fig.~\ref{fig:nmr_otoc_numerical_examples}(a, b) systematically deviates from the data, accumulating relative errors up to ${\sim}30\%$ as shown by the residuals directly below. Including a 2nd-order cumulant---corresponding to a Gaussian-shaped reactivity---substantially improves the fit, and the 3rd-order correction provides only marginal further improvement.

The corresponding effective reactivities are shown in Fig.~\ref{fig:nmr_otoc_numerical_examples}(c, d). The binned raw reactivity (gray staircase) reveals a broad distribution of Pauli path weights with fine-grained oscillations. Although the exact $R(w)$ contains many features, most are not resolved at finite noise rates and have negligible influence on $C(\gamma)$ for $0 < \gamma < 0.02$. To make this precise, we convolve $R(w)$ with a Gaussian kernel of width $\gamma_G = 0.04$ (dark curve, $R(w) * G$), which removes features below the scale $\Delta w \sim 1/\gamma_G$ at the cost of a relative error $e^{\gamma^2/(2\gamma_G^2)}-1$ on $C(\gamma)$. The resulting smoothed distribution closely tracks the effective Gaussian and Airy reactivity models (blue and green curves), justifying the use of low-order cumulant fits. The failure of the delta-function model (red) to capture the width of the distribution directly explains the large residuals of the 1st-order fit in panels~(a, b).

\subsection{Stark localized XY model}

\begin{figure}
    \centering
    \includegraphics[scale=1]{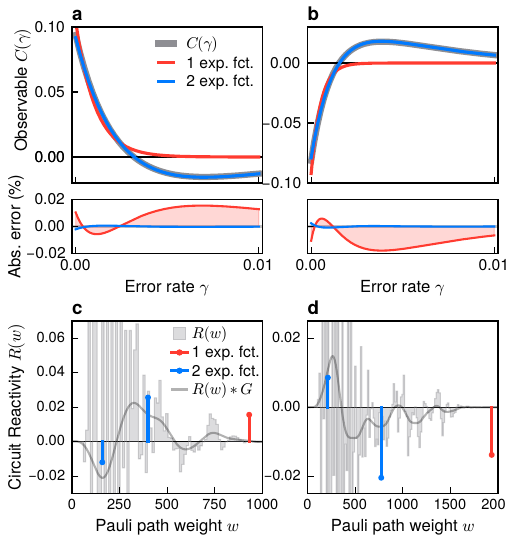}
    \caption{Noise response of the $Z$ expectation value for a $4\times 4$ $XY$ model with linear potential at time steps $30$ (left) and $48$ (right), Eq.~\eqref{eq:xy_model_def}. Top: Decay of the signal under single qubit depolarizing noise with rate $\gamma$ (gray). While one exponential $\hat{C}_1(\gamma)=\hat{R}_1 e^{-w_1 \gamma}$ (red) is insufficient to describe $C(\gamma)$, the fit $\hat{C}_2(\gamma)=\hat{R}_1 e^{-w_1 \gamma}+\hat{R}_2 e^{-w_2 \gamma}$  (blue) agrees well with $C(\gamma)$. Bottom: Binned Reactivity (filled gray) and Gaussian convolution of $R(w)$ with a Gaussian kernel with rate $\gamma_G=0.02$. The $\delta$-peaks denote the fit results ($\hat{R}_i$,$w_i$) for the single exponential (red) and double exponential (blue). While the single exponential fit (red) disregards too much information in the reactivity to accurately describe $C(\gamma)$, the double exponential fit (blue) captures the separation of the positive and negative contributions and, conversely, the transition of $C(\gamma)$ across $0$.}
    \label{fig:bimodal_numerical_examples}
\end{figure}

Next, we consider the influence of noise in models with a separation of noise scales, causing effectively different parts of the expectation values to decay with different rates. As an example, we consider the trotterized XY model in a linear potential on a square lattice
\begin{align}
    H = \sum_{\langle i,j\rangle } X_iX_j+Y_i Y_j + \sum_{x,y}  y g Z_{i(x,y)}. \label{eq:xy_model_def}
\end{align}
We simulate a $4\times 4$ system with open boundary conditions, $g=4$, a Trotter step size of $0.1$ and a domain wall initial state $|\Psi_\mathrm{init}\rangle$ with $\langle \Psi_\mathrm{init}|Z_{i(x,y)}|\Psi_\mathrm{init}\rangle = 1-2\Theta(x)$.

The $Z$ expectation value in this setting constitutes different parts oscillating at various frequencies. Conversely, we find the reactivity to separate into different parts. Around $30$ and $48$ time steps, we find parts with opposite signs, as shown in Fig.~\ref{fig:bimodal_numerical_examples} (c, d). At $30$ time steps, we find one predominantly negative contribution to the signal with Hamming weight $w \lesssim 225$ and a positive contribution with $w \gtrsim 225$. As a result, with noise, $C(\gamma)$ transitions from positive to negative, Fig.~\ref{fig:bimodal_numerical_examples} (a). This behavior cannot be captured by a single exponential decay as shown by attempting to fit a single exponential (red) to $C(\gamma)$ (gray). For both examples shown, the fitted shapes of the reactivity yield too large decay rates of about $900$ and $2000$ Fig.~\ref{fig:bimodal_numerical_examples} (c and d, red) compared to the reactivity $R(w)$. This in turn causes $\hat{C}(\gamma)$ to decay faster than $C(\gamma)$ and to never cross $0$. Using a single exponential model for error mitigation cannot be expected to yield accurate results.
This motivates to model $C(\gamma)$ as a multi-exponential model; here, we fit two decay rates, based on the shape of the reactivity $\hat{C}(\gamma)=ae^{-w_1 \gamma}+be^{-w_2 \gamma}$. This may be generalized to a sum over $M$ exponentials to capture arbitrary complex $C(\gamma)$ at the cost of a more complex fitting procedure. Indeed, the effective reactivity description with two delta peaks Fig.~\ref{fig:bimodal_numerical_examples} (bottom, blue) captures the essential features, in particular the two parts of the reactivity with different signs. As a result, $\hat{C}(\gamma)$ is in good agreement with $C(\gamma)$, Fig.~\ref{fig:bimodal_numerical_examples} (top) blue, justifying the use of this model for this case study.

\subsection{2D TFIM}
The third case study is the second order trotterized two-dimensional TFIM as probed before in experiment~\cite{haghshenas2026digitalquantummagnetismtrappedion}. The Hamiltonian is
\begin{align}
H = \sum_{\langle i,j\rangle } J Z_i Z_j +\sum_j  h X_j, \label{eq:2d_tfi}
\end{align}
with $h/J=2$, periodic boundary conditions and Trotter step size $J/4$. We initialize all qubits to be in the product state of $\exp(i\theta \sigma^y/2)|0\rangle$ with $\theta = -\pi /18$ and measure the non-trivial part of the disconnected Ising order parameter $O=N^{-2} \sum_{i\neq j} Z_i Z_j$, which we aim to error mitigate as a whole.
\begin{figure}
    \includegraphics[scale=1]{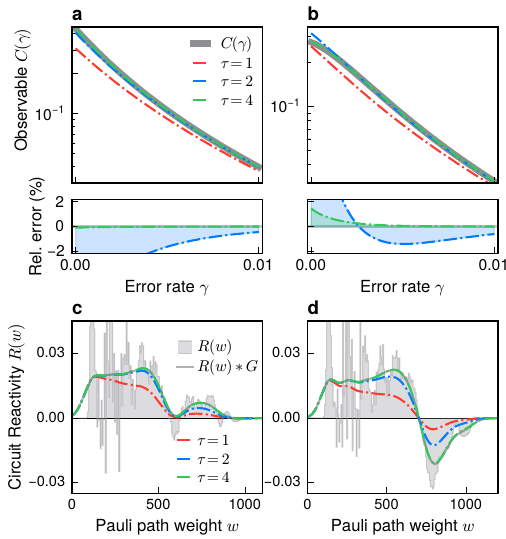}
    \caption{The decay of the disconnected Ising order parameter $C(\gamma)$ with single qubit depolarizing error rate $\gamma$ (a,b) and the corresponding reactivities $R(w)$ (c, d) for the $4\times 4$ 2D TFIM, Eq.~\eqref{eq:2d_tfi}, at $20$ (left) and $23$ (right) time steps. Since the observable is a sum of many operators, the reactivity $R(w)$ becomes broad and features at large $w\sim 750$ are relevant for precise error mitigation: At $23$ ($20$) time steps, the negative (positive) part of $R(w)$ at $w\sim 750$ causes $C(\gamma)$ to bend downwards (upwards) at small $\gamma$. This motivates the application of the universal filter method. We use $C(\gamma_r)$ with noise rates $\gamma_r=r_j \gamma_0$ to reconstruct the reactivity using three different hyperparameter values $\tau=1,2,4$ (red, blue, green), see Eq.~\eqref{eq:tec_main}. Restricting the filter function optimization to small weight $w$ by setting $\tau=1$ yields a good agreement of the reconstruction of $R(w)$ only for small $w$, but still manages to capture some aspects of $R(w)$ for larger $w$. This provides the starting point for tunable error mitigation and allows one to improve the accuracy continuously with an increasing computational cost up to high accuracy at hyperparameter value $\tau=4$.}
    \label{fig:reactivity_tfim}
\end{figure}

We find that the fact that the observable $O$ is a summation over many operators leads it to exhibit broad reactivities, as shown in Fig.~\ref{fig:reactivity_tfim}(c, d). 
Error mitigation with such broad reactivities can be more challenging for the cumulant expansion and the multi-exponential model ansatz, while the filter method is less restrictive. Indeed, with the filter function we can successfully reconstruct the reactivity with tunable accuracy by varying the tuning parameter $\tau$. The computationally cheapest choice shown, $\tau=1$, captures $R(w)$ and $C(\gamma)$ roughly. Increasing $\tau$ yields an accurate description of $R(w)$ with a bias of $|\hat{C}(0)-C(0)|$ less than $2\cdot 10^{-2}$. This small bias is achievable by focusing on $w<1000$ with the selected $\tau=4$, Fig.~\ref{fig:TEC}, which is noticeably smaller than the circuit volume ($V=1280$ and $V=1472$). This enables performing error mitigation more efficiently than PEC: The overhead of TEC for $\tau=4$ for $t=23$ is $(\sum_r |h_r|)^2=10^{4.2}$, which is four times more efficient than PEC, which has a sampling overhead of $\sim \! 10^{4.8}$.

\section{Implementation and comparison of error mitigation methods} \label{sec:num-mitigation} %

Having illustrated the physical intuition behind each of our  mitigation methods, in this section, we embark on a more systematic evaluation of their performance.
We compare our methods both against each other, and against the most common existing methods, probabilistic error cancellation (PEC) and single-exponential zero-noise extrapolation (ZNE).
We perform this comparison by implementing each method in numerical experiments, and plotting the root-mean-square error (RMSE) they achieve against the number of experimental shots.

Our results are supportive of broader trends that we have remarked on before.
For every circuit and observable considered, we find that the RMSE of single-exponential ZNE quickly approaches a strict bias floor that cannot be improved on  by adding further shots.
All three of our methods (cumulant PP-ZNE, multi-exponential PP-ZNE, and TEC) outperform this floor at sufficient numbers of shots owing to their  tunability.
Our methods also, in most circumstances, drastically outperform PEC.
Comparing between our methods, we find that cumulant PP-ZNE and multi-exponential PP-ZNE yield the most efficient error mitigation on certain circuits, where their ans\"atze are a good fit to the experimental data.
Meanwhile, we find that TEC features the most reliable improvements over existing methods across all circuits, particularly at small desired RMSE.
Across all circuits and methods, we find that mitigation using discrete noise insertions yields order-of-magnitude improvements in number of shots compared to mitigation using amplified noise.

\subsection{Details of our comparison}

Our study proceeds as follows.
For each circuit and mitigation method, we assume that one has access to noisy expectation values from a quantum experiment with either discrete noise insertions or amplified noise.
We set the device noise rate to $\gamma_0 \equiv 0.0025$, comparable to current leading quantum processors.
We then simulate applying each method multiple times to each circuit, incorporating shot noise fluctuations, and utilizing our adaptive schemes of selecting the tuning parameter of each method based on the convergence of the mitigated result.
We conclude by plotting the root-mean-square difference between the mitigated result and the exact result over these applications, as a function of the number of experimental shots $N_\mathrm{shots}$.

For both PP-ZNE methods, we proceed as follows.
For mitigation using amplified noise $\gamma$, we use $n_r$ equidistant noise rates between $\gamma_0$ and $4\gamma_0$.
We allocate an equal number of experimental shots, $N_\mathrm{shots}/n_r$, to each noise rate.
For mitigation using discrete noise insertions $k$, we use $n_k$ different discrete noise event numbers, from zero to $n_k-1$.
We allocate an equal number of experimental shots, $N_\mathrm{shots}/n_k$, to each number.
The values of $n_r$ and $n_k$ are specified in the figure captions for each study.
For TEC, for simplicity, we consider only discrete noise insertions $k$.
We allow all possible values of $k$ and allocate shots according to our adaptive algorithm described in Section~\ref{sec:TEC-numerical} and the Appendix.

We include three additional curves for comparison in each study.
First, we consider conventional ZNE, which we perform identically to PP-ZNE above but using a single-exponential ansatz.
Second, we consider PEC, which has an MSE equal to its variance, $(3e^{\gamma_0}/2-1/2)^{2V}/N_{\text{shots}} \approx e^{3\gamma_0 V}/N_{\text{shots}}$, by definition.
Third, we consider a rough benchmark for the ``optimal'' performance that any error mitigation method could hope to achieve, defined as follows.
To achieve a small bias $\epsilon$ in a reliable manner, we assume that one must be able to resolve the reactivity function at least up to a weight $w_*$, such that the cumulative reactivity, $\max_{w_*' \geq w_*}\left| \sum_{w > w_*'} R(w)\right| \leq \epsilon$, above $w_*$ is at most $\epsilon$.
Since the contribution from information paths of weight $w_*$ is damped by $e^{-\gamma_0 w_*}$, this requires a number of shots $e^{2\gamma_0 w_*}/\epsilon'^2$ to resolve within standard deviation $\epsilon'$.
Setting the standard deviation and bias equal, $\epsilon' = \epsilon$, yields the number of shots required to resolve the signal to within RMSE $\epsilon\sqrt{2}$, $N_{\text{shots,optimal}} = e^{2\gamma_0 w_*}/\epsilon^2$.

\subsection{Error mitigation of simulated OTOCs in nuclear spin systems}

\begin{figure}
    \centering
    \includegraphics[scale=1]{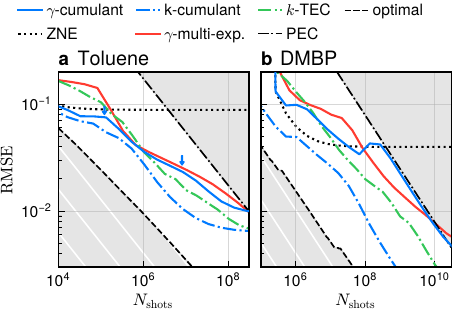}
    \caption{Accuracy results for error mitigation of OTOCs in the toluene molecule after $6$ time steps (left) and DMBP molecule after $9$ time steps (right) evaluated using $1000$ shot noise realizations. We show results for the cumulant method from fitting Eq.~\eqref{eq:cumulant_exp} with $M\leq 3$ to four values $C(\gamma_r)$ with $\gamma_r=0.0025r$ and $r=1,2,3$ or $4$. The three predictions $\hat{C}_M(0)$ are combined using the telescopic sum Eq.~\eqref{eq:telesc_sum} (blue). This yields a step like behavior whenever the next more accurate $\hat{C}_M(0)$ becomes relevant (blue arrows), see Appendix~\ref{app:comp_akaike_telesc}. We compare the result to the $k$-cumulant method (dash-dotted blue) and the $k$-TEC method (dash-dotted green); for the former,  we fit to $C(\gamma_0,k)$ with $k\leq 3$. We also show the Akaike weighted result, Eq.~\eqref{eq:akaike_weight}, for fitting the multi-exponential model (red), for which we used $10$ values of $C(\gamma_r)$ with $\gamma_r$ equidistantly spaced in the range $0.0025\leq \gamma_r\leq 0.01$ to account for the larger amount of fitting parameters. For reference, we also show the performance of single-exponential ZNE (dotted), PEC (gray), and our estimate of the optimal achievable error mitigation cost $N_\text{shots, optimal}$ (dashed gray).}%
    \label{fig:error_mit_results_otoc_nmr}
\end{figure}

We first study the performance of our three mitigation methods for quantum simulations of OTOCs in nuclear spin dynamics~\cite{zhang2025quantumcomputationmoleculargeometry}. The results are shown in Fig.~\ref{fig:error_mit_results_otoc_nmr}.

Due to their tunability, all methods are able to error mitigate the noisy measurements with high accuracy when increasing the complexity of the method.
For the PP-ZNE methods, which feature a discrete tuning parameter, this manifests in a step-like pattern in the accuracy-cost curve, e.g., Fig.~\ref{fig:error_mit_results_otoc_nmr} (left, blue solid): With increasing number of shots $N_\mathrm{shots}$, the RMSE decreases until it saturates for a given order of the method (marked with blue arrows). Once the shot requirements are met for the next order, the accuracy increases further, allowing for a systematic improvement of the accuracy with an increasing number of shots (see Appendix~\ref{app:comp_akaike_telesc} for additional detailed studies).
For the TEC method, the improvement of accuracy with the number of shots is more gradual, owing to its continuous tuning parameter.
Overall, the performance of the $\gamma$-cumulant method agrees well with the expectations from the analysis of the reactivity $R(w)$. For example, for the molecule toluene, a simple exponential fit yields an RMSE of $\sim \! 10^{-1}$; including a quadratic correction improves the RMSE to $\sim \! 3\times 10^{-2}$; and including a cubic correction further reduces it to less than $10^{-2}$. This shows the importance of higher-order cumulant corrections, and their ability to systematically improve the mitigation accuracy.

Comparing the two methods that use noise amplification (abbreviated ``$\gamma$-''), we find that the number of shots $N_\mathrm{shots}$ needed to achieve a target accuracy varies by a factor $\sim 3$ for a given system. For toluene, the $\gamma$-cumulant method performs slightly better, Fig.~\ref{fig:error_mit_results_otoc_nmr} (left, blue solid). For DMBP, there is no systematic difference in the performance of the $\gamma$-cumulant and $\gamma$-multi-exponential method. Both methods need $\sim \! 5\cdot 10^9$ shots to achieve a root-mean-square error below $10^{-2}$. %

Meanwhile, in both systems studied, error mitigation based on inserting discrete noise events (abbreviated ``$k$-'') is significantly more effective compared to noise amplification. In particular, the $k$-cumulant expansion requires $\sim \! 10\times$ fewer shots to achieve a comparable RMSE to the other $\gamma$-methods, and $\sim \! 30$ or $100\times$ fewer shots compared to PEC. %
As discussed in Section~\ref{sec:k_error}, the efficiency gain when using discrete noise insertion follows from discrete noise damping larger weights $w$ more compared to the damping of smaller weights, making them more distinct than in continuous noise amplification (see also Appendix~\ref{ssec:cost_estimate_sum_exp}).

Compared to the $k$-TEC method, the $k$-cumulant method is significantly more efficient. This advantage is expected given the localized nature of the reactivity, for which the cumulant method is particularly well suited.

We note that the performance of PEC is expected to scale worse with the number of qubits; for systems larger than the $16$ qubits simulated here, we expect the advantage of each of our methods compared to PEC to be larger. We find error mitigation based on discrete noise events to be also more efficient in other studied examples, as we will discuss next.

\subsection{Error mitigation of the XY model}

\begin{figure}
    \centering
    \includegraphics[scale=1]{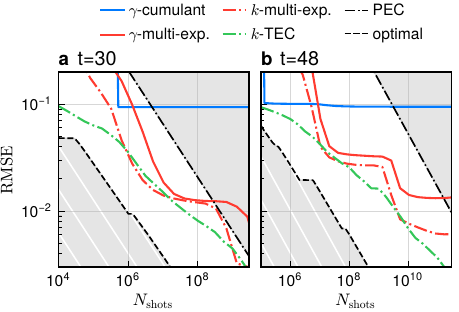}
    \caption{%
    Error mitigation results for the XY model, Eq.~\eqref{eq:xy_model_def}, after $30$ time steps (left) and $48$ time steps (right) analogous to Fig.~\ref{fig:error_mit_results_otoc_nmr}. The results for the multi-exponential method are obtained by fitting Eq.~\eqref{eq:sum_exp_ansatz} with $M\leq 3$ to ten values $C(\gamma_r)$ with $\gamma_r$ equidistantly spaced in the range $0.0025\leq \gamma_r\leq 0.01$. The three predictions $\hat{C}_M(0)$ are combined using the Akaike weights Eq.~\eqref{eq:akaike_weight} (red line). We compare the result to the telescopic sum result, Eq.~\eqref{eq:telesc_sum}, for the cumulant error mitigation (blue line) with $4$ values of $C(\gamma_r)$ like before. We also show the results for the $k$-multi-exponential method (dash-dotted red) and $k$-TEC method (dash-dotted green); for the former, we fit to $10$ values of $C(\gamma_0,k)$, $k\leq 9$. For reference, we also show the performance of PEC (gray), and an estimate for the optimal achievable error mitigation cost (dashed gray). In this example, conventional ZNE using a single-exponential fit is limited identically to the $\gamma$-cumulant method (blue).}%
    \label{fig:error_mit_results_bimod}
\end{figure}

In Fig.~\ref{fig:error_mit_results_bimod}, we study the performance of our mitigation methods in the Stark-localized XY model~\cite{Morong_2021}.

We recall that this system has a more intricate reactivity structure: the noisy signal $C(\gamma)$ changes sign with increasing noise rate $\gamma$, due to the reactivity containing large tracts of different sign that decay at different rates.
This prohibits effective error mitigation via cumulant expansions, which are best-suited to modeling positive reactivity functions.
We observe correspondingly that the prediction accuracy of the cumulant expansion is limited to $\gtrsim 10^{-1}$.
The multi-exponential and filter method approaches are more suitable for error mitigation in this setting, yielding RMSEs below $10^{-2}$ at sufficiently large numbers of shots.
Similar to the results for nuclear-spin OTOCs (Fig.~\ref{fig:error_mit_results_otoc_nmr}), the error of the multi-exponential method plateaus until the number of shots is sufficient and the next order correction becomes statistically significant.
As before, error mitigation based on discrete noise events is more efficient.
Due to the plateaus in the multi-exponential fit (and the stability of TEC), these two trade position for being the ``best method'' depending on the target accuracy.
Both approaches consistently improve over PEC; when targeting an RMSE of about $10^{-2}$ at $30$ and $48$ Trotter steps, the number of shots is reduced by $\sim \! 10\times$ and $\sim \! 100\times$, respectively.

\subsection{Error mitigation of the 2D TFIM}\label{sec:num_ex_2d_tfi}

Finally we study the disconnected Ising order parameter in the 2D TFIM~\cite{haghshenas2026digitalquantummagnetismtrappedion}.
We recall that the reactivity function of the model displayed qualitatively different behavior between $n_t=20$ and $n_t=23$ time steps (Fig.~\ref{fig:reactivity_tfim}).
At $n_t=20$, the reactivity is broad yielding a convex decay of the noisy signal $C(\gamma)$, while at $n_t=23$, a small negative contribution in $R(w)$ at large $w$ causes $C(\gamma)$ to be concave at small $\gamma$.
This poses a challenge to more heuristic error mitigation methods, as we will see.

In Fig.~\ref{fig:error_mit_results_2d_tfi}, we study the performance of our mitigation methods at these time steps.
At $n_t=20$ steps, when $C(\gamma)$ is convex, we find that the RMSE of all methods decays as the number of shots increases.
For most of the target error range, PP-ZNE with a $\gamma$-cumulant function is more efficient than the $\gamma$-multi-exponential method. The advantage of the cumulant expansion is to be expected, as it describes the broadly peaked and otherwise featureless reactivity function well.
Meanwhile, at $n_t=23$ steps, where $C(\gamma)$ features a concave bend, PP-ZNE with a $\gamma$-cumulant function cannot improve below a residual bias $\sim \! 0.05$.
Notably, the multi-exponential fit also demonstrates a slow convergence with similar initial bias, requiring $M=3$ and around $N_\mathrm{shots} \gtrsim 10^7$ for a further reduction of the bias.
By contrast, the $k$-TEC method has similar performance at $t=23$ as it does at $t=20$, and displays a continuous systematic improvement in the RMSE as the number of shots is increased over the entire range considered.

\begin{figure}
    \centering
    \includegraphics[scale=1]{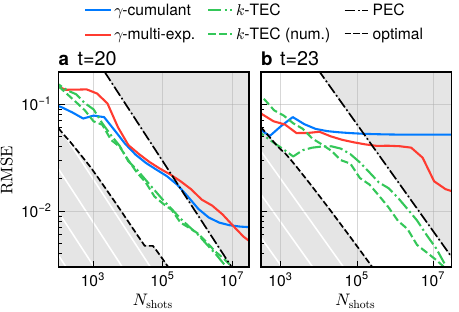}
    \,\,
    \caption{Error mitigation results for the 2D TFIM. \textbf{a}, \textbf{b}: Comparison of the accuracies when increasing the number of shots for the different error mitigation methods, analogous to Fig.~\ref{fig:error_mit_results_otoc_nmr}, for the $\gamma$-cumulant method (blue), the $\gamma$-multi-exponential method (red), and the $k$-TEC method using either the analytic filter function (dash-dotted green) or numerically-optimized filter functions (dashed green).
    For either filter function, the threshold weight is chosen via the adaptive algorithm described in the main text. Note that for this case study the observable is an average over $N(N-1)$ Z-basis operators making its measurement shot-efficient. %
    For comparison, the PEC scaling (dash-dotted black) and the theoretically optimal result are shown (dashed black). %
    }
    \label{fig:error_mit_results_2d_tfi}
\end{figure}

As discussed in Section~\ref{sec:TEC-numerical}, in some cases, the efficiency of TEC can be moderately improved upon by numerically optimizing the filter function coefficients, $h_k$, to reduce the sampling overhead.
We show the performance of this approach in Fig.~\ref{fig:error_mit_results_2d_tfi} (dashed green).
We find that it yields comparable performance to the analytic filter function at $n_t = 20$ steps, and an improvement by $\sim 3\times$ in the number of shots at $n_t = 23$ steps.
Both the numerical and analytic filter functions reduce the number of shots compared to PEC by $3\times$ to $10\times$ for all values of RMSE considered; as mentioned previously, we expect this advantage to improve for larger system sizes, beyond the reach of our numerical simulations, where PEC becomes less efficient.
We provide further discussions and numerical comparisons between the analytic and numerical implementations of TEC in Appendix~\ref{app:tec}.

\section{Conclusions \& Outlook} \label{sec:outlook}

Our work aims to provide quantum error mitigation with both a physical guide and a means of systematic refinement.
The effective reactivity function identifies the components of quantum information dynamics that are resolvable at finite experimental precision.
Linear mitigation protocols act as filters on these components, with their profiles determining which information is restored under mitigation, and their coefficients determining the sampling overhead.
Reconstructing the effective reactivity function of a quantum circuit, as in Pauli-path zero-noise extrapolation, or engineering its filter, as in tunable error cancellation, thus provide complementary ways to trade the bias of mitigation methods against their sampling cost.
Our hope is that, as the entanglement and bond dimension currently guide tensor network simulations, the reactivity function and tunable mitigation methods introduced here can help guide quantum error mitigation moving forward. %

These ideas open a potentially broad experimental program.
Pauli-path ZNE has already enabled first gains in quantum simulations of molecular OTOCs for NMR structure determination~\cite{zhang2025quantumcomputationmoleculargeometry}.
Our numerical results suggest further gains in accuracy and sampling efficiency from the refinable models, discrete noise insertions, and tunable error cancellation developed here  for broad classes of experiments, even in regimes where conventional methods may fail.
Testing these protocols on hardware, including their adaptation to coherent and non-unital noise, is a natural next step.

From a theoretical perspective, our work motivates  fundamental investigations of the physical behavior of reactivity functions.
What is the functional form of the reactivity function in different classes of quantum circuit dynamics, and how does its support grow over time?
Our framework shows that answers to these questions would help inform quantum error mitigation methods, and allow more efficient and higher-precision experimental studies of such systems.
Developing a stronger theory of effective reactivity functions, in particular, may assist these investigations, as well as their applications to approximate mitigation methods.

\vspace{0.5em}
\textbf{Acknowledgments.}
We thank R. Cortinas, A. Karamlou, D. Kafri, D. Abanin, X. Mi, O. Shtanko,  V. N. Smelyanskiy, R. Somma and the broader Google Quantum AI team for valuable discussions.

\vspace{0.5em}
\textbf{Author contributions.}
TS conceived of the reactivity function as a central object in error mitigation; related ideas involving Pauli paths were independently developed by TO and ZM. TS and PS developed the effective reactivity function and advantage of discrete noise insertions, TO and ZM developed the cumulant expansion method, PS developed the multi-exponential method, and TS developed the tunable error cancellation method, with input from all authors. ZM, TO, and NN led error mitigation of the NMR-OTOC experiment with contribution from TS. PS and MR developed the numerical simulations of the reactivity. PS developed the implementation and comparison of all methods, with input from ZM, TO, and TS. PS and ZM developed the extension to general noise models. All authors contributed to the refinement and presentation of the results.

\vspace{0.5em}
\textbf{Statement of AI use.}
Generative AI tools were used during the preparation of this manuscript to assist with copyediting and text polishing, and to assist with portions of the analytical derivations and calculations presented in this work. 
We acknowledge conversations with GPT 5.6 Pro and Gemini 3.1 Pro for the derivation of the TEC filter function, the design of the adaptive algorithm for implementing TEC, and the connection between filter functions and PP-ZNE.
All AI-assisted derivations and calculations were independently verified by the authors; all scientific claims, conclusions, and content of this manuscript remain the sole responsibility of the authors.

\vspace{0.5em}
\textbf{Data availability.}
The numerical simulations, error-mitigation code, and analysis scripts developed for this work are
available upon reasonable request from the authors.
The nuclear-spin-echo OTOC, Stark-many-body-localization, and two-dimensional transverse-field
Ising model experimental data analyzed in this work were previously published in
Refs.~\cite{zhang2025quantumcomputationmoleculargeometry}, \cite{Morong_2021}, and
\cite{haghshenas2026digitalquantummagnetismtrappedion}, respectively.

\onecolumngrid
\raggedbottom

\newpage
\appendix
\addtocontents{toc}{\protect\endgroup}
\renewcommand{\tocname}{Appendix: Table of Contents}
\tableofcontents

\section{Circuit reactivity and error dynamics}\label{app:react_Pauli paths}

\noindent
Section~\ref{sec:reactivity} in the main text derived the reactivity representation for Pauli noise only.
This was for exposition purposes, rather than any limitations of the reactivity representation, which is applicable to a broad range of noise models.
In this appendix we demonstrate this, by extending the reactivity representation to a broad class of non-Pauli noise models.
We show that the relation $C(\gamma)=\sum_w e^{-\gamma w}R(w)$ (Eq.~\eqref{eq:laplace}) holds whenever the noise-strength dependence admits a decomposition condition that is relatively easy to satisfy experimentally.
We first define this noisy-circuit setting, then we develop the corresponding generalized operator-path construction and derive the functional form of the reactivity in this setting.
We finish by expanding the form of the reactivity to the discrete noise insertion case, yielding the results used in Sec.~\ref{sec:k_error} of the main text.

\subsection{Noisy circuit model}\label{app:circuit_noise_model}

\begin{figure}[ht]
	\begin{centering}
		(a)
		\begin{quantikz}[align equals at=1, column sep=2ex]
			& \inputD[style=gateaccent2]{\rho_{\mathrm{init}} \vphantom{O}}
			&
			& \qwbundle{N}
			&[1ex] \gate[style=gateaccent2]{U_1}
			& \gate[style=gateaccent2]{U_2}
			& \push{\;\cdots\;}
			& \gate[style=gateaccent2]{U_T}
			& \meterD[style=gateaccent2]{O}
			& \setwiretype{c} \rstick{$\ \langle O \rangle =: C(0)$}
		\end{quantikz}

		\vspace{1em}
		(b)
		\begin{quantikz}[align equals at=1, column sep=2ex]
			& \inputD[style=gateaccent2]{\rho_{\mathrm{init}} \vphantom{O}}
			&
			& \qwbundle{N}
			&[1ex] \gate[style=gateaccent2]{U_1}
			\gategroup[wires=1, steps=2, style={dashed, rounded corners, inner sep=0pt, draw=gray}]{}
			&[-1ex] \gate[style=gatenoisy]{\mathcal{N}_1(\gamma)}
			&[2ex] \gate[style=gateaccent2]{U_2}
			\gategroup[wires=1, steps=2, style={dashed, rounded corners, inner sep=0pt, draw=gray}]{}
			&[-1ex] \gate[style=gatenoisy]{\mathcal{N}_2(\gamma)}
			& \push{\;\cdots\;}
			& \gate[style=gateaccent2]{U_T}
			\gategroup[wires=1, steps=2, style={dashed, rounded corners, inner sep=0pt, draw=gray}]{}
			&[-1ex] \gate[style=gatenoisy]{\mathcal{N}_T(\gamma)}
			&[1ex] \meterD[style=gateaccent2]{O}
			& \setwiretype{c} \rstick{$C(\gamma)$}
		\end{quantikz}
		\par\end{centering}
	\caption{\label{fig:gen-circ-setup}
		(a)~Ideal circuit. An initial $N$-qubit state $\rho_{\mathrm{init}}$ evolves
		through $T$ unitary layers $U_1,\ldots,U_T$, after which the observable $O$
		is measured, giving $C(0)$.
		(b)~Noisy circuit. Each layer is followed by a noise channel
		$\mathcal{N}_t(\gamma)$, giving the experimentally observed signal
		$C(\gamma)$. State-preparation and measurement errors are discussed below.}
\end{figure}
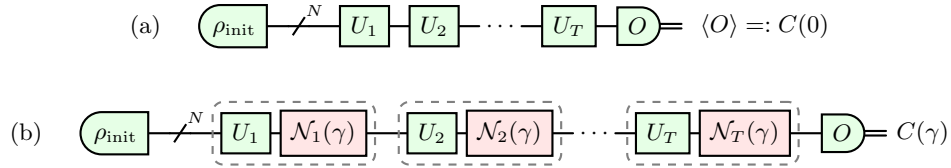

Though one cannot define a (physically-motivated) reactivity model for all possible circuits, the class of circuits which admit such a description is quite broad.
Following the main text, we require a Markovian noise model where layers of unitary gates $\mathcal{U}_t$ are interleaved with layers of noise $\mathcal{N}_t$ (see Fig.~\ref{fig:gen-circ-setup}).

%

\begin{align}
    C(\gamma)
    &= \mathrm{Tr}\!\left(O\,\mathcal{U}_{\mathrm{noisy}}(\rho_{\mathrm{init}})\right)
    = 2^N \llangle O|\,\mathcal{U}_{\mathrm{noisy}}\,|\rho_{\mathrm{init}}\rrangle
    \nonumber\\
    &= 2^N \llangle O|\,\mathcal{N}_T(\gamma_T)\,\mathcal{U}_T\,
    \mathcal{N}_{T-1}(\gamma_{T-1})\,\mathcal{U}_{T-1}\cdots
    \mathcal{N}_1(\gamma_1)\,\mathcal{U}_1\,|\rho_{\mathrm{init}}\rrangle .
    \label{eq:app_noisy_circuit}
\end{align}
Without loss of generality, we write the noise $\mathcal{N}_t$ after $\mathcal{U}_t$, and absorb state preparation and measurement (SPAM) error into the first and last channel (this can also be mitigated separately).
This allows the description of decoherence and dephasing, coherent and correlated errors, crosstalk, and leakage, among many other sources of noise.
It also allows for the description of analog noise, by dividing time into sufficiently short intervals and taking the limit $dt\rightarrow 0$.

Key to this work is the notion of parameterized noise that can be increased in strength but not decreased below some experiment-specific minimum value.
In practice, this can be achieved to a good approximation via methods such as probabilistic error amplification, even when one has no ability to adjust parameters of the underlying gates.
We consider here two separate parameterizations of the noise strength: a global parameterization where we write $\mathcal{N}_t=\mathcal{N}_t(\gamma)$, and a layer-specific parameterization where we have a unique noise parameter $\gamma_t$ for each $\mathcal{N}_t$.
(In App.~\ref{app:generalized_reactivity} we give conditions under which a layer-specific parameterization can be reduced to a global parameterization for our purposes.)
In both cases, we define $\gamma=0$ ($\gamma_t=0$) to imply that no noise exists; $\mathcal{N}_t(0)=I$, and rescale so that $\gamma=1$ ($\gamma_t=1$) is the best achievable noise rate.

%
%

\providecommand{\figScaleWide}{0.85}

\begin{figure*}[t]
    \centering
    \scalebox{\figScaleWide}{%
    \begin{tikzpicture}[
        font=\small,
        >=stealth,
        box/.style={draw, rounded corners=2pt, minimum height=0.82cm,
                    align=center, inner xsep=6pt, inner ysep=4pt},
        arrow/.style={->, thick},
        lbl/.style={font=\scriptsize, align=center, inner sep=1.5pt},
        panel/.style={font=\small\bfseries},
    ]

    \node[panel] at (-0.15,2.05) {(a)};

    \node[box, fill=blue!6] (circ) at (1.75,1.10)
        {\textbf{circuit dynamics}\\[-1pt]
         $\rho_{\mathrm{init}},\,\{U_t,\mathcal N_t\},\,O$};
    \node[box, fill=blue!3] (R) at (6.05,1.10) {$R(w)$};

    \node[box, fill=red!6] (control) at (1.75,-1.10)
        {\textbf{modification}\\[-1pt] $\lambda$};
    \node[box, fill=red!3] (F) at (6.05,-1.10) {$F(\lambda,w)$};

    \draw[arrow] (circ)    -- node[above=2pt, lbl] {information\\dynamics}     (R);
    \draw[arrow] (control) -- node[below=2pt, lbl] {controlled\\interrogation} (F);

    \node[circle, draw, minimum size=0.62cm, inner sep=0pt] (mult) at (7.60,0.00) {$\times$};
    \node[circle, draw, minimum size=0.72cm, inner sep=0pt] (sum)  at (8.75,0.00) {$\sum_w$};
    \node[box, fill=green!7] (C) at (10.05,0.00) {$C(\lambda)$};

    \draw[arrow] (R.east) -- (mult.north west);
    \draw[arrow] (F.east) -- (mult.south west);
    \draw[arrow] (mult) -- (sum);
    \draw[arrow] (sum)  -- (C);

    \draw[gray!40] (10.85,-2.15) -- (10.85,2.25);

    \node[panel] at (11.15,2.05) {(b)};

    \node[lbl, text=gray!70!black] at (15.55,2.02)
        {one reactivity, two interrogations};

    \node[box, fill=blue!3] (Rb) at (11.95,0.00) {$R(w)$};

    \node[box, fill=red!5] (amp) at (15.55,1.15)
        {\textbf{noise amplification}\\[3pt]
         $F_r(w)=e^{-\gamma_0 r w}$};

    \node[box, fill=red!5] (disc) at (15.55,-1.20)
        {\textbf{discrete insertion}\\[3pt]
         $F_k(w)=e^{-\gamma_0 w}\bigl(1-4w/3\circV\bigr)^{k}$};

    \node[box, fill=green!6] (Cr) at (19.30,1.15) {$C(r\gamma_0)$};
    \node[box, fill=green!6] (Ck) at (19.30,-1.20) {$C(\gamma_0,k)$};

    \coordinate (fork) at (12.85,0.00);
    \draw[thick] (Rb.east) -- (fork);
    \draw[arrow] (fork) |- (amp.west);
    \draw[arrow] (fork) |- (disc.west);
    \draw[arrow] (amp)  -- (Cr);
    \draw[arrow] (disc) -- (Ck);

    \end{tikzpicture}}

    \caption{\label{fig:response_problem}
    \textbf{Error mitigation as a response problem.}
    \textbf{(a)}~The circuit dynamics determine the reactivity $R(w)$; an experimentally
    controlled modification $\lambda$ determines a known modification function
    $F(\lambda,w)$. Their overlap gives the measured response
    $C(\lambda)=\sum_w F(\lambda,w)R(w)$.
    \textbf{(b)}~The same reactivity can therefore be interrogated in more than one way.
    Two are used here: amplifying the noise rate by a ratio $r$, and inserting $k$ discrete
    noise events. Both retain the device's baseline rate $\gamma_0$ --- inserting
    events multiplies on top of it rather than replacing it.}
\end{figure*}
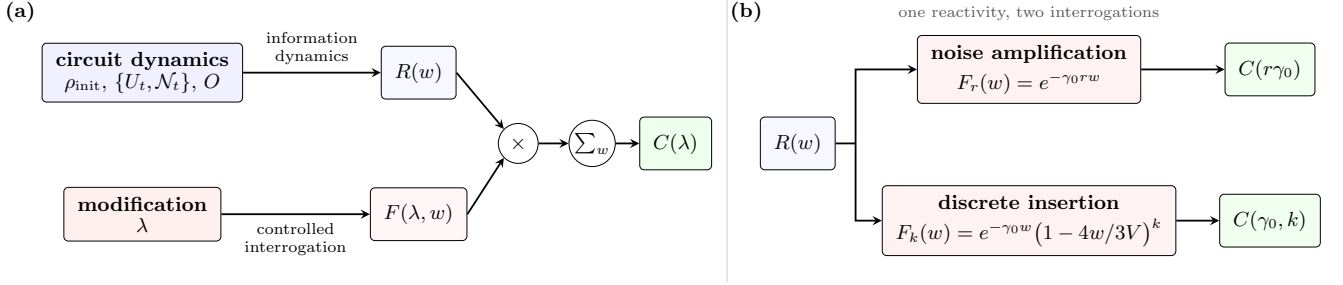

\subsection{Continuous noise amplification}\label{app:generalized_reactivity}

As described in Sec.~\ref{sec:reactivity}, we require throughout this work that the noise be
amplifiable, i.e.\ that its strength can be increased in a controlled, consistent manner.
To extend that construction beyond Pauli noise, assign each layer a
noise strength $\gamma_t$ and assume that its channel can be written as
\begin{equation}
    \mathcal{N}_t(\gamma_t)
    =
    \sum_{\mathcal{S}_t}
    e^{-\gamma_t w_{\mathcal{S}_t}}
    \mathcal{S}_t ,
    \label{eq:app_channel_ansatz}
\end{equation}
where the components $\mathcal{S}_t$ and weights $w_{\mathcal{S}_t}$ are independent of
$\gamma_t$. Thus changing the noise strength modifies only the scalar weight of each component.

%
%

\providecommand{\figScaleWide}{0.92}

\begin{figure*}[t]
    \centering
    \scalebox{\figScaleWide}{%
    \begin{tikzpicture}[
        font=\small, >=stealth,
        comp/.style={circle, draw=red!55!black, fill=red!10, minimum size=0.54cm,
                     inner sep=0pt, font=\scriptsize},
        endc/.style={draw, rounded corners=2pt, fill=green!10, draw=green!55!black,
                     minimum height=0.62cm, inner xsep=5pt, font=\small},
        faint/.style={draw=gray!30, line width=0.3pt},
        pathA/.style={draw=blue!65!black,   line width=1.3pt},
        pathB/.style={draw=violet!70!black, line width=1.3pt},
        pathC/.style={draw=orange!80!black, line width=1.3pt},
        lbl/.style={font=\scriptsize, align=center, inner sep=1.5pt},
        panel/.style={font=\small\bfseries},
    ]

    \node[panel] at (-0.55,2.35) {(a)};

    \def\ya{1.10} \def\yb{0.00} \def\yc{-1.10}
    \foreach \x/\t in {1.6/1, 3.4/2, 6.0/T}{
        \node[comp] (n\t a) at (\x,\ya) {$0$};
        \node[comp] (n\t b) at (\x,\yb) {$1$};
        \node[comp] (n\t c) at (\x,\yc) {$2$};
    }
    \node[lbl] at (1.6,1.92) {$t=1$};
    \node[lbl] at (3.4,1.92) {$t=2$};
    \node[lbl] at (4.7,1.92) {$\cdots$};
    \node[lbl] at (6.0,1.92) {$t=T$};

    \node[endc] (init) at (-0.15,0.00) {$|\rho_{\mathrm{init}}\rrangle$};
    \node[endc] (obs)  at (7.55,0.00) {$\llangle O|$};

    \foreach \u in {a,b,c}{ \draw[faint] (init) -- (n1\u); }
    \foreach \u in {a,b,c}{ \foreach \v in {a,b,c}{ \draw[faint] (n1\u) -- (n2\v); }}
    \foreach \u in {a,b,c}{ \foreach \v in {a,b,c}{ \draw[faint] (n2\u) -- (nT\v); }}
    \foreach \u in {a,b,c}{ \draw[faint] (nT\u) -- (obs); }

    \draw[pathA] (init) -- (n1a) -- (n2b) -- (nTb) -- (obs);   
    \draw[pathB] (init) -- (n1b) -- (n2a) -- (nTb) -- (obs);   
    \draw[pathC] (init) -- (n1c) -- (n2b) -- (nTc) -- (obs);   

    \node[fill=white, inner sep=1.5pt, font=\scriptsize, text=orange!80!black]
        at (5.02,-1.03) {$e^{-5\gamma}$};

    \draw[gray!40] (8.75,-3.05) -- (8.75,2.55);

    \node[panel] at (9.05,2.35) {(b)};

    \node[lbl, anchor=west] at (9.75,1.92) {accumulated weight $w_{\vec{\mathcal{S}}}$};
    \draw[pathA] (9.85,1.55) -- (10.75,1.55);
        \node[anchor=west, font=\scriptsize] at (10.85,1.55) {$0{+}1{+}1=2$};
    \draw[pathB] (9.85,1.15) -- (10.75,1.15);
        \node[anchor=west, font=\scriptsize] at (10.85,1.15) {$1{+}0{+}1=2$};
    \draw[pathC] (9.85,0.75) -- (10.75,0.75);
        \node[anchor=west, font=\scriptsize] at (10.85,0.75) {$2{+}1{+}2=5$};

    \draw[->, gray!60!black] (12.3,0.28) -- (12.3,-0.08);

    \def\ox{10.15} \def\oy{-2.15}
    \draw[->, gray!70!black] (\ox-0.3,\oy) -- (\ox+4.9,\oy)
        node[right, font=\scriptsize] {$w$};
    \draw[->, gray!70!black] (\ox,\oy-0.62) -- (\ox,\oy+1.82)
        node[above, font=\scriptsize] {$R(w)$};
    \foreach \w/\h in {0/0.80, 1/1.32, 3/0.50, 4/-0.38}{
        \draw[line width=2.6pt, gray!55] (\ox+0.78*\w+0.42,\oy) -- ++(0,\h);
    }
    \draw[line width=2.6pt, blue!65!black]   (\ox+0.78*2+0.42,\oy)      -- ++(0,0.92);
    \draw[line width=2.6pt, violet!70!black] (\ox+0.78*2+0.42,\oy+1.00) -- ++(0,0.62);
    \draw[line width=2.6pt, orange!80!black] (\ox+0.78*5+0.42,\oy) -- ++(0,0.72);
    \foreach \w in {0,1,2,3,4,5}{
        \node[font=\scriptsize] at (\ox+0.78*\w+0.42,\oy-0.85) {$\w$};
    }

    \end{tikzpicture}}

    \caption{\label{fig:generalized_paths}
    \textbf{Generalized path expansion under continuous noise amplification.}
    \textbf{(a)}~Each layer noise channel is resolved into components $\mathcal{S}_t$, drawn as
    circles and labelled by their weight $w_{\mathcal{S}_t}$. A route through the lattice, choosing
    one component per layer, is a generalized path
    $\vec{\mathcal{S}}=(\mathcal{S}_1,\ldots,\mathcal{S}_T)$; three are highlighted against the
    full set of routes in gray. Every path carries an amplitude $A_{\vec{\mathcal{S}}}$ fixed by
    the circuit dynamics and is attenuated under homogeneous amplification by
    $e^{-\gamma w_{\vec{\mathcal{S}}}}$, with $w_{\vec{\mathcal{S}}}=\sum_t w_{\mathcal{S}_t}$ the
    accumulated weight; the orange path, with $w_{\vec{\mathcal{S}}}=5$, is labelled with the
    factor $e^{-5\gamma}$ it carries. Turning $\gamma$ up changes only the attenuation; which
    paths exist, and their amplitudes, are untouched.
    \textbf{(b)}~Many microscopically distinct paths share the same accumulated weight --- two of
    the three highlighted here both reach $w=2$, and the bar at $w=2$ is drawn split into their
    two contributions. Summing the amplitudes within each bin defines the reactivity
    $R(w)=\sum_{\vec{\mathcal{S}}}A_{\vec{\mathcal{S}}}\,\delta_{w_{\vec{\mathcal{S}}},w}$. It is
    therefore a property of a group of paths rather than of any single one, and, unlike a
    probability distribution, it may take negative values.}
\end{figure*}
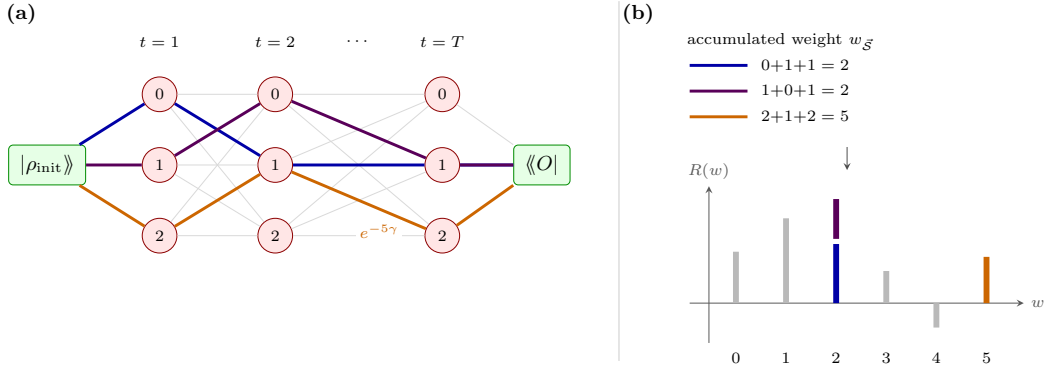

Inserting Eq.~\eqref{eq:app_channel_ansatz} at every noisy layer generalizes the path-integral
formalism studied in the main text,
\begin{align}
    C(\vec{\gamma})
    &=
    \sum_{\vec{\mathcal{S}}}
    e^{-\sum_t\gamma_t w_{\mathcal{S}_t}}
    A_{\vec{\mathcal{S}}},
    \label{eq:app_generalized_paths}
    \qquad
    A_{\vec{\mathcal{S}}}
    =
    2^N \llangle O|
    \prod_t \mathcal{S}_t\mathcal{U}_t
    |\rho_{\mathrm{init}}\rrangle ,
\end{align}
where
$\vec{\mathcal{S}}=(\mathcal{S}_1,\ldots,\mathcal{S}_T)$
is a path through the superoperator components of the noise,
Fig.~\ref{fig:generalized_paths}(a).
This generalizes the paths of the main text: $\mathcal{S}_t$ need not be a Pauli operator, or even
orthogonal to the other components at time $t$; it need only carry a single weight
$w_{\mathcal{S}_t}$ through the decay rate $\gamma_t$, as above.

For homogeneous amplification, $\gamma_t=\gamma$, only the accumulated weight
$
    w_{\vec{\mathcal{S}}}
    =
    \sum_t w_{\mathcal{S}_t}$
matters, and the response reduces to
\begin{align}
    C(\gamma)
    &=
    \sum_{\vec{\mathcal{S}}}
    e^{-\gamma w_{\vec{\mathcal{S}}}}
    A_{\vec{\mathcal{S}}}
    =
    \sum_w e^{-\gamma w}R(w),
    \label{eq:app_generalized_reactivity}\\
    R(w)
    &=
    \sum_{\vec{\mathcal{S}}}
    A_{\vec{\mathcal{S}}}
    \delta_{w_{\vec{\mathcal{S}}},w}.
\end{align}
Thus paths that differ microscopically but have the same accumulated noise weight contribute to
the same reactivity bin, Fig.~\ref{fig:generalized_paths}(b). Equation~\eqref{eq:app_generalized_reactivity}
is the Laplace relation of the main text, now obtained without assuming Pauli noise.

\subsection{Spectral paths: Biorthogonal eigendecomposition}
\label{app:biorthogonal}

Equation~\eqref{eq:app_channel_ansatz} does not require the components
$\mathcal{S}_t$ to be spectral modes. A particularly useful realization, however, is obtained from
the eigenmodes of the noise channel itself. This covers a broad class of physical noise processes
and gives the generalized paths a direct interpretation as paths through operator space.

\zpar{Left and right noise modes}
Consider a diagonalizable noise channel $\mathcal{N}_t(\gamma)$ on
$\mathcal{L}(\mathcal{H})$. Since a general channel need not be normal, its eigenoperators need
not be orthogonal. We therefore introduce biorthogonal right and left modes
$|L_{\mathcal{S}_t}\rrangle$ and $\llangle R_{\mathcal{S}_t}|$ satisfying
\begin{align}
    \mathcal{N}_t(\gamma)|L_{\mathcal{S}_t}\rrangle
    &=
    e^{-\gamma w_{\mathcal{S}_t}}
    |L_{\mathcal{S}_t}\rrangle,\\
    \llangle R_{\mathcal{S}_t}|\mathcal{N}_t(\gamma)
    &=
    e^{-\gamma w_{\mathcal{S}_t}}
    \llangle R_{\mathcal{S}_t}|,\\
    \llangle R_{\mathcal{S}_t}|L_{\mathcal{S}'_t}\rrangle
    &=
    \delta_{\mathcal{S}_t,\mathcal{S}'_t}.
\end{align}
The channel then takes the form
\begin{equation}
    \mathcal{N}_t(\gamma)
    =
    \sum_{\mathcal{S}_t}
    e^{-\gamma w_{\mathcal{S}_t}}
    |L_{\mathcal{S}_t}\rrangle
    \llangle R_{\mathcal{S}_t}|.
    \label{eq:app_merge}
\end{equation}
Thus each mode has a fixed direction in operator space and a scalar response
$e^{-\gamma w_{\mathcal{S}_t}}$. For normal channels the left and right modes may be chosen to
coincide; Pauli channels are the simplest example, with Pauli eigenoperators and real weights.
For non-normal channels the two mode families differ, while complex eigenvalues naturally produce
complex weights. In general,
\begin{equation}
    \llangle R_{\mathcal{S}_t}|R_{\mathcal{S}'_t}\rrangle
    \neq
    \llangle L_{\mathcal{S}_t}|L_{\mathcal{S}'_t}\rrangle
    \neq
    \delta_{\mathcal{S}_t,\mathcal{S}'_t}.
\end{equation}


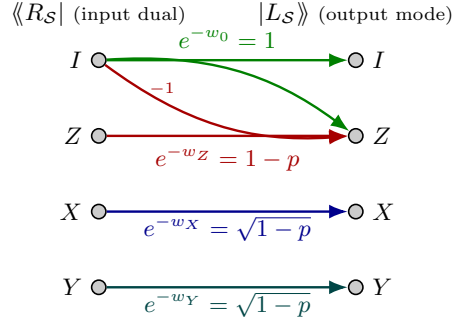
\begin{figure}[t]
    \centering
    \begin{tikzpicture}[
            font=\small,
            dot/.style={circle, draw=black, fill=black!20, line width=0.6pt, inner sep=1.9pt},
            arrx/.style={-Latex, line width=0.9pt, draw=blue!55!black},
            arry/.style={-Latex, line width=0.9pt, draw=teal!55!black},
            arrg/.style={-Latex, line width=0.9pt, draw=green!50!black},
            arrr/.style={-Latex, line width=0.9pt, draw=red!65!black},
            lab/.style={inner sep=1pt},
        ]
        \def\dy{10mm}
        \def\dx{34mm}

        \node[dot, label=left:$I$] (SI) at (0,0)       {};
        \node[dot, label=left:$Z$] (SZ) at (0,-\dy)    {};
        \node[dot, label=left:$X$] (SX) at (0,-2*\dy)  {};
        \node[dot, label=left:$Y$] (SY) at (0,-3*\dy)  {};

        \node[dot, label=right:$I$] (OI) at (\dx,0)      {};
        \node[dot, label=right:$Z$] (OZ) at (\dx,-\dy)   {};
        \node[dot, label=right:$X$] (OX) at (\dx,-2*\dy) {};
        \node[dot, label=right:$Y$] (OY) at (\dx,-3*\dy) {};

        \node[above=0.7em of SI] {$\llangle R_{\mathcal{S}}|$ \scriptsize(input dual)};
        \node[above=0.7em of OI] {$|L_{\mathcal{S}}\rrangle$ \scriptsize(output mode)};

        \draw[arrx] (SX) -- node[midway, below, lab, text=blue!55!black]
        {$e^{-w_X}=\sqrt{1-p}$} (OX);
        \draw[arry] (SY) -- node[midway, below, lab, text=teal!55!black]
        {$e^{-w_Y}=\sqrt{1-p}$} (OY);

        \draw[arrg] (SI) to[bend left=0]   (OI);
        \draw[arrg] (SI) to[bend left=20]  (OZ);
        \node[lab, text=green!50!black] at ($(SI)!0.5!(OI) + (0,+3mm)$) {$e^{-w_0}=1$};

        \draw[arrr] (SZ) to[bend right=0] (OZ);
        \draw[arrr] (SI) to[bend right=20]
        node[pos=0.15, right, text=red!65!black] {$\scriptstyle -1$} (OZ);
        \node[lab, text=red!65!black] at ($(SZ)!0.5!(OZ) + (0,-3mm)$) {$e^{-w_Z}=1-p$};

    \end{tikzpicture}

    \caption{\label{fig:amp-damping-flow}
        \textbf{Mode-resolved action of the amplitude-damping channel.}
        For a damping probability $p$, the channel is diagonalizable but non-normal, so its left and
        right modes need not coincide. Following the convention of Eq.~\eqref{eq:app_merge}, the
        left column shows the operators sampled by $\llangle R_{\mathcal S}|$, while the right column
        shows the corresponding output directions $|L_{\mathcal S}\rrangle$; each colored set of arrows
        represents one spectral mode, with unlabeled arrows carrying unit coefficient. The transverse
        $X$ and $Y$ modes coincide on the two sides and decay as
        $e^{-w_X}=e^{-w_Y}=\sqrt{1-p}$. The non-normal structure appears in the population sector:
        the invariant mode, $e^{-w_0}=1$, connects the identity component to the steady-state direction,
        whereas the decaying population mode, $e^{-w_Z}=1-p$, has distinct left and right structure.
        Thus amplitude damping provides a simple example in which a fixed biorthogonal eigenbasis
        captures noise scaling even though an orthogonal mode decomposition does not, with
        $w_X=w_Y=w_Z/2$.}
\end{figure}

\zpar{A non-normal example}
Amplitude damping makes the biorthogonal structure concrete. Writing
$p=1-e^{-\gamma}$, the spectral factors are
$1$, $e^{-\gamma}$, and $e^{-\gamma/2}$, the latter twice in the transverse sector. Hence the
channel has the form of Eq.~\eqref{eq:app_merge}, with weights $w=0,1,$ and $1/2$.
The parameter $\gamma=-\log(1-p)$ is simply the noise coordinate in which these factors are
exponential; the physical damping probability need not itself vary linearly with $\gamma$.
The channel is diagonalizable but non-normal: left and right modes coincide for the transverse
coherences but differ in the population sector. Figure~\ref{fig:amp-damping-flow} shows this
mode-resolved action.

\zpar{Paths through operator space}
Inserting Eq.~\eqref{eq:app_merge} at every layer gives the spectral analogue of the main text's
Pauli path expansion, Eq.~\eqref{eq:pauli_paths}, now written in the generalized eigenbasis,
\begin{align}
    C(\gamma)
    &=
    \sum_{\vec{\mathcal{S}}}
    e^{-\gamma w_{\vec{\mathcal{S}}}}
    A_{\vec{\mathcal{S}}},
    \label{eq:app_spectral_paths}\\
    A_{\vec{\mathcal{S}}}
    &=
    \llangle O|
    \prod_t
    |L_{\mathcal{S}_t}\rrangle
    \llangle R_{\mathcal{S}_t}|
    \mathcal{U}_t
    |\rho_{\mathrm{init}}\rrangle,\\
    w_{\vec{\mathcal{S}}}
    &=
    \sum_t w_{\mathcal{S}_t}.
\end{align}
The path
$\vec{\mathcal{S}}=(\mathcal{S}_1,\ldots,\mathcal{S}_T)$
records the sequence of noise modes through which the circuit propagates information: the unitary
layers transfer amplitude between modes, while the noise attenuates or phases them.

Grouping paths by accumulated weight gives
\begin{align}
    R(w)
    &=
    \sum_{\vec{\mathcal{S}}}
    A_{\vec{\mathcal{S}}}
    \delta_{w_{\vec{\mathcal{S}}},w},\\
    C(\gamma)
    &=
    \sum_w e^{-\gamma w}R(w).
    \label{eq:app_native_reactivity}
\end{align}
The reactivity therefore retains the net contribution at each accumulated weight while discarding
the microscopic sequence of modes. In the Pauli limit, the modes become Pauli strings and
Eq.~\eqref{eq:app_native_reactivity} reduces directly to the Pauli path reactivity of the main
text.

\zpar{Remark: Why spectral modes?}
The useful decomposition is one whose modes remain fixed as the noise strength $\gamma$ is varied.
This is generally not true for an SVD: for non-normal channels the singular vectors may depend on
$\gamma$, while for coherent unitary errors the singular values are all unity and discard the
accumulated phase. A left-right spectral decomposition avoids both problems when the noise is
generated by a fixed diagonalizable generator,
\begin{equation}
    \mathcal{N}_t(\gamma)
    =
    e^{\gamma\mathcal{L}_t}
    =
    \sum_{\mathcal{S}_t}
    e^{-\gamma w_{\mathcal{S}_t}}
    |L_{\mathcal{S}_t}\rrangle
    \llangle R_{\mathcal{S}_t}|,
    \label{eq:app_spectral_amplification}
\end{equation}
so that varying $\gamma$ changes only the eigenvalues, while the modes---and hence the path
amplitudes---remain fixed.

\zpar{Remark: Scope of the exponential form}
The spectral construction above is a sufficient, but not necessary, realization of
Eq.~\eqref{eq:app_channel_ansatz}. What is required for the path expansion is only that varying
$\gamma$ changes scalar factors $e^{-\gamma w_{\mathcal{S}_t}}$ multiplying fixed,
$\gamma$-independent superoperators $\mathcal{S}_t$. These components need not themselves be
spectral projectors.

A particularly important sufficient condition is a fixed diagonalizable generator,
$\mathcal{N}_t(\gamma)=e^{\gamma\mathcal{L}_t}$, for which the decomposition above follows
directly. If the generator is nondiagonalizable, Jordan blocks additionally produce factors of the
form $\gamma^m e^{-\gamma w}$; such cases require an enlarged modification kernel and are outside
the scope considered here.

\zpar{Example: a coherent-noise error}
A coherent rotation illustrates a different feature of the same spectral construction. Consider
the single-qubit noise channel generated by
\begin{equation}
    U_\gamma
    =
    R_z(\gamma\theta_0)
    =
    e^{-i\gamma\theta_0 Z/2},
    \qquad
    \mathcal{N}_{\mathrm{coh}}(\gamma)(\rho)
    =
    U_\gamma\rho U_\gamma^\dagger .
\end{equation}
The channel is normal, so its left and right modes may be chosen to coincide. The operators $I$
and $Z$ are fixed, while the transverse combinations satisfy
\begin{equation}
    \mathcal{N}_{\mathrm{coh}}(\gamma)
    \left(
        \frac{X\mp iY}{\sqrt{2}}
    \right)
    =
    e^{\pm i\gamma\theta_0}
    \left(
        \frac{X\mp iY}{\sqrt{2}}
    \right).
\end{equation}
The corresponding generalized weights are therefore $w=0$ for $I$ and $Z$, and
$w=\mp i\theta_0$ for the two transverse modes. In contrast with amplitude damping, these modes
are not attenuated: their weights are purely imaginary and describe coherent phase accumulation.
This simple example motivates the complex weights considered in the next subsection.

\subsection{Reality of the generalized reactivity under complex-valued spectral paths}

Complex spectral weights arise naturally when noise contains coherent dynamics. Individual terms
in the path expansion may then be complex, even though the measured expectation value is real.
The required reality follows directly from Hermiticity preservation.

\zpar{Conjugation symmetry}
Let $\rho_{\mathrm{init}}$ and $O$ be Hermitian. For a Hermiticity-preserving circuit map,
complex spectral modes occur in conjugate pairs: a mode with factor
$e^{-\gamma w}$ is accompanied by one with factor $e^{-\gamma w^*}$.
The corresponding operator paths likewise have conjugate weights and amplitudes. Grouping them
by accumulated weight therefore gives
\begin{equation}
    R(w^*) = R(w)^*.
    \label{eq:app_reactivity_conjugation}
\end{equation}
In particular, $R(w)$ is real whenever $w$ is real. Complete positivity is not required for this
statement; Hermiticity preservation alone is sufficient.

\zpar{Decay and oscillation}
A conjugate pair contributes
\begin{align}
    e^{-\gamma w}R(w)
    +
    e^{-\gamma w^*}R(w^*)
    &=
    2e^{-\gamma\,\mathrm{Re}(w)}
    \mathrm{Re}\!\left[
        R(w)e^{-i\gamma\,\mathrm{Im}(w)}
    \right].
    \label{eq:app_manifestly_real}
\end{align}
Thus $\mathrm{Re}(w)$ describes attenuation, while $\mathrm{Im}(w)$ describes coherent
oscillation, with conjugate pairs ensuring a real measured signal. The same reactivity framework
therefore extends naturally from real to complex weights.

For example, the transverse modes of the coherent $Z$ rotation above have conjugate imaginary
weights $w=\mp i\theta_0$, and their contributions combine according to
Eq.~\eqref{eq:app_manifestly_real} to give a real signal.

\begin{figure*}[thb]
    \centering
    \includegraphics[width=\textwidth]{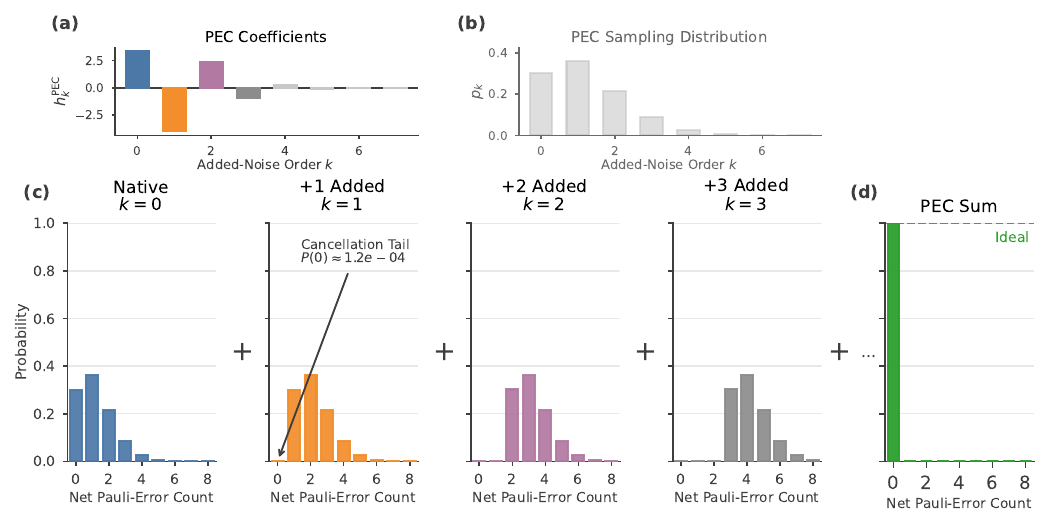}
    \caption{\label{fig:pec-discrete-insertion}
    \textbf{Discrete-noise-insertion picture of probabilistic error cancellation.}
    Illustrative single-rate depolarizing model with $V=1000$ noise locations and
    $\bar{k}=3V\gamma_0/4=1.2$, where $\gamma_0$ is the native noise rate and $N\gamma_0$ the
    circuit's total noise exposure.
    \textbf{(a)}~The PEC coefficients $h_k^{\mathrm{PEC}}$ of Eq.~\eqref{eq:pec_filter_main},
    which alternate in sign and decay factorially.
    \textbf{(b)}~The distribution actually sampled in PEC, obtained by normalizing the magnitudes
    in (a), $p_k=|h_k^{\mathrm{PEC}}|/\sum_j|h_j^{\mathrm{PEC}}|$. Because
    $\sum_k|h_k^{\mathrm{PEC}}|=e^{2\bar{k}}$, this is exactly a Poisson distribution of mean
    $\bar{k}$, $p_k=e^{-\bar{k}}\bar{k}^k/k!$, and its normalization fixes the sampling overhead
    $X_{\mathrm{PEC}}=\big(\sum_k|h_k^{\mathrm{PEC}}|\big)^2=e^{4\bar{k}}$ of
    Eq.~\eqref{eq:pec_cost_main}.
    \textbf{(c)}~Each experiment $C(\gamma_0,k)$ with $k$ inserted Pauli errors yields a
    distribution over net Pauli-error count; increasing $k$ shifts weight to larger counts, with a
    small zero-error tail because an inserted Pauli can cancel a native one at the same location.
    \textbf{(d)}~Combining the experiments in (c) with the signed weights of (a) cancels every
    nonzero-error sector and isolates the ideal noise-free contribution.}
\end{figure*}

\begin{figure*}[hbt]
    \centering
    \includegraphics[width=\textwidth]{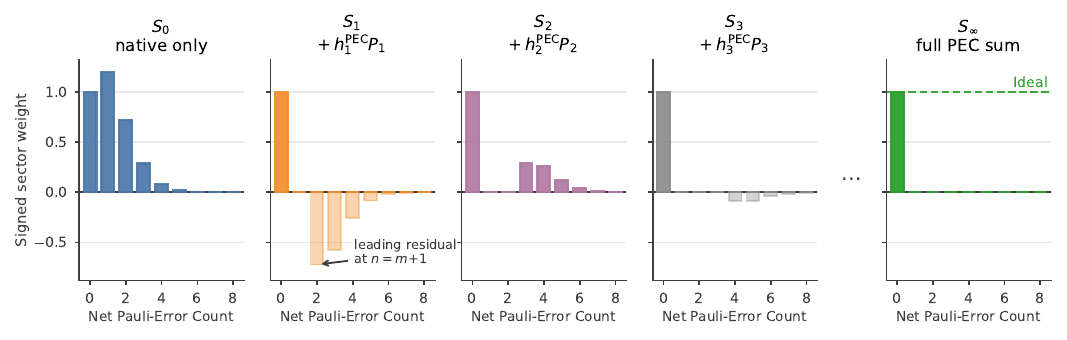}
    \caption{\label{fig:pec-convergence}
    \textbf{Convergence of the PEC reconstruction.}
    Partial sums $S_m(n)=\sum_{k=0}^{m}h_k^{\mathrm{PEC}}P_k(n)$ of the experiments in
    Fig.~\ref{fig:pec-discrete-insertion}(c), where $P_k(n)$ is the distribution over net
    Pauli-error count $n$ underlying $C(\gamma_0,k)$. Bars are signed weights rather than
    probabilities: solid for positive, faded for negative. The zero-error sector is already exact
    at $m=0$, since $h_0^{\mathrm{PEC}}P_0(0)=e^{\bar{k}}e^{-\bar{k}}=1$. Each additional order
    cancels one further sector, so $S_m$ vanishes on $n=1,\dots,m$ and leaves a leading residual at
    $n=m+1$ of magnitude $\bar{k}^{m+1}/(m+1)!$ with sign $(-1)^m$. The residual thus moves to
    larger $n$ and shrinks factorially, converging to the ideal contribution (green). Same model
    as Fig.~\ref{fig:pec-discrete-insertion}.}
\end{figure*}

\subsection{Discrete noise insertion}
\label{app:discrete_insertion}

Continuous amplification probes the reactivity by scaling the noise already present in the
device. A complementary control is to leave the native channels
$\mathcal{N}_t(\gamma_t)$ unchanged and deliberately insert additional noise events. The same
reactivity $R(w)$ describes the circuit in both cases; only the kernel through which it is probed
changes. We derive that kernel here for stochastic Pauli noise
\begin{equation}
    \mathcal{N}_t
    =
    e^{-\gamma_t\mathcal{V}_t},
    \qquad
    \mathcal{V}_t(P)
    =
    \frac{3 N}{4}
    \left(
        P-\sum_E q_t(E)EPE^\dagger
    \right),
\end{equation}
which is diagonal in the Pauli basis.

\zpar{Normalization of the Pauli weight}
For a stochastic Pauli channel, define
    $s_{E,P}
    =
    \frac{1}{2^N}
    \mathrm{Tr}\!\left(EPE^\dagger P\right)
    =
    \pm1,$
and write its mode weight as
\begin{equation}
    \bar w_{P_t}
    =
    \frac{3 N}{4}
    \left(
        1-\sum_E q_t(E)s_{E,P_t}
    \right).
    \label{eq:app_pauli_weight}
\end{equation}
The normalization $3N/4$ is chosen so that $\bar w_{P_t}$ reduces to the ordinary Pauli Hamming
weight for uniform single qubit depolarizing noise. Indeed, for a Pauli string $P_t$ of Hamming weight
$w_{P_t}$,
\begin{equation}
    \sum_E q_t(E)s_{E,P_t}
    =
    1-\frac{4w_{P_t}}{3 N},
    \label{eq:app_mean_sign}
\end{equation}
and hence $\bar w_{P_t}=w_{P_t}$. For a general Pauli noise model, $\bar w_{P_t}$ remains real but need
not be integer valued.

At the native layer strengths $\gamma_t$, the accumulated noise exposure of a Pauli path is
\begin{equation}
    w_{\vec P}
    =
    \sum_t \gamma_t \bar w_{P_t},
    \label{eq:app_native_path_weight}
\end{equation}
so that
\begin{equation}
    C_{\mathrm{noisy}}
    =
    \sum_{\vec P}
    e^{-w_{\vec P}}A_{\vec P}
    =
    \sum_w e^{-w}R(w).
    \label{eq:pauli_path_v}
\end{equation}
Under a homogeneous amplification $\gamma_t\rightarrow r\gamma_t$, the same reactivity is probed
as $C(r)=\sum_w e^{-rw}R(w)$. Thus the normalization above fixes what is meant by one unit of
weight without changing any physical prediction.

\zpar{One inserted noise event}
We now insert one additional Pauli error sampled from the same learned noise model. Define
\begin{equation}
    \gamma_{\Sigma}
    :=
    \frac{1}{T}\sum_t \gamma_t,
\end{equation}
and sample the layer $t$ and Pauli error $E$ with probability
\begin{equation}
    p(E,t)
    =
    \frac{\gamma_t q_t(E)}{T\gamma_{\Sigma}}.
    \label{eq:app_insert_probability}
\end{equation}
For a fixed path $\vec P$, the insertion multiplies its amplitude by $s_{E,P_t}$. Averaging over
the inserted event therefore gives
\begin{align}
    f_1(w_{\vec P})
    &=
    \sum_{t,E}
    \frac{\gamma_tq_t(E)}{T\gamma_{\Sigma}}
    s_{E,P_t}
    \nonumber\\
    &=
    1-
    \frac{4}{3NT\gamma_{\Sigma}}
    \sum_t\gamma_t\bar w_{P_t}
    \nonumber\\
    &=
    1-\frac{4w_{\vec P}}{3NT\gamma_{\Sigma}}.
    \label{eq:app_one_insertion}
\end{align}
After averaging, the microscopic path dependence has disappeared: the inserted event acts only
through the accumulated weight $w_{\vec P}$. Consequently,
\begin{equation}
    C_{\mathrm{noisy}}(1)
    =
    \sum_w
    e^{-w}
    \left(
        1-\frac{4w}{3NT\gamma_{\Sigma}}
    \right)
    R(w).
\end{equation}

%
%

\providecommand{\figScaleWide}{0.92}

\begin{figure*}[t]
    \centering
    \scalebox{\figScaleWide}{%
        \begin{tikzpicture}[
                >=Latex, font=\small,
                wire/.style={line width=1.0pt},
                tick/.style={gray!55, line width=0.6pt},
                ins/.style={cross out, draw, line width=1.0pt, minimum size=4.4pt, inner sep=0pt},
                lbl/.style={font=\scriptsize, align=center, inner sep=1.5pt},
                panel/.style={font=\small\bfseries},
                kplus/.style={line width=1.2pt, draw=blue!65!black},
                kk2/.style={line width=1.2pt, draw=violet!70!black},
                kk3/.style={line width=1.2pt, draw=orange!80!black},
            ]

            \node[panel] at (-0.35,2.35) {(a)};

            \draw[wire] (0.55,0) -- (7.15,0);
            \node[font=\small] at (0.15,0) {$\rho$};
            \node[font=\small] at (7.55,0) {$O$};

            \foreach \x/\t in {1.2/1, 2.2/2, 3.2/3, 4.2/4, 5.2/5, 6.2/6}{
                    \draw[tick] (\x,-0.22) -- (\x,0.22);
                    \node[lbl, gray!60!black] at (\x,-0.5) {$\t$};
                }
            \node[lbl, gray!60!black] at (3.7,-0.95) {layer $t$};

            \node[ins, red!65!black] (e1) at (2.2,0) {};
            \node[ins, red!65!black] (e2) at (4.2,0) {};
            \node[ins, red!65!black] (e3) at (5.2,0) {};
            \node[lbl, red!65!black] at (2.2,0.95) {$E$\\$s_{E,P_2}=+1$};
            \node[lbl, red!65!black] at (4.2,1.55) {$E'$\\$s_{E',P_4}=-1$};
            \node[lbl, red!65!black] at (5.2,0.95) {$E''$\\$s_{E'',P_5}=-1$};
            \draw[red!65!black, dashed, line width=0.5pt] (e2) -- (4.2,1.15);

            \node[lbl, anchor=north, text width=7.4cm] at (3.6,-1.35)
            {one error sampled at $(E,t)$ with probability
                $p(E,t)=\gamma_t q_t(E)/(T\gamma_{\Sigma})$; each choice flips the path
                amplitude by its own sign $s_{E,P_t}=\pm1$};

            \draw[->, line width=1.1pt, gray!55!black] (8.15,0) -- (9.45,0);
            \node[lbl, gray!45!black, anchor=south, text width=2.2cm] at (8.8,0.18)
            {average over\\$(E,t)$};
            \node[lbl, gray!45!black, anchor=north, text width=2.6cm] at (8.8,-0.22)
            {microscopics drop out; only $w$ survives};

            \node[panel] at (10.05,2.35) {(b)};

            \def\ox{10.9} \def\oy{-1.15} \def\sx{2.35} \def\sy{1.45}
            \draw[->, gray!70!black] (\ox-0.15,\oy+\sy) -- (\ox+2.25*\sx,\oy+\sy)
            node[right, font=\scriptsize] {$w/w_0$};
            \draw[->, gray!70!black] (\ox,\oy-0.55) -- (\ox,\oy+2.25*\sy)
            node[above, font=\scriptsize] {$f_k(w)$};
            \foreach \u in {0,1,2}{
                    \draw[gray!55] (\ox+\u*\sx,\oy+\sy-0.06) -- (\ox+\u*\sx,\oy+\sy+0.06);
                    \node[font=\scriptsize] at (\ox+\u*\sx,\oy+\sy-0.3) {$\u$};
                }
            \node[font=\scriptsize, anchor=east] at (\ox-0.1,\oy+2*\sy) {$1$};
            \node[font=\scriptsize, anchor=east] at (\ox-0.1,\oy) {$-1$};
            \draw[gray!55] (\ox-0.06,\oy+2*\sy) -- (\ox+0.06,\oy+2*\sy);
            \draw[gray!55] (\ox-0.06,\oy) -- (\ox+0.06,\oy);

            \draw[gray!60, dashed] (\ox+\sx,\oy-0.35) -- (\ox+\sx,\oy+2.2*\sy);
            \node[lbl, gray!40!black, anchor=south] at (\ox+\sx,\oy+2.2*\sy)
            {$w_0=\tfrac{3}{4}NT\gamma_{\Sigma}$};

            \draw[kplus, domain=0:2, samples=120, smooth]
            plot (\ox+\x*\sx, {\oy+\sy+(1-\x)*\sy});
            \draw[kk2, domain=0:2, samples=160, smooth]
            plot (\ox+\x*\sx, {\oy+\sy+pow(1-\x,2)*\sy});
            \draw[kk3, domain=0:2, samples=160, smooth]
            plot (\ox+\x*\sx, {\oy+\sy+pow(1-\x,3)*\sy});

            \node[lbl, text=blue!65!black]   at (\ox+1.52*\sx, \oy+\sy-0.92*\sy) {$k=1$};
            \node[lbl, text=violet!70!black] at (\ox+1.52*\sx, \oy+\sy+0.62*\sy)  {$k=2$};
            \node[lbl, text=orange!80!black] at (\ox+1.52*\sx, \oy+\sy-0.30*\sy)  {$k=3$};

            \node[lbl, anchor=north, text width=5.6cm] at (\ox+1.1*\sx, \oy-0.62)
            {$f_k(w)=\bigl(1-w/w_0\bigr)^{k}$: the same reactivity, read through a
                tunable kernel};

        \end{tikzpicture}}

    \caption{\label{fig:discrete-insertion-kernel}
        \textbf{From microscopic noise insertions to a weight-selective response.}
        \textbf{(a)}~For a single insertion, an error $E$ and layer $t$ are sampled jointly according to
        $p(E,t)=\gamma_t q_t(E)/(T\gamma_{\Sigma})$. The crosses illustrate alternative possible insertion
        events. For a particular Pauli path $\vec P$, each realization contributes the local factor
        $s_{E,P_t}=\pm1$, which depends on both the sampled error and the Pauli operator carried by the path
        at that layer.
        \textbf{(b)}~For the Pauli noise model considered here, averaging over $(E,t)$ collapses this
        microscopic dependence to the accumulated path weight $w$. One insertion produces
        $f_1(w)=1-w/w_0$, where $w_0=3NT\gamma_{\Sigma}/4$, and $k$ independently sampled insertions give
        $f_k(w)=(1-w/w_0)^k$. These curves are the insertion factors multiplying the native attenuation
        $e^{-w}$, so that
        $C_{\mathrm{noisy}}(k)=\sum_w e^{-w}f_k(w)R(w)$.
        The zero-weight sector is unchanged, all contributions at $w=w_0$ are removed for $k\geq1$, and
        for $w>w_0$ the insertion factor changes sign for odd $k$. Discrete insertion therefore leaves the
        underlying reactivity $R(w)$ unchanged while controllably reweighting the different weight sectors
        through which it is observed.}
\end{figure*}
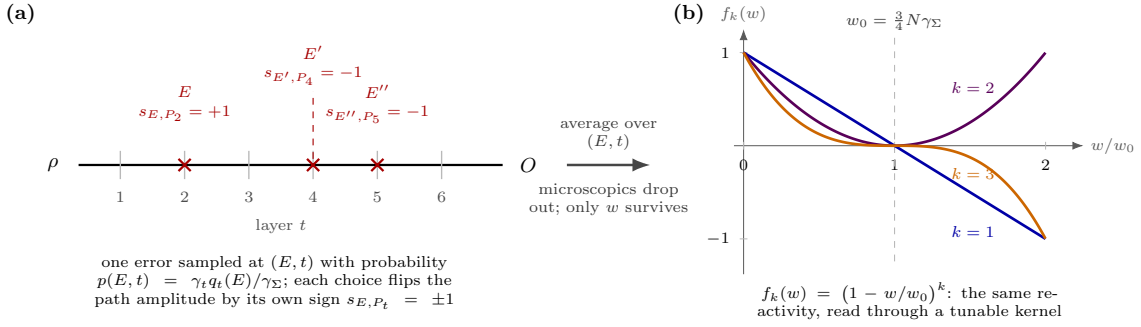

\zpar{A fixed number of inserted events}
If exactly $k$ events are sampled independently according to
Eq.~\eqref{eq:app_insert_probability}, the factors multiply and
\begin{equation}
    C_{\mathrm{noisy}}(k)
    =
    \sum_w
    e^{-w}
    \left(
        1-\frac{4w}{3NT\gamma_{\Sigma}}
    \right)^k
    R(w).
    \label{eq:app_discrete_kernel}
\end{equation}
Thus the discrete modification function is
\begin{equation}
    F_{\mathrm{disc}}(k,w)
    =
    e^{-w}
    \left(
        1-\frac{4w}{3NT\gamma_{\Sigma}}
    \right)^k.
    \label{eq:app_F_discrete}
\end{equation}
This is the discrete counterpart of the exponential kernel for continuous noise amplification.
Importantly, fixing the total number $k$ of inserted events correlates the circuit modification
globally, so the modified ensemble need not itself factorize into independent layer channels.
Nevertheless, it acts on the same reactivity through the known kernel
Eq.~\eqref{eq:app_F_discrete}.%

The relation between the discrete and exponential kernel is stronger than the small-weight approximation $[1-(4w)/(3NT\gamma_{\Sigma})]^k \approx \exp[-(4wk)/(3NT\gamma_{\Sigma})]$. In fact,
\begin{align}
    e^{-rw}
    &=
    e^{-w}
    \sum_{k=0}^{\infty}
    \exp\!\left[-\frac{3NT\gamma_{\Sigma}}{4}(r-1)\right]
    \frac{
        \left[\frac{3NT\gamma_{\Sigma}}{4}(r-1)\right]^k
    }{k!}
    \left(
        1-\frac{4w}{3NT\gamma_{\Sigma}}
    \right)^k .
    \label{eq:app_poisson_insertion}
\end{align}
Thus homogeneous amplification by a factor $r$ can be reproduced exactly by drawing the number of
inserted noise events from a Poisson distribution with mean
$3NT\gamma_{\Sigma}(r-1)/4$. Continuous amplification is therefore a particular probabilistic
mixture of the fixed-$k$ insertion experiments.

Setting $r=0$ in Eq.~\eqref{eq:app_poisson_insertion} reproduces exactly the PEC coefficients
$h_k^{\mathrm{PEC}}=e^{\bar{k}}(-\bar{k})^k/k!$, with $\bar{k}=3NT\gamma_{\Sigma}/4$ fixed by the
circuit's total noise exposure; for the single native rate $\gamma_0$ of the main text this is
$\bar{k}=3V\gamma_0/4$, recovering Eq.~\eqref{eq:pec_filter_main}. Figure~\ref{fig:pec-discrete-insertion} illustrates this
discrete-insertion picture of PEC in a simplified single-rate depolarizing model.

Truncating the PEC sum at order $m$ has a simple structure in this picture. Because
$h_0^{\mathrm{PEC}}$ and the native zero-error weight are reciprocal, the noise-free sector is
already reproduced exactly at $m=0$; each subsequent order then removes one further error sector,
leaving a residual at net error count $m+1$ of size $\bar{k}^{m+1}/(m+1)!$ with alternating sign.
The residual therefore retreats to higher error counts and shrinks factorially with the number of
retained terms, as shown in Fig.~\ref{fig:pec-convergence}.

\section{Numerical calculation of reactivity functions}\label{app:numerics_calc_react}

In this section we discuss the numerical procedures used to obtain the reactivities $R(w)$ for single qubit depolarizing channels.
The techniques developed here generalize to arbitrary noise channels in analogy to the generalization of the reactivity representation in App.~\ref{app:react_Pauli paths}.
We explore two different approaches, direct Pauli propagation simulation~\cite{rudolph2025pauli,beguvsic2025real} of the Pauli paths and density matrix-based simulation.
In the main text, we solely present results obtained via density matrix simulations as they are exact and efficient with circuit depth, but they are restricted to system sizes of about $16$ qubits.
Pauli propagation can be used for larger systems but only for circuits of moderate depth.

\subsection{Pauli propagation simulation}
In order to obtain a reactivity function, we can apply Pauli propagation to simulate the Heisenberg time evolution of the measured operator $O$. We first superficially introduce Pauli propagation and then explain how to modify it to obtain the reactivity $R(w)$.

We start by decomposing the measurement operator $O$ into Pauli strings
\begin{align}
    O(0) = \sum_P c_{0,P} P,
\end{align}
storing only terms with non-vanishing coefficients $|c_{0,P}|> 0$. Often, $O$ is itself a single Pauli string or a sum over only a few Pauli strings.
Then we iterate over circuit layers indexed by $t$,
\begin{align}
    O(t+1) = \sum_P c_{t,P} U_t PU_t^\dagger = \sum_P c_{t+1,P} P,
\end{align}
with gate update rules described in Refs.~\cite{rudolph2025pauli,beguvsic2025real}.
At any point in time, we store the (sparse) Pauli string decomposition of $O(t)$ and employ minimally invasive pruning of the Pauli sum by discarding the Pauli strings with the smallest coefficients. Using the decomposition of $O(t)$ in terms of Pauli strings, the expectation value with respect to a state $|\Psi\rangle$ can be calculated as
\begin{align}
    \langle \Psi| O(t)|\Psi\rangle  = \sum_P c_{t,P} \langle \Psi| P|\Psi\rangle.
\end{align}

In order to calculate the reactivity $R(w)$, we now further separate the Pauli strings according to the path weight $w$. Remember that the path weight is the cumulative sum of Pauli weights, i.e., the cumulative number of non-identity Paulis in a Pauli string throughout the circuit. We start at
\begin{align}
    O(0) = \sum_w \sum_{P} c_{0,w,P} P,
\end{align}
where we added an index $w$ to $c$ to denote the path weight, which is initialized as the Pauli weight $|P|$.
One granular approach to track path weight is to attach it to every Pauli string inside the Pauli sum and adapt the gate update rules accordingly. That is, in every iteration we calculate
\begin{align}
    O(t+1) = \sum_w \sum_{P} c_{t,w,P} U_t PU_t^\dagger = \sum_w \sum_{P} c_{t+1,w,P}  P,
\end{align}
but update coefficients in a weight-resolved way,
\begin{align}
    c_{t+1,w,P} = \sum_{w'} \delta_{w,w'+|P|} \sum_{P'} c_{t,w',P'} \mathrm{Tr}[P U_t P'U_t^\dagger]/2^N.
\end{align}
Here we group the new set of Pauli strings by the new path weights, which is their old path weight plus the Pauli weight of the newly generated Pauli string $w=w'+|P|$.
In practice, this means that, instead of adding together coefficients of identical Pauli strings after gates create duplicates, we only add their coefficients if $w$ is also identical.
This then yields the reactivity
\begin{align}
    R(t,w) =  \sum_P c_{t,w,P}  \langle \Psi| P|\Psi\rangle.
\end{align}

Note that there is a choice in when to update the path weight. Above we increment it per layer by adding the current Pauli weight to it. To be consistent with the definition of the noise channels used in the main text,
we can instead increment the path weight at every gate by the number of non-identity Paulis that the gate acted on.

This Pauli propagation approach for computing the reactivity is in principle scalable to an arbitrary number of qubits. However, it scales with an additional linear factor in the number of gates worse than conventional Pauli propagation, which already scales possibly exponentially with the number of non-Clifford gates. For shallow to moderate-depth quantum circuits with many qubits, this is a feasible trade-off. At scales where exact simulation is possible, however, the following Fourier transform approach scales more favorably for deeper circuits.

\subsection{Fourier Transformation}\label{app:density_matrix_fourier_trafo}
The reactivity $R(w)$ can be straightforwardly calculated as a Fourier transform. Recall Eq.~\eqref{eq:laplace}
\begin{equation}
    C(\gamma)=\sum_w e^{-\gamma w}R(w),
\end{equation}
which shows that $C(\gamma)$ is the Laplace transform of the reactivity. By making $\gamma$ complex $\gamma \rightarrow i \nu$, $R(w)$ becomes the Fourier transform of $C(i \nu)$
\begin{equation}
    C(i\nu)=\sum_w e^{-i\nu w}R(w).
\end{equation}
Hence, we can simulate $C(i\nu)$ in a density matrix simulation and obtain $R(w)$ as its Fourier transform.

For the single qubit depolarizing channel, we perform the density matrix simulation replacing the single qubit depolarizing channel with its complex variant
\begin{equation}
    \mathcal{N}_C(\nu)[\rho] = \frac{1}{4}\left( \rho + X \rho X  + Y \rho Y + Z \rho Z\right) + \frac{e^{-i\nu}}{4}\left( 3\rho - X \rho X  - Y \rho Y - Z \rho Z\right).
\end{equation}
This is just the projector on $\rho=I$ plus $e^{-i \nu}$ times the projector on the $X,Y,Z$ subspace --- the two subspaces of the depolarizing channel with different eigenvalues.
Inserting this channel after every two qubit gate (or where the Hamming distance $w$ is to be measured) yields
\begin{align}
    C(i \nu) = \sum_{w} e^{-i \nu w} R(w),
\end{align}
where $R(w)$ is now the amplitude of the oscillation of $C(i \nu)$ with frequency $w$. We simulate $C(i \nu)$ for $\nu=v 2\pi/V$ for $v$ in $\{0,1,\ldots,V-1\}$, where $V$ is the expected maximal circuit volume. Note that $C(i \nu)^*=C(-i \nu)=C(i(2\pi-\nu))$ allowing us to reuse simulation results. Then
\begin{align}
    R(w) = \frac{1}{V}\sum_{v=0}^{V-1} e^{i w v 2 \pi/V} C(iv 2 \pi/ V)
\end{align}
is the reactivity used to calculate $C(\gamma)$ in the main text.

\subsection{Model selection and combination in a realistic setting}\label{sec:optimal_combination_estimates}
In the main text we present a range of error mitigation algorithms that trade between two sources of noise: model bias and statistical variance. We seek to combine these different estimates to yield a final estimate without prior knowledge of the noiseless answer. In this section, we present three separate methods to do so. We find the best result combines the estimates with different weights based on their significance, similar to a telescopic sum~\cite{giles2007a}. The two discussed alternatives are to identify the most faithful estimate, which has a similar efficiency to weighting, or to use the goodness-of-fit~\cite{akaike_new_1974}, which is more rigorous but less efficient in most cases.

\subsubsection{Selecting estimates based on significance}\label{sec:optimization_selected}
In this section we discuss combining estimates of different accuracies using a weighted sum and derive optimal weighting parameters under certain assumptions.

The cumulant and multi-exponential methods yield different noisy estimates $\hat{C}_M(0)$ for the exact noiseless signal $C(0)$. Each of them is measured with a finite number of shots on a quantum computer and, hence, $\hat{C}_M(0)$ is subject to a statistical error $\sigma_M$. In this paper, for numerical efficiency, we obtain $\sigma_M$ for the cumulant method using error propagation based on the Jacobian $J_{r,m} = \partial_{\hat{\kappa}_m} \hat{C}(\gamma_r)$ yielding the variance matrix on $\hat{\kappa}_m$ as $(J^T J)^{-1} \sigma_{C(\gamma_r)}^2$, where we used that we sampled all $\sigma_{C(\gamma_r)}^2$ with the same number of shots and, hence, have equal $\sigma_{C(\gamma_r)}$. The variance matrix is then used to estimate the error $\sigma_M$ on $\hat{C}_M(0)$ via error propagation. For the multi-exponential method, we consider the exponents $\hat{w}_m$ as fixed and only perform error propagation to get error estimates to the prefactors $\hat{R}_m$ using the Jacobian $J_{r,m} = \partial_{\hat{R}_m} \hat{C}(\gamma_r)$ and proceed like for the cumulant method.

Using $\hat{C}_M(0)$ and $\sigma_M$, we model the noisy estimate as $\hat{C}_M(0) = \mu_M + \sigma_M z_M$, where $z_M$ is a normally distributed random variable. We want to choose $\hat{C}_M(0)$ over $\hat{C}_{M-1}(0)$ if it is in expectation over shot noise realizations, denoted by $\mathbb{E}$, closer to the exact $C(0)$.
\begin{align}
     \mathbb{E} [(\hat{C}_M(0)-C(0))^2] &< \mathbb{E} [(\hat{C}_{M-1}(0)-C(0))^2] \\
    \Rightarrow  (\mu_M-C(0))^2 + \sigma_M^2 &<(\mu_{M-1}-C(0))^2 + \sigma_{M-1}^2.
\end{align}
We assume that $\hat{C}_M(0)$ has a significantly smaller bias than $\hat{C}_{M-1}(0)$, i.e., $(\mu_M-C(0))^2 \ll (\mu_{M-1}-C(0))^2$, yielding
\begin{align}
    (\mu_{M-1}-C(0))^2 \approx (\mu_{M}-\mu_{M-1})^2 = \mathbb{E} [(\hat{C}_M(0)-\hat{C}_{M-1}(0))^2] - \sigma_M^2 - \sigma_{M-1}^2.
\end{align}
Then the condition for choosing $\hat{C}_M(0)$ is
\begin{align}
    2\sigma_M^2 &< \mathbb{E} [(\hat{C}_M(0)-\hat{C}_{M-1}(0))^2].
\end{align}
As we do not have access to $\mathbb{E} [(\hat{C}_M(0)-\hat{C}_{M-1}(0))^2]$, we simplify further and choose $\hat{C}_M(0)$ over $\hat{C}_{M-1}(0)$ if
\begin{align}
    \sqrt{2} \sigma_M &< |\hat{C}_M(0)-\hat{C}_{M-1}(0)|. \label{eq:cond_disc}
\end{align}
We start with the lowest order, evaluate the condition above until we found the order up to which the condition is fulfilled and then stop and select that $\hat{C}_M$ as the prediction.

In practice, this can be made more systematic using other means of error estimation like bootstrapping. However, for numerical efficiency to evaluate the accuracy of our method on many examples, we use error propagation.

The disadvantage of selecting $\hat{C}_M(0)$ this way is that it yields discrete jumps in the estimate when increasing the number of shots $N_\mathrm{shots}$. Hence, we derive in the following a weighted combination of the $\hat{C}_M$.

\subsubsection{Weighted combination of estimates}\label{sec:optimization_weighting}

In this part we build upon the previous section and develop a weighted combination of the estimates $\hat{C}_M(0)$ with error estimate $\sigma_M$. It is based on the telescopic sum~\cite{giles2007a}. For that, we calculate the difference $\hat{D}_M(0) = \hat{C}_M(0)-\hat{C}_{M-1}(0)$. We start with the lowest order estimate $\hat{C}_{M=1}(0)$. Then we seek to find the optimal weight $a_M$ to add $\hat{D}_M(0)$. This is achieved by minimizing the expected error of the combined estimate
\begin{equation}
    \hat{C}_{\leq 2}(0) = \hat{C}_{1}(0)+a_{2}\hat{D}_{2}(0)
\end{equation}
We optimize the estimate $\hat{C}_{\leq 2}$ to yield the smallest expected deviation to the exact result $C(0)$ by minimizing the mean square error (MSE) of $\hat{C}_{\leq 2}(0)$. This is done by finding the value of $a$ where the variance is minimized relative to the true expectations.

Again, we model $\hat{C}_{1}(0) = \mu_1 + z_1 \sigma_1$ and $\hat{D}_{2}(0)=\mu_{\hat{D}_2} + z_{\hat{D}_2} \sigma_{\hat{D}_2}$, where $z_1$ and $z_{\hat{D}_2}$ are normally distributed random variables with covariance $\mathbb{E}[\sigma_1 z_1 \sigma{\hat{D}_2} z_{\hat{D}_2}]=\mathbb{E}[(\hat{C}_{1}(0) - \mu_1)(\hat{D}_{2}(0) - \mu_{\hat{D}_2})]=-\sigma_1^2$, neglecting correlations between $\hat{C}_M(0)$. We expect the covariance to be small as we expect $\sigma_2\gg \sigma_1$. In order to solve the approximation problem, we approximate $C(0) \approx \mu_1+\mu_{\hat{D}_2}$.

Then, the mean square error (MSE) over shot noise realizations is
\begin{align}
    \mathrm{MSE}[\hat{C}_{\leq 2}(0)] &= \mathbb{E}[ (\hat{C}_{\leq 2}(0) - C(0))^2] \\
    &= \mathbb{E}[ (\hat{C}_{1}(0)+a_{2}\hat{D}_{2}(0) - C(0))^2] \\
    &= (\mu_1 + a_2 \mu_{\hat{D}_2} - C(0))^2 + \sigma_1^2 + a_2^2 \sigma_{\hat{D}_2}^2 - 2a_2 \sigma_1^2\\
    &= (a_2-1)^2 \mu_{\hat{D}_2}^2 + \sigma_1^2 + a_2^2 \sigma_{\hat{D}_2}^2 - 2a_2 \sigma_1^2
\end{align}
Now we can find the optimal $a_2^*$ through minimization of the MSE
\begin{align}
    a_2^* = \frac{\mu_{\hat{D}_2}^2+\sigma_1^2}{\mu_{\hat{D}_2}^2 + \sigma_{\hat{D}_2}^2}
\end{align}
We use
\begin{align}
    \mathbb{E}[\hat{D}_2^2]=\mu_{\hat{D}_2}^2 + \sigma_{\hat{D}_2}^2
\end{align}
and get
\begin{align}
     a_2^* = \frac{\mathbb{E}[(\hat{C}_2(0)-\hat{C}_{1}(0))^2]-\sigma_2^2}{\mathbb{E}[(\hat{C}_2(0)-\hat{C}_{1}(0))^2]}.
\end{align}
In practice, we have only access to a single instance of $\hat{C}_M(0)-\hat{C}_{M-1}(0)$ and, hence, use it as a proxy for the expected difference. We then iterate the procedure for all $M$ yielding the final formula for the weight
\begin{align}
     a_M^* = \frac{(\hat{C}_M(0)-\hat{C}_{M-1}(0))^2-\sigma_M^2}{(\hat{C}_M(0)-\hat{C}_{M-1}(0))^2}.
\end{align}

\subsubsection{Deficiency of error based estimate combinations}
While the derivations suggest that the derived combination of estimates $\hat{C}_M$ is optimal and reduces the MSE, we found that it can fail in rare instances, when the error mitigation method is unstable, $\sigma_M$ is large, and the prediction is unphysical $|\hat{C}_M(0)|\gg 1$.
In these cases there is about a $16\%$ chance that the condition Eq.~\eqref{eq:cond_disc} to choose $\hat{C}_M$ is fulfilled despite the estimate being unphysical.
Similarly, the correction $a_M^* \hat{D}_M$ can be unphysically large in such cases. Hence, we choose a more stable
\begin{align}
     a_M^\mathrm{stable} = \max\left[\frac{\min[(\hat{C}_M(0)-\hat{C}_{M-1}(0))^2,4]-\sigma_M^2-\sigma_{M-1}^2}{\min[(\hat{C}_M(0)-\hat{C}_{M-1}(0))^2,4]},0\right].
\end{align}

\subsubsection{Goodness-of-fit based combinations}\label{sec:optimization_aic}
Last, we discuss using the goodness-of-fit $\chi^2 = -\sum_r (C(\gamma_r)-\hat{C}(\gamma_r))^2/(2\sigma_{C(\gamma_r)}^2)$ to choose $\hat{C}_M(0)$. This has been developed in Ref.~\cite{akaike_new_1974} and subsequent work. Here, we will focus on the version presented in Refs.~\cite{akaike_new_1974,Symonds2010ABG,burnham2002model}.

This combination of parameters is based on calculating the information theoretical criterion
\begin{align}
    \mathrm{AIC}_M = -2 \chi^2 + \frac{ 2 n_r d_{M}}{n_r-d_M-1},
\end{align}
where $d_{M}$ is the number of fitting parameters in the model, $n_r$ is the number of noisy values $C(\gamma_r)$ fitted to and $\chi^2$ is the log-likelihood using that the shot noise is approximately Gaussian
\begin{align}
    \chi^2 = -\sum_r \frac{(\hat{C}(\gamma_r) - C(\gamma_r))^2}{2 \sigma_{C(\gamma_r)}^2}.
\end{align}
Then, we assign weights to each estimate $\hat{C}_M(0)$
\begin{align}
    a_M = \left. e^{-\mathrm{AIC}_M/2}\middle/\sum_{M'} e^{-\mathrm{AIC}_{M'}/2}\right.
\end{align}
This yields the combined prediction
\begin{align}
    \hat{C}_{\leq M}(0) = \sum_{M_i\leq M} a_{M_i} \hat{C}_{M_i}(0). \label{eq:app_aic_comb}
\end{align}

\subsubsection{Numerical comparison}\label{app:comp_akaike_telesc}

We discuss a numerical comparison of the different combination methods for some examples for brevity.

\begin{figure}[H]
    \centering
    \includegraphics[scale=1]{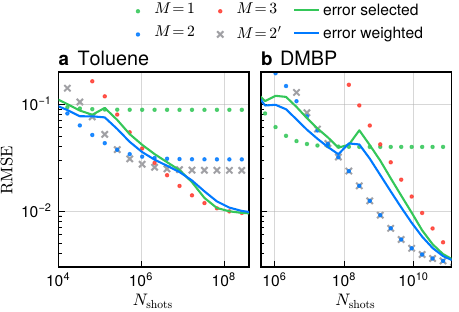}
    \includegraphics[scale=1]{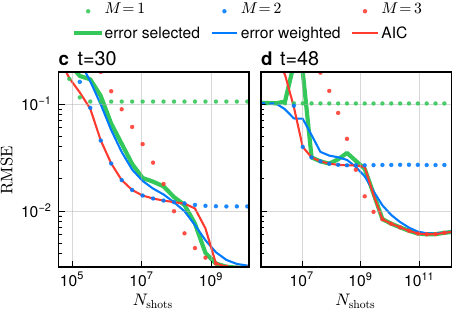}
    \caption{Comparison of different method estimates $\hat{C}_M(0)$ and different combination strategies for (a)  toluene, (b) DMBP, (c) $t=30$ XY and (d) $t=48$ example, analogous to the results shown in the main text. The comparison is made for both variants and both methods: In (a) and (b), we compare results for the $\gamma$-cumulant method using $n_r=4$ while in (c) and (d) we consider the $k$-multi-exponential method with $n_r=10$. The reason for choosing different $n_r$ is that the multi-exponential method has more fitting parameters, hence, we fit to more $\hat{C}(\gamma_r)$ to yield reasonable fitting results. We combine the results of $\hat{C}_M(0)$ for $M=1$ to $M=3$, either by selecting the most accurate significant prediction (green, App.~\ref{sec:optimization_selected}), weighting the different predictions using the error estimates (blue, App.~\ref{sec:optimization_weighting}) or using the AIC (red, App.~\ref{sec:optimization_aic}). With $M=2'$, we refer to fitting the true Gaussian with correct normalization $R(w)=(2\pi\sigma_2^2)^{-1/2} \exp(-(w-w_\mu)^2/2\sigma_2^2)$ on $w\geq0$ rather than $M=2$, corresponding to the unnormalized Gaussian $R(w)=\exp(-(w-w_\mu)^2/2\sigma_2^2)$ on all $w$. Results vary for other examples, but have a similar overall behavior: The AIC tends to perform best, while selecting the most significant estimate $\hat{C}_M(0)$ (green) tends to be less stable than the weighting (blue).}
    \label{fig:comparison_different_combination_methods}
\end{figure}

We use the following procedure to obtain the shown data. First, we generate the reactivity curves as described in App.~\ref{app:numerics_calc_react}. Using the reactivity curves, we calculate $C(\gamma_r)$ for $\gamma_r$ equidistantly spaced in $[0.0025,0.01]$ for $n_r$ different values to evaluate the $\gamma$-methods. For the $k$-methods, we consider $C(k)$ with $k=0,1,\ldots,n_r-1$ and $\gamma_0=0.0025$. We simulate the effect of shot noise when measuring $N_\mathrm{shots}$ by adding normally distributed random numbers with zero mean and standard deviation $\sigma_{C(\gamma_r)}$ using the upper bound $\sigma_{C(\gamma_r)} = (\sqrt{N_\mathrm{shots}})^{-1}$ of the standard deviation of a binomial distribution taking values $\pm 1$. We then fit the different functional forms for the cumulant method and the multi-exponential method and evaluate their prediction $\hat{C}_M(0)$. We show the results in the main text and in Fig.~\ref{fig:comparison_different_combination_methods} for several examples.

The behavior of the different methods is similar for other examples to the extent also discussed in the main text. As we only took $n_r=4$ datapoints $C_M(\gamma_r)$ for the cumulant method, the AIC combination Eq.~\eqref{eq:app_aic_comb} is not applicable, indicating that in principle, it might be preferable to fit to more $C(\gamma_r)$.

Focusing on the different orders of fitting methods $M$, we find in general the higher order fits to yield more accurate results (red dots in Fig.~\ref{fig:comparison_different_combination_methods} a-c), but they are less stable compared to lower order fits (green dots in Fig.~\ref{fig:comparison_different_combination_methods} a-c). This is the essence of trading the complexity of the method against its variance. To perform the most efficient and accurate error mitigation, we aim for a low complexity error mitigation when restricted to a small number of shots $N_\mathrm{shots}$ and gradually transition to higher complexity as we increase the number of shots $N_\mathrm{shots}$. As the error mitigation is based on the same $C(\gamma_r)$ for all $M$, this is indeed possible by combining the different estimates $\hat{C}_M(0)$, as shown in the main text, which we discuss in the following.

Comparing the different combination strategies, we find the AIC to be overall the most stable and also tends to be the most efficient method to combine the estimates $\hat{C}_M(0)$ for different $M$: The predictions of AIC fluctuate the least across the range of $N_\mathrm{shots}$ considered, Fig.~\ref{fig:comparison_different_combination_methods} (c and d, red). The error based selected prediction, Fig.~\ref{fig:comparison_different_combination_methods} (c and d, green), yields sometimes predictions which are significantly worse, causing the RMSE to spike. We found the error based weighted combination of results, Fig.~\ref{fig:comparison_different_combination_methods} (c and d, blue), to be more stable than the error based selected prediction but we found it in rare occasions also to be unstable and it is overall less efficient than the AIC combination. These observations motivate choosing the AIC for the multi-exponential method and the weighted error based combination for the cumulant method in the main text.

\section{Cumulant error mitigation and fitting theory}
\label{app:cumulant_details}

The main text introduces cumulant PP-ZNE, its fitting model, and the procedure
used to combine different orders [Sec.~\ref{sec:cumulant}].
Here we develop its physical interpretation, focusing on how finite reactivity
width produces systematic bias in single-exponential extrapolation and on the
range over which a low-order cumulant description remains useful.

\subsection{Finite reactivity width and extrapolation bias}
\label{app:cumulant_bias}

Recall that under homogeneous noise amplification,
$C(\gamma)=\sum_w e^{-\gamma w}R(w)$.
To understand what is learned from measurements beginning at the native noise rate $\gamma_0$, we factor out the native attenuation:
\begin{equation}
    C(\gamma)
    =
    \sum_w
    e^{-\gamma_0 w}R(w)
    e^{-(\gamma-\gamma_0)w}.
    \label{eq:app_cumulant_reweighted}
\end{equation}
Thus, varying the noise above $\gamma_0$ probes moments of the
\emph{native-noise-weighted reactivity},
$e^{-\gamma_0 w}R(w)$.
This is the reactivity that remains visible through the native device noise: large-$w$ contributions that are already strongly suppressed at $\gamma_0$ have correspondingly little influence on the subsequent noise-amplification data.

Expanding the logarithm of Eq.~\eqref{eq:app_cumulant_reweighted} gives the cumulant model introduced in Eq.~\eqref{eq:cumulant_exp}.
For later reference, we denote its order-$M$ truncation by
\begin{equation}
    C^{(M)}(\gamma)
    =
    \exp\!\left[
        \sum_{m=0}^{M}
        \frac{(\gamma_0-\gamma)^m}{m!}\kappa_m
    \right].
    \label{eq:app_cumulant_truncated}
\end{equation}
In an exact cumulant expansion, the $\kappa_m$ are fixed by the derivatives
of $\log C(\gamma)$ at $\gamma_0$.
In practice, however, PP-ZNE fits a finite noise interval rather than an
infinitesimal neighborhood of $\gamma_0$.
The fitted coefficients $\hat{\kappa}_m$ should therefore be viewed as
effective parameters over that interval: they can absorb contributions from
higher orders, as discussed in Sec.~\ref{sec:cumulant}.
Their usefulness does not require $(\gamma-\gamma_0)w\ll1$ throughout the
fitting window; rather, the truncated model must accurately describe the
experimentally resolved response over the range used for extrapolation.

Consider first the simple case $R(w)\geq0$.
Then
$e^{-\gamma_0 w}R(w)/C(\gamma_0)$
is a probability distribution over path weights.
Its first two cumulants, $\kappa_1$ and $\kappa_2$, are the mean weight and variance.
The first-order model therefore retains only a single characteristic weight and reduces to single-exponential ZNE.
The second order additionally resolves the reactivity width through curvature of $\log C(\gamma)$, while higher orders capture asymmetry and finer structure.
For signed reactivities the same expansion remains algebraically valid, but this probabilistic interpretation is lost.

\zpar{Connection to effective reactivity}
The same hierarchy follows from the finite-resolution picture of Sec.~\ref{sec:effective}.
Gaussian smoothing of the reactivity over a weight scale $\gamma_G^{-1}$ gives
\begin{equation}
    R_{\gamma_G}(w)
    =
    \int_{-\infty}^{\infty}
    \frac{\gamma_G\,\mathrm{d}w'}{\sqrt{2\pi}}
    \exp\!\left[
        -\frac{\gamma_G^2}{2}(w-w')^2
    \right]
    R(w'),
    \label{eq:app_gaussian_convolution}
\end{equation}
whose noise response is
\begin{equation}
    C_{\gamma_G}(\gamma)
    =
    e^{\gamma^2/(2\gamma_G^2)}C(\gamma),
    \qquad
    \log C_{\gamma_G}(\gamma)
    =
    \log C(\gamma)
    +
    \frac{\gamma^2}{2\gamma_G^2}.
    \label{eq:app_gaussian_response}
\end{equation}
This identity uses the full-line convolution, including the small tail of $R_{\gamma_G}$ at $w<0$; restricting to $w\geq0$ introduces boundary corrections.
Coarse graining therefore first modifies the quadratic curvature of the log-response, or equivalently the second cumulant.
For $\gamma\ll\gamma_G$,
\begin{equation}
    \frac{C_{\gamma_G}(\gamma)}{C(\gamma)}
    =
    1+
    \frac{1}{2}
    \left(
        \frac{\gamma}{\gamma_G}
    \right)^2
    +
    \mathcal{O}\!\left[
        \left(
            \frac{\gamma}{\gamma_G}
        \right)^4
    \right].
    \label{eq:app_gaussian_resolution}
\end{equation}
Thus, fine structure in $R(w)$ below the resolution scale $\gamma_G^{-1}$ can change substantially while producing little change in the accessible signal.
Reactivity width, the second cumulant, and curvature of $\log C(\gamma)$ are therefore three descriptions of the same experimentally resolved structure.

\begin{figure*}[t]
    \includegraphics[width=\textwidth]{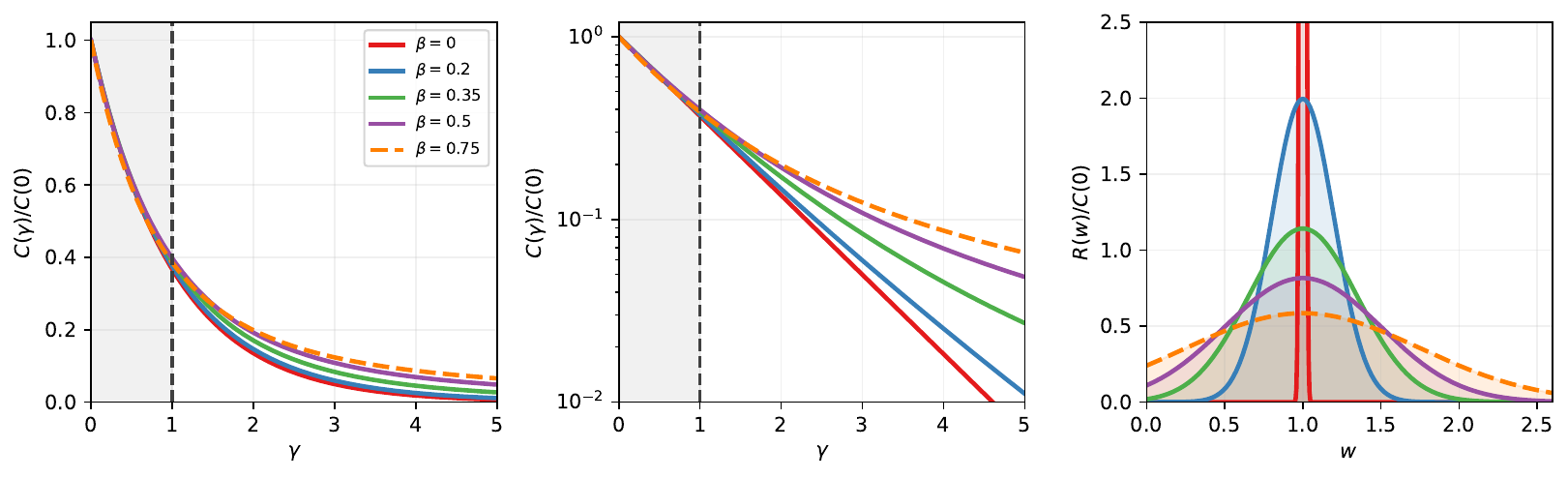}
    \caption{
    \label{fig:cumulant_model}
    \textbf{Finite reactivity width produces curvature in the noise response.}
    \textbf{Left}: normalized response of the positive Gaussian reactivity,
    Eq.~\eqref{eq:app_truncated_gaussian_R}, for fixed $\bar w$ and increasing
    relative width $\beta=\sigma_w/\bar w$.
    \textbf{Middle}: the same response on a logarithmic scale.
    A sharply localized reactivity gives approximately single-exponential decay, while increasing width produces increasing curvature.
    The dashed line denotes the native noise strength $\gamma_0$.
    \textbf{Right}: the corresponding reactivity functions.
    Axes are in units with $\bar w=1$; the shaded region $\gamma<\gamma_0$ is
    experimentally inaccessible.
    }
\end{figure*}

\zpar{Positive finite-width model}
To make the connection concrete, we consider a Gaussian reactivity restricted to the
physical domain $w\geq0$.
We parameterize the underlying Gaussian by its center $\bar w$ and width $\sigma_w$.
Without this truncation, $\bar w$ and $\sigma_w^2$ are the first two cumulants of
the zero-noise reactivity $R(w)/C(0)$.
The native-noise weighting $e^{-\gamma_0 w}$ shifts the first cumulant to
$\kappa_1=\bar w-\gamma_0\sigma_w^2$, while leaving
$\kappa_2=\sigma_w^2$.
The truncation at $w=0$ introduces the additional error-function corrections below.
\begin{equation}
    R_{\mathrm{G}}(w)
    =
    C(0)
    \frac{
        \sqrt{2/\pi}
    }{
        \sigma_w
        \left[
            1+
            \operatorname{erf}
            \!\left(
                \frac{\bar w}{\sqrt{2}\sigma_w}
            \right)
        \right]
    }
    \exp\!\left[
        -\frac{(w-\bar w)^2}{2\sigma_w^2}
    \right],
    \qquad
    w\geq0,
    \label{eq:app_truncated_gaussian_R}
\end{equation}
whose noise response is
\begin{align}
    C_{\mathrm{G}}(\gamma)
    &=
    C(0)
    \exp\!\left[
        -\gamma\bar w
        +
        \frac{\gamma^2\sigma_w^2}{2}
    \right]
    \frac{
        1+
        \operatorname{erf}
        \!\left[
            \frac{
                \bar w-\gamma\sigma_w^2
            }{
                \sqrt{2}\sigma_w
            }
        \right]
    }{
        1+
        \operatorname{erf}
        \!\left[
            \frac{\bar w}{\sqrt{2}\sigma_w}
        \right]
    }.
    \label{eq:app_truncated_gaussian_C}
\end{align}
Figure~\ref{fig:cumulant_model} shows the response for increasing relative width
$\beta=\sigma_w/\bar w$.
As the reactivity broadens, its different weight components decay at different rates,
producing curvature in $\log C(\gamma)$.
For the untruncated Gaussian this curvature is exactly set by
$\kappa_2=\sigma_w^2$.

\begin{figure*}[t]
    \includegraphics[width=0.8\textwidth]{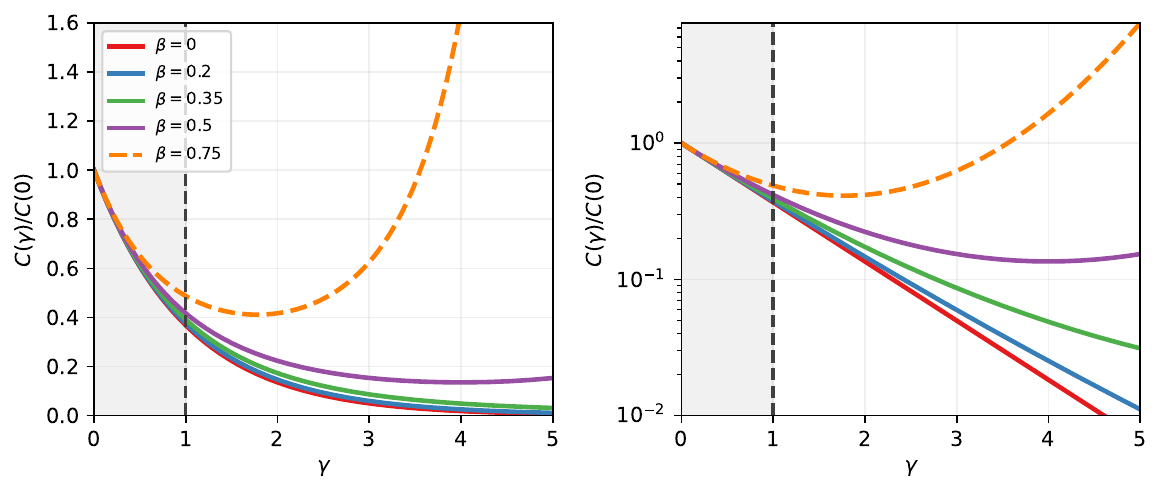}
    \caption{
    \label{fig:cumulant_model_unnorm}
    \textbf{Unnormalized Gaussian reactivity model.}
    The quadratic cumulant response
    $C^{(2)}(\gamma)/C(0)
    =
    \exp[-\gamma\bar w+\gamma^2\sigma_w^2/2]$
    for fixed $\bar w$ and increasing relative width
    $\beta=\sigma_w/\bar w$.
    \textbf{Left}: linear scale.
    \textbf{Right}: the same curves on a logarithmic scale.
    Unlike the truncated Gaussian response of
    Eq.~\eqref{eq:app_truncated_gaussian_C}, this model does not impose the
    physical boundary $w\geq0$ and therefore contains no associated
    error-function correction.
    The resulting negative-$w$ tail eventually produces an unphysical upturn
    at large noise, with broader reactivities turning upward sooner.
    Axes are in units with $\bar w=1$; the dashed line marks the native noise
    strength, $\gamma_0\bar w=1$, and the shaded region $\gamma<\gamma_0$ is
    experimentally inaccessible.
    }
\end{figure*}

\zpar{Quadratic cumulant model}
The connection becomes especially transparent if we remove the $w\geq0$
truncation and its associated error-function corrections.
The resulting unnormalized Gaussian model has
\begin{equation}
    \frac{C^{(2)}(\gamma)}{C(0)}
    =
    \exp\!\left[
        -\gamma\bar w
        +
        \frac{\gamma^2\sigma_w^2}{2}
    \right].
    \label{eq:app_unconstrained_second_cumulant}
\end{equation}
Equivalently, when expanded about the native noise rate,
$\kappa_1=\bar w-\gamma_0\sigma_w^2$ and
$\kappa_2=\sigma_w^2$.
Thus, the first cumulant sets the local decay rate, while the second captures
the curvature associated with finite reactivity width.
Without the boundary $w\geq0$, however, the Gaussian has a negative-$w$ tail
that the Laplace kernel amplifies as $\gamma$ grows, so the response eventually
turns upward [Fig.~\ref{fig:cumulant_model_unnorm}].
The quadratic model is therefore an effective description over the noise
range used for extrapolation, not a global model of the reactivity.

\begin{figure*}[t]
    \includegraphics[width=0.8\textwidth]{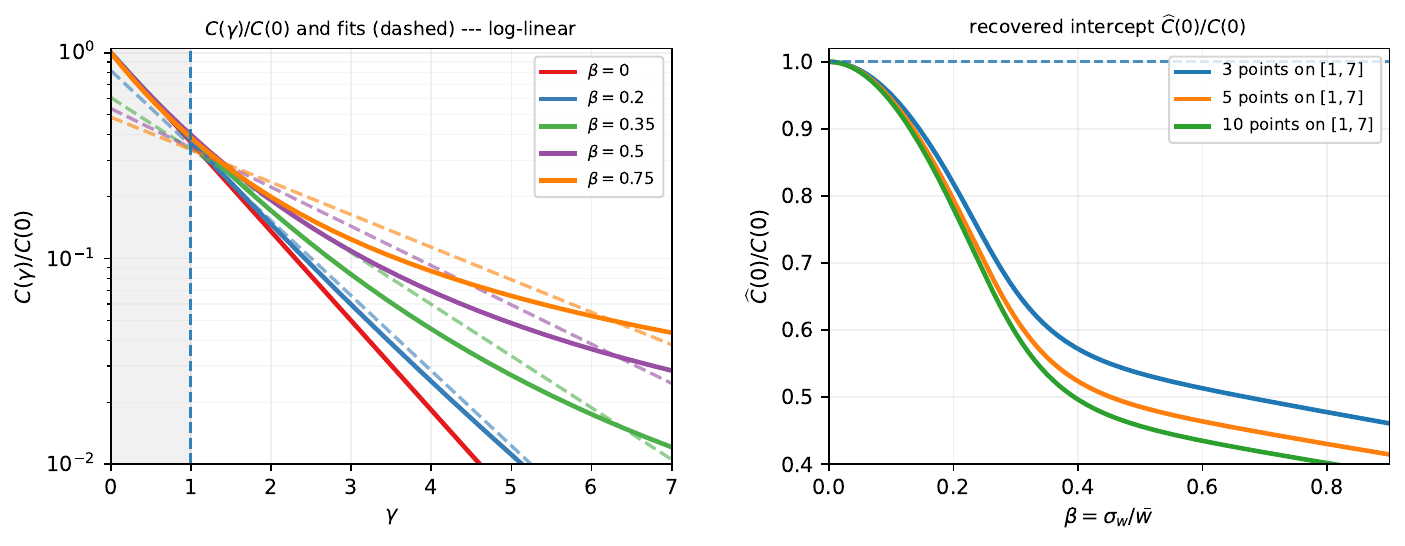}
    \caption{
    \label{fig:bias_log_linear}
    \textbf{Finite reactivity width biases single-exponential extrapolation.}
    \textbf{Left}: exact positive-Gaussian responses (solid) and
    single-exponential fits (dashed) over the accessible noise range
    $\gamma\geq\gamma_0$.
    \textbf{Right}: recovered $\widehat C(0)/C(0)$ versus relative width
    $\beta$, for fits using 3, 5, and 10 noise rates on
    $[\gamma_0,7\gamma_0]$.
    Since the data are noiseless, the deviation from unity is entirely
    deterministic model bias; the number of fit points changes how the fit
    weights the curvature, not whether the bias vanishes.
    Axes are in units with $\bar w=1$, so $\gamma_0\bar w=1$ (dashed line).
    }
\end{figure*}

\zpar{Deterministic extrapolation bias}
Finite reactivity width directly produces a systematic bias in
single-exponential ZNE.
To isolate this effect from sampling noise, we generate noiseless data from
Eq.~\eqref{eq:app_truncated_gaussian_C}, fit over
$\gamma\in[\gamma_0,7\gamma_0]$, and extrapolate to zero noise.
As shown in Fig.~\ref{fig:bias_log_linear}, the bias grows as the reactivity
becomes broader and the curvature of the response becomes more pronounced.
Because the data are noiseless, this error cannot be removed by taking
additional measurements: more measurements only determine the inadequate
single-exponential model more precisely.

This example gives a simple physical interpretation of the cumulant
hierarchy.
For a localized effective reactivity, first order captures its characteristic
weight, second order resolves its width, and higher orders progressively
refine its shape.
The aim is not to reconstruct microscopic detail in $R(w)$, but to retain
the structure that remains experimentally resolvable and matters for
extrapolation to zero noise.
In practice, the useful order is selected from the data using the
model-combination procedures of Sec.~\ref{sec:cumulant} and
App.~\ref{sec:optimal_combination_estimates}.

\subsection{Convergence of the cumulant expansion}
\label{app:cumulant_expansion_examples}

The discussion above assumes that the cumulant series itself is locally convergent.
Since it is a Taylor expansion of $\log C(\gamma)$, its radius of convergence is set by the nearest
singularity in the complex $\gamma$ plane; in particular, every zero of $C(\gamma)$ produces a
logarithmic singularity.

For orientation, consider a normalized positive reactivity expanded about zero noise,
$\log C(\gamma)=\sum_{m\geq1}(-\gamma)^m\kappa_m/m!$. The same conclusions apply to the
noise-weighted effective reactivity when expanding about $\gamma_0$.

\zpar{Representative cases}
\begin{enumerate}\itemsep4pt

    \item \textbf{Gaussian.}
    For $\mathcal{N}(\mu,\sigma)$,
    $\kappa_1=\mu$, $\kappa_2=\sigma^2$, and $\kappa_m=0$ for $m\geq3$.
    The second-order expansion is exact for the untruncated Gaussian; the
    physical boundary $w\geq0$ adds the error-function corrections of
    Eq.~\eqref{eq:app_truncated_gaussian_C}.

    \item \textbf{Poisson.}
    For parameter $\eta$, $\kappa_m=\eta$ for all $m$.
    The $m$th term scales as $\eta\gamma^m/m!$, so the expansion converges for every finite
    $\gamma$.

    \item \textbf{Exponential.}
    For $R(w)=\eta e^{-\eta w}$ on $w\geq0$,
    $C(\gamma)=1/(1+\gamma/\eta)$.
    The nearest singularity is at $\gamma=-\eta$, giving the finite radius
    $|\gamma|<\eta$.

    \item \textbf{Two separated components.}
    For $R(w)=[\delta(w)+\delta(w-w_*)]/2$,
    $C(\gamma)=[1+e^{-\gamma w_*}]/2$.
    Its nearest zeros satisfy $\gamma w_*=\pm i\pi$, so the expansion about zero converges only
    for $|\gamma|<\pi/w_*$. Separated weight scales can therefore limit a local cumulant
    description even when the real positive response is smooth.

    \item \textbf{Many weak contributions.}
    For a sum of many weak, approximately independent contributions, normalized higher cumulants
    are progressively suppressed. The effective reactivity approaches a Gaussian and a low-order
    description becomes increasingly accurate, consistent with the Gaussian operator-size
    distributions found in 1D Hamiltonian dynamics without conserved quantities~\cite{schuster2023operator}.

\end{enumerate}

Three limitations therefore bound a useful cumulant order: the amount of reactivity structure
resolved by the experiment, the statistical precision available to fit it, and, as these examples
illustrate, the analytic distance over which the response must be continued. The cumulant order
should be increased only while all three permit it.

\section{Multi-exponential mitigation}
In this appendix we provide further details for the multi-exponential error mitigation method used in the main text.
For this method, we also combined different models using the techniques presented in Section~\ref{sec:optimal_combination_estimates}.
We first investigate the impact of choosing the spacing $\Delta w$, by providing additional numerical results and an analytical calculation showing that the relative discretization error scales as $\gamma_0 \Delta w$ under reasonable assumptions.
Then, we discuss an alternative to the multi-exponential fitting mentioned in the main text, namely the Lasso method, in more detail.
Finally, we analyze the error mitigation cost for fitting two exponentials, obtaining the scaling result $e^{2\gamma_0 w_2}$ reported in the main text.
The analytical calculation additionally provides intuition for the efficiency gain when using discrete noise insertion compared to continuous noise amplification.

\subsection{Spacing of exponentials in the multi-exponential method}\label{app:details_multi_exp}
We provide numerical results on the systematic error due to the chosen spacing for the bimodal example and argue that we expect in general the systematic error due to the discretization to scale with the spacing $\Delta w$ as $|\hat{C}(0)-C(0)| \sim \gamma_0 (\Delta w) C(0)$.
\begin{figure}[ht]
    \centering
    \includegraphics[scale=0.8]{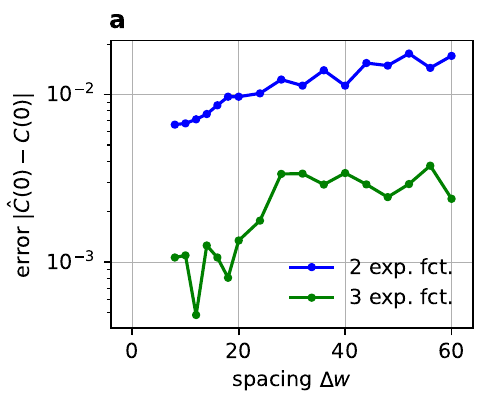}
    \qquad
    \includegraphics[scale=0.8]{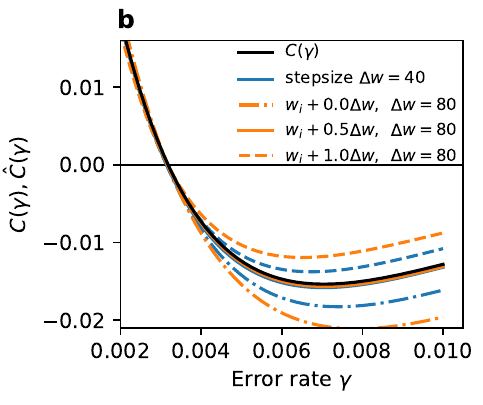}
    \caption{Results for the discretization $\Delta w$ on the noisy estimate $\hat{C}$ for the XY model at $30$ Trotter steps. In (a) we use the multi-exponential method with different spacing $\Delta w$ and fit either two or three exponentials to $100$ values $C(\gamma_r)$ with $\gamma_r \in [0.0025,0.01]$ equidistant. The fit is performed as outlined in the main text: We discretize $w_i = w_0 + i \Delta w$ with $i \in \mathbb{N}$ and iterate through all possible combinations of either $2$ (blue) or $3$ (green) decay rates out of the set of $w_i$. For every possible combination, we determine the $\hat{R}_m$ by fitting the multi-exponential sum to $C(\gamma)$. We then choose the combination of $w_i$ which minimizes the residual $\sum_r |\hat{C}(\gamma_r)-C(\gamma_r)|^2$. In order to make the method more stable we use that the choice of $w_0$ is not expected to change the performance. Hence, we perform this procedure for all $w_0 \in[1,\Delta w+1]$ and select the $w_0$ for which the stability of the fit as evaluated through error propagation is the largest. (b) We show the impact of binning $R_{\gamma_G}(w)$ with bin width $\Delta w$, i.e., spacing on $\hat{C}(\gamma)$. Here, $\hat{C}(\gamma)$ is obtained by binning $R_{\gamma_G}(w)$ with $\gamma_G=0.08\gg \gamma_\mathrm{max}=0.01$ yielding the bin value $R_i$ and calculating $\hat{C}(\gamma)=\sum_i e^{-\gamma (w_i+d \Delta w)} R_i$. Results are shown for $\Delta w=40$ (blue) and $\Delta w=80$ (orange) for $d=0.0$ (dash dotted), $d=0.5$ (solid) and $d=1.0$ (dashed). This shows that for noise rates $\gamma \ll 1/\Delta w$, indeed $d=0$ and $d=1$ yield $\hat{C}(\gamma)$ based on the discretized $R_{\gamma_G}(w)$ above and below the true $C(\gamma)$. }
    \label{fig:numerical_results}
\end{figure}
The numerical results on the systematic error or bias induced by choosing a finite spacing $\Delta w$ are shown in Fig.~\ref{fig:numerical_results} a. Overall, we only find a weak dependence on the  spacing $\Delta w$. For the two exponential fit, the error decreases from about $0.02$ for large $\Delta w=60$ by about a factor $3$ to $0.007$ for $\Delta w=8$. Increasing the number of exponentials yields a larger reduction in the systematic error to about $0.0025$ at $\Delta w=60$, which is reduced again by a factor of about three decreasing $\Delta w$ to $8$. Hence, we expect that one could improve on the results in the main text by further improving the choice of $\Delta w$, which we leave for future research. %
In practice, one might optimize $\Delta w$ and the number of fitted exponentials $M$ simultaneously in a similar fashion: increasing $M$ and reducing $\Delta w$ until the estimated error due to the shot noise limitations exceeds the estimated systematic error, i.e., bias.

For now, we proceed by discussing the choice of $\Delta w$ from an analytical perspective. By choosing a spacing $\Delta w$, we effectively error mitigate the whole reactivity between $w_i$ and $w_i+\Delta w$ by amplifying it with one uniform factor between $e^{\gamma_0 w_i}$ and $e^{\gamma_0 (w_i+\Delta w)}$. When choosing (A) $e^{\gamma_0 w_i}$ the error mitigated result becomes
\begin{align}
    \hat{C}^{(A)}(0) &= \sum_i e^{\gamma_0 w_i} \int_{w_i}^{w_i+\Delta w}\!\mathrm{d}w \, e^{-\gamma_0w} R(w) = \sum_i \int_{w_i}^{w_i+\Delta w}\!\mathrm{d}w \, e^{-\gamma_0(w-w_i)} R(w)\\
    &= C(0) + \sum_i \int_{w_i}^{w_i+\Delta w}\!\mathrm{d}w \, \left(1-\gamma_0(w-w_i) + \mathcal{O}(\gamma_0^2 \Delta w^2)\right) R(w)\\
    &\approx C(0) + \sum_i \int_{w_i}^{w_i+\Delta w}\!\mathrm{d}w \, (-\gamma_0)(w-w_i) R(w).
\end{align}
Similarly, choosing (B) the right edge $e^{\gamma_0 (w_i+\Delta w)}$
\begin{align}
    \hat{C}^{(B)}(0) & = \sum_i \int_{w_i}^{w_i+\Delta w}\!\mathrm{d}w \, e^{-\gamma_0(w-w_i-\Delta w)} R(w)\\
    &\approx C(0) + \sum_i \int_{w_i}^{w_i+\Delta w}\!\mathrm{d}w \, (-\gamma_0)(w-w_i-\Delta w) R(w).
\end{align}
Hence, the difference between both estimates is
\begin{align}
    \hat{C}^{(B)}(0)-\hat{C}^{(A)}(0) & =  \sum_i \int_{w_i}^{w_i+\Delta w}\!\mathrm{d}w \, \gamma_0\Delta w \,R(w) = C(0) \gamma_0 \Delta w.
\end{align}
Note that we amplified in (A) all $R(w)$ with a factor which is too small $e^{\gamma_0 w_i} \leq e^{\gamma_0 w}$ while in (B) we amplified with a factor being generally too large $e^{\gamma_0 (w_i+\Delta w)} \geq e^{\gamma_0 w}$. This way, (A) and (B) correspond to two extreme cases and we expect for $\gamma \ll 1/\Delta w$ either $\hat{C}^{(B)}(\gamma)\leq C(\gamma) \leq \hat{C}^{(A)}(\gamma)$ or $\hat{C}^{(B)}(\gamma)\geq C(\gamma) \geq \hat{C}^{(A)}(\gamma)$. This expectation is confirmed numerically in Fig.~\ref{fig:numerical_results} b. Hence, we expect
\begin{align}
    |\hat{C}(0)-C(0)| \leq |\hat{C}^{(B)}(0)-\hat{C}^{(A)}(0)| = |C(0)| \gamma_0 \Delta w.
\end{align}
This does not incorporate the error of fitting, and, hence, only provides an estimate for the discretization error itself and motivates the choice $\Delta w\sim 1/\gamma_0$. In particular we choose for $\gamma_0=0.0025$ $\Delta w=1/(10 \gamma_0)=40$ in the main text.

\subsection{Lasso Method}
\label{app:spacing_lasso}

An alternative to the multi-exponential method as presented in the main text is the Lasso method. It consists of fitting the effective reactivity
\begin{align}
    \hat{R}(w)=\sum_m^M \hat{R}_m \delta_{w,w_m},
\end{align}
where $w_m = m \Delta w $ for some small $\Delta w$ and $M=V/\Delta w$ with the circuit volume $V$, i.e., the number of space-time points noise may occur. Fitting $M$ variables to $n_r \ll V/\Delta w=M$ datapoints is an ill-conditioned optimization problem (although, as discussed before, this may not pose a problem to predict $\hat{C}(0)$ since singularities may cancel). Therefore, we introduce an $L_1$ regularization term to the least square optimization function
\begin{align}
    \mathrm{Cost}(\hat{R}_1,\hat{R}_2,\ldots,\hat{R}_M) = \sum_r [C(\gamma_r) - \hat{C}(\gamma_r,\hat{R}_1,\ldots)]^2 + \alpha_{L_1} \sum_m |\hat{R}_m|, \label{eq:lass_cost}
\end{align}
where $\alpha_{L_1}$ is a hyperparameter. The challenge is choosing the right value for $\alpha_{L_1}$. A value too large yields an estimate $\hat{C}(0)$ with large bias, a value too small yields overfitting.

\begin{figure}
    \centering
    \includegraphics[scale=0.9]{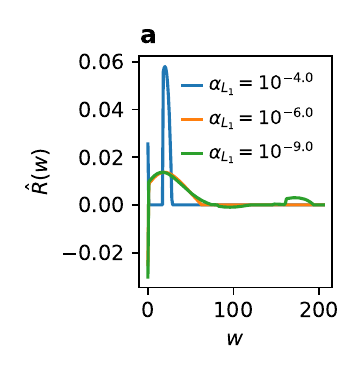}
    \includegraphics[scale=0.9]{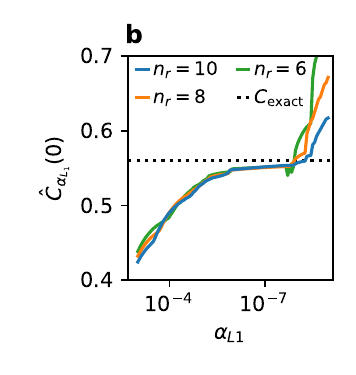}
    \includegraphics[scale=0.9]{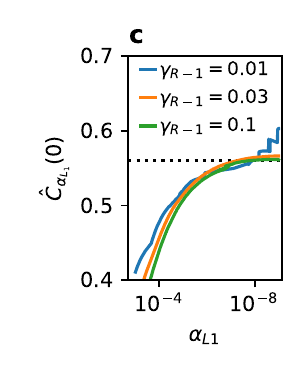}
    \caption{Results for learning the reactivity using the Lasso method. We consider the toluene example also presented in the main text. Left: We generate $10$ noisy measurements $\hat{C}(\gamma_r)$ equidistantly spaced between $\gamma_0=0.0025$ and $\gamma_{n_r-1} = 0.01$ neglecting shot noise. We use scikit-learn to find the $\hat{R}(w)=\sum_m \hat{R}_m \delta_{10\cdot m,w}$ minimizing the cost function, Eq.~\eqref{eq:lass_cost}, at different values of $\alpha_{L_1}$. This shows how overfitting yields artificial features in $\hat{R}(w)$ at small $\alpha_{L_1}$. Center: Out of the $10$ noisy measurements, we remove $N_\mathrm{rem}$ points and fit $\hat{R}(w)$ to the remaining $n_r$  measurements as a function of $\alpha_{L_1}$. Overfitting causes the prediction $\hat{C}_{\alpha_{L_1}}(0)$ to diverge after initial convergence. Right: We take $n_r=50$ measurement results equidistantly spaced between $0.0025$ and varying $\gamma_{n_r-1}$. This shows that the range $\gamma_{n_r-1}$ used is also an important factor for overfitting.}
    \label{fig:lasso_method_results}
\end{figure}

We show results for different $\alpha_{L_1}$ in Fig.~\ref{fig:lasso_method_results} for the toluene example also discussed in the main text; we observe similar trends for other examples. For large $\alpha_{L_1}$ the method tends to produce a $\hat{R}(w)$ which is localized in a single region with most $\hat{R}(w)=0$, see Fig.~\ref{fig:lasso_method_results} (left, blue). As we decrease $\alpha_{L_1}$, this region becomes larger (orange) until other regions appear where $|\hat{R}(w)|>0$ for very small $\alpha_{L_1}=10^{-9}$ (green). This is one of the signatures of overfitting we observe.

Indeed, considering the estimate $\hat{C}_{\alpha_{L_1}}(0)$ as a function of $\alpha_{L_1}$, we find the estimate to converge further to the exact value until about $\alpha_{L_1}=10^{-8}$, at which it starts to diverge, Fig.~\ref{fig:lasso_method_results} (center, blue).

Consequently, a crucial aspect of such machine learning based error mitigation techniques is to prevent overfitting by selecting the right amount of regularization, here the value of $\alpha_{L_1}$. We checked different methods and signatures of overfitting for choosing $\alpha_{L_1}$ without prior knowledge. This includes checking for convergence, tracking the number of regions where $|\hat{R}(w)|>0$ and removing ${C}(\gamma_r)$ in the learning dataset. We found removing ${C}(\gamma_r)$ beginning with the largest $r$ to work most reliably and show the results in Fig.~\ref{fig:lasso_method_results} (center) when removing $N_\mathrm{rem}$ of the ${C}(\gamma_r)$. Removing datapoints does not change the results significantly in the regime of convergence, compare $n_r=6$ to $n_r=10$ in Fig.~\ref{fig:lasso_method_results} (center). Once we observe overfitting, the results for different $n_r$ deviate significantly from each other, which is expected from overfitting: The Lasso method becomes susceptible to details in the ${C}(\gamma_r)$ rather than predicting overall trends. We note that the overfitting is not a consequence of having too few datapoints $n_r$, but rather of having a too small region of $\gamma$ values. Increasing the number of points $n_r$ without changing the $\gamma$ range does not prevent overfitting, see Fig.~\ref{fig:lasso_method_results} (right). Hence, in our scenario, overfitting is a consequence of insufficient or inaccessible information.

While detecting overfitting with ad-hoc methods could work, we deemed it too unreliable in this study and focused on the multi-exponential fitting in the main text.

\subsection{Theoretical case study of error mitigation scaling}\label{app:cost}\label{ssec:cost_estimate_sum_exp}

In this section we analyze the scaling of the sampling cost to perform the multi-exponential error mitigation scheme through a simplified analytical calculation. We expect the resulting insights to hold more generally, albeit obtained through an (over-) simplified calculation. The key insights are:

1. The optimal noise amplification corresponds to an additional signal decay by about a factor $3$ (Eqs.~\eqref{eq:optimal_damping_one},~\eqref{eq:2norm_transcendental}),

2. The cost of error mitigation scales as $e^{2\gamma_1 w_2}$, where $\gamma_1$ is the minimal accessible noise rate, and singularities do not necessarily propagate to the error mitigation estimate (Eqs.~\eqref{eq:final_scaling_result},~\eqref{eq:final_scaling_result2}) and

3. Discrete noise insertion is more efficient than continuous noise amplification (Eqs.~\eqref{eq:compare_cost_di_c},~\eqref{eq:compare_cost_di_c2}).

In more detail, we consider the case of the reactivity being highly peaked around two values $w_1, w_2$ with $w_1<w_2$, so that we can approximate $R(w)\approx R_1 \delta_{w,w_1} + R_2 \delta_{w,w_2}$. However, we expect the results to hold more generally: We find the error mitigation cost to be closely related to the distinguishability of $w_1$ and $w_2$ and thus we expect similar scaling results for each pair $w_i,w_j$ in the more general case $R(w)=\sum_i R_i \delta_{w,w_i}$.

We then calculate the sampling cost assuming $R(w)=R_1 \delta_{w,w_1} + R_2 \delta_{w,w_2}$. We consider either continuously modifying the noise rate $\gamma$ yielding the estimate $\hat{C}(0) = a C(\gamma_1) + b C(\gamma_2)$ for the exact measurement result or inserting $k$ discrete noise events yielding $\hat{C}(0) = a C(\gamma_1,0) + b C(\gamma_1,k)$. If $a,b$ are (roughly) known before sampling, we can use importance sampling yielding a sampling cost overhead of $(|a|+|b|)^2$. If $w_1,w_2$ are unknown, we sample and measure $C(\gamma_1)$ and $C(\gamma_2)$ with the same accuracy, so that the cost overhead is $2(a^2+b^2)$.

Optimizing the cost, we find that the optimal $\gamma_2$ or number $k$ of inserted noise events corresponds to an additional damping $e^{-w_2 (\gamma_2-\gamma_1)}$ (similar for inserting $k$ noise events) of the signal by a factor of about $3.2-3.5$, depending on the cost model.

In the case, that $w_1$ and $w_2$ are known (or unknown), the first order in $w_2/(w_2-w_1)$ correction to the cost for the continuous case compared to the discrete noise event insertion is larger by a factor $1+\frac{2w_1}{3N}$ (or $1+\frac{4w_1}{3N}$),
where we also expanded in $w_1/N$ and $w_2/N$ with $N$ being the total number of possible noise events. Hence, in this setting, the insertion of discrete noise events is more efficient than the continuous amplification of noise; the gain scales with $w_1/N$. An intuitive interpretation is that the discrete noise insertion enhances noise by $\propto \log(1-4w/(3N))$ rather than $\propto w$, the concavity of the logarithm makes the noise response of different $w$ more distinct compared to the linear case $\propto w$.

We will proceed presenting all the calculations underlying these statements.

\subsubsection{Optimal error mitigation using continuous noise amplification}

We start by analyzing the cost for the error mitigation for continuous noise amplification (as opposed to discrete noise insertion) assuming we have a rough estimate for the values $w_1$ and $w_2$, and, hence, the cost is given by the 1-norm as explained in the following.

\emph{Error mitigation via matrix inversion}
To extract the zero-noise estimate $\hat{C}(0) = R_1+R_2$, we use the relation between the two measurements $C(\gamma_1)$ and $C(\gamma_2)$ to $R_1,R_2$ yielding the matrix equation:
\begin{equation}
    \begin{pmatrix} R_1 \\ R_2 \end{pmatrix} =
    M^{-1} \begin{pmatrix} C(\gamma_1) \\ C(\gamma_2) \end{pmatrix}
    \quad \text{where} \quad
    M = \begin{pmatrix} e^{-\gamma_1 w_1} & e^{-\gamma_1 w_2} \\ e^{-\gamma_2 w_1} & e^{-\gamma_2 w_2} \end{pmatrix}
\end{equation}
Expanding the solution to the inversion problem yields the estimate in terms of the measured values $C(\gamma_1), C(\gamma_2)$
\begin{equation}
    \hat{C}(0) = R_1+R_2 = a C(\gamma_1) + b C(\gamma_2).
\end{equation}
Now $a$ and $b$ are obtained by inverting $M$ %
\begin{align}
    a &= \frac{e^{-\gamma_2 w_2} - e^{-\gamma_2 w_1}}{\det(M)}  = \frac{e^{\gamma_1 w_2}\left(1 - e^{-\gamma_2 \Delta w}\right)}{1 - e^{-\Delta \gamma \Delta w}}\\
    b &= \frac{e^{-\gamma_1 w_1} - e^{-\gamma_1 w_2}}{\det(M)} = -\frac{e^{\gamma_1 w_2 + \Delta \gamma w_1}\left(1 - e^{-\gamma_1 \Delta w}\right)}{1 - e^{-\Delta \gamma \Delta w}}
\end{align}
Note that $\det(M) < 0$ as $\Delta \gamma \Delta w > 0$ so that $a>0$ and $b<0$.

\emph{1-Norm cost overhead for error mitigation via continuous noise amplification}
Assuming optimal allocation of Monte Carlo shots (taking shots proportional to the magnitude of the coefficients), the sampling variance overhead is given by the square of the 1-norm:
\begin{equation}
    \text{Cost} \propto (|a| + |b|)^2
\end{equation}

Then, because $a > 0$ and $b < 0$, %
\begin{align}
    |a| + |b| &= a-b%
    = e^{\gamma_1 w_2} \frac{\left(1 - e^{-\gamma_2 \Delta w}\right) + e^{\Delta \gamma w_1}\left(1 - e^{-\gamma_1 \Delta w}\right)}{1 - e^{-\Delta \gamma \Delta w}}
\end{align}

We now aim for the most efficient error mitigation scheme and optimize $\gamma_2$ to minimize the cost.

Introducing $C = 1 - e^{-\gamma_1 \Delta w}$ we get %
\begin{equation}
    \frac{|a| + |b|}{e^{\gamma_1 w_2}} = 1 + C \frac{e^{-\Delta \gamma \Delta w} + e^{\Delta \gamma w_1}}{1 - e^{-\Delta \gamma \Delta w}}.
\end{equation}

Expanding with $e^{\Delta \gamma \Delta w}$ we isolate the $\Delta \gamma$ dependence %
\begin{equation}
    |a| + |b| = e^{\gamma_1 w_2} \left[ 1 + \left(1 - e^{-\gamma_1 \Delta w}\right) \underbrace{\frac{1 + e^{\Delta \gamma w_2}}{e^{\Delta \gamma \Delta w} - 1}}_{g(\Delta \gamma)} \right].
\end{equation}
Reducing the cost is equivalent to minimizing $g(\Delta \gamma)$, which is independent of $\gamma_1$. %
Before proceeding with optimizing this expression rigorously, we note that approximating $\Delta w \ll 1$ and $e^{\Delta \gamma w_2}\gg 1$ yields a cost $e^{\gamma_1 w_2}(1+\gamma_1e^{\Delta \gamma w_2}/\Delta \gamma)$, which is optimized by $\Delta \gamma=1/w_2$. This highlights that the optimal spacing $\Delta \gamma$ yields an additional damping of the signal in the order of $\exp(1)$: $e^{-\gamma_2 w_2} = e^{-\gamma_1 w_2}e^{-1}$, which is close to the more rigorous solution as we will show next.

\emph{More rigorous calculation of the asymptotic 1-Norm Cost in the Limit $w_2 \gg \Delta w$}

Minimizing $g(\Delta \gamma)$ with respect to $\Delta \gamma$ yields the transcendental equation

\begin{equation}
    w_1 - w_2 e^{-\Delta \gamma \Delta w} = \Delta w e^{-\Delta \gamma w_2}
\end{equation}

In order to expand in $\Delta w/w_2$, we first show that $\Delta w/w_2 =\mathcal{O}(\epsilon)$ implies $\Delta \gamma \Delta w =\mathcal{O}(\epsilon)$.

We rewrite the transcendental equation introducing $\epsilon = \frac{\Delta w}{w_2} \ll 1$
\begin{equation}
    1 - \epsilon - e^{-\Delta \gamma \Delta w} = \epsilon e^{-\Delta \gamma \Delta w/\epsilon} \label{eq:transcendental_x}
\end{equation}
Because $\Delta \gamma > 0$ and weights are positive, the exponential $e^{-\Delta \gamma \Delta w/\epsilon}$ is strictly bounded: $0 < e^{-\Delta \gamma \Delta w/\epsilon} < 1$. We can use these limits to bound the left side of Equation \ref{eq:transcendental_x}:
\begin{itemize}
    \item {Lower Bound:} Since $\epsilon e^{-\Delta \gamma \Delta w/\epsilon} > 0$, it must be that $1 - \epsilon - e^{-\Delta \gamma \Delta w} > 0$, which means $e^{-\Delta \gamma \Delta w} < 1 - \epsilon$.
    \item {Upper Bound:} Since $\epsilon e^{-\Delta \gamma \Delta w/\epsilon} < \epsilon(1) = \epsilon$, it must be that $1 - \epsilon - e^{-\Delta \gamma \Delta w} < \epsilon$, which implies $e^{-\Delta \gamma \Delta w} > 1 - 2\epsilon$.
\end{itemize}

Taking the natural logarithm of the two bounds yields:
\begin{equation}
    -\log(1 - \epsilon) < \Delta \gamma \Delta w < -\log(1 - 2\epsilon)
\end{equation}
For infinitesimally small $\epsilon$, the logarithms expand to $\epsilon < \Delta \gamma \Delta w < 2\epsilon$, hence, $\Delta \gamma \Delta w = \mathcal{O}(\epsilon)$. %

We rewrite Eq.~\eqref{eq:transcendental_x}, defining $y=\Delta \gamma \Delta w/\epsilon$:
\begin{equation}
    1 - \epsilon - e^{-y \epsilon} = \epsilon e^{-y}
\end{equation}

Because we have shown $y\epsilon \ll 1$, we expand $e^{-y \epsilon}$ to second order in $\epsilon$: %
\begin{equation}\label{eq:optimal_damping_one}
    1 - \epsilon - \left( 1 - y \epsilon + \frac{1}{2} y^2 \epsilon^2 \right) \approx \epsilon e^{-y} \implies (y - 1) - \frac{1}{2} y^2 \epsilon \approx e^{-y}
\end{equation}
For now, we only need the optimal $y$ to zeroth order in $\epsilon$, and assume a perturbative solution $y \approx y_0 + \mathcal{O}(\epsilon)$. Setting $\epsilon = 0$ leaves the fundamental equation $y_0 - 1 = e^{-y_0} \Leftrightarrow e^{y_0} = \frac{1}{y_0 - 1}$. The unique root in the interval $(1,2)$ is $y_0 \approx 1.27846$ (as $1<y<2$). %

The optimal $y\approx y_0+\mathcal{O}(\epsilon)$ implies $\Delta \gamma \approx \frac{y_0}{w_2} $. Since this $y$ minimizes $g$ we have $g(\epsilon y/\Delta w) = g(\epsilon y_0/\Delta w) + \mathcal{O}(\epsilon^2)$.

We expand
\begin{equation}
    g(\Delta \gamma) = \frac{1 + e^{y_0}}{e^{{y_0} \epsilon} - 1} \approx  \frac{1}{\epsilon} \left( \frac{1 + e^{y_0}}{{y_0}} \right) - \frac{1 + e^{y_0}}{2}=\frac{1}{\epsilon(y_0-1)} - \frac{y_0}{2(y_0-1)}
\end{equation}
and substitute $g(\Delta \gamma)$ back into our global 1-norm formula $|a| + |b|$.
\begin{equation}
    |a| + |b| \approx e^{\gamma_1 w_2} \left[ 1 + \left(1 - e^{-\gamma_1 \Delta w}\right) \left( \frac{1}{y_0 - 1} \frac{w_2}{\Delta w} - \frac{y_0}{2(y_0 - 1)} \right) \right]
\end{equation}

Expanding $(1 - e^{-\gamma_1 \Delta w}) \approx \gamma_1 \Delta w$, the $\Delta w$ cancels in the leading term:
\begin{equation}\label{eq:final_scaling_result}
    |a| + |b| \approx e^{\gamma_1 w_2} \left[ 1 + \gamma_1 w_2 \left(\frac{1}{y_0 - 1}\right) - \gamma_1 \Delta w \left( \frac{y_0}{2(y_0 - 1)} \right) \right]
\end{equation}
Interestingly, this implies that the singularity of the matrix inversion for small $\Delta w$ has no influence on estimating $a+b$.

\subsubsection{\texorpdfstring{Inverting signals with $k$ noise events inserted}{Inverting signals with k noise events inserted}}
To evaluate the protocol where $k$ noise events are inserted, we define
\begin{equation}
    \alpha = -\log\left(1 - \frac{4w_1}{3V}\right), \quad \beta = -\log\left(1 - \frac{4w_2}{3V}\right)
\end{equation}
Assuming $w_2 > w_1$, it strictly follows that $\beta > \alpha > 0$. %

Now the measurements $C(\gamma_1,k=0)$ and $C(\gamma_1,k)$ are related to the parameters $R_1$ and $R_2$ via the matrix $M$ as
\begin{align}
    \begin{pmatrix} C(\gamma_1,k=0) \\ C(\gamma_1,k) \end{pmatrix} = M \begin{pmatrix} R_1 \\ R_2 \end{pmatrix}, \qquad M = \begin{pmatrix}
        e^{-\gamma_1 w_1} & e^{-\gamma_1 w_2} \\
        e^{-\gamma_1 w_1} e^{-\alpha k} & e^{-\gamma_1 w_2} e^{-\beta k}.
    \end{pmatrix}.
\end{align}
This corresponds to replacing $\gamma_2 w_1 \rightarrow k \alpha + \gamma_1 w_1$ and $\gamma_2 w_2 \rightarrow k \beta + \gamma_1 w_2$ compared to the continuous amplification. Hence, the results apply in the same way upon this replacement, as we will derive in the following.

Again, we want to calculate the zero-noise estimate $\hat{C}(0)=R_1+R_2 = a C(\gamma_1,0) + b C(\gamma_1,k)$. %

Inverting the matrix yields the coefficients $a$ (for the baseline measurement) and $b$ (for the $k$-insertion measurement):
\begin{align}
    a = \frac{e^{-\gamma_1 w_2} e^{-\beta k} - e^{-\gamma_1 w_1} e^{-\alpha k}}{\det(M)}, \qquad b = \frac{e^{-\gamma_1 w_1} - e^{-\gamma_1 w_2}}{\det(M)}
\end{align}
with $\det(M) = e^{-\gamma_1(w_1+w_2)}\left(e^{-\beta k} - e^{-\alpha k}\right)$. Because $\beta > \alpha$, we have $e^{-\beta k} < e^{-\alpha k}$, so $\det(M) < 0$ and so that with $w_2 > w_1$, it follows $a>0$ and $b<0$.

\paragraph{1-Norm cost overhead for error mitigation for discrete noise insertion}
The cost overhead is
\begin{align}
    |a| + |b| &= a-b %
    = e^{\gamma_1 w_2} \left[ 1 + \left(1 - e^{-\gamma_1 \Delta w}\right) \underbrace{\frac{1 + e^{\beta k}}{e^{(\beta-\alpha)k} - 1}}_{g(k)} \right]
\end{align}
This equation is mathematically isomorphic to the continuous tuning equation. Hence, the same optimization results apply replacing $\Delta \gamma\rightarrow k$, $w_2\rightarrow \beta$  and $\Delta w\rightarrow (\beta-\alpha)$

\begin{equation}
    |a| + |b| \approx e^{\gamma_1 w_2} \left[ 1 + \left(1 - e^{-\gamma_1 \Delta w}\right) \left( \frac{1}{y_0 - 1} \frac{\beta}{\beta-\alpha} - \frac{y_0}{2(y_0 - 1)} \right) \right]
\end{equation}

\subsubsection{Cost Comparison: Discrete Insertion vs. Continuous Tuning}

We established that the cost difference between discrete noise insertion and continuous noise amplification are the factors:
\begin{align}
    F_{\text{cost},{\text{cont}}} &\propto  \frac{w_2}{w_2 - w_1} \\
    F_{\text{cost},{\text{disc}}} &\propto  \frac{\beta}{\beta - \alpha}
\end{align}
where the discrete decay rates were defined as:
\begin{equation}
    \alpha = -\log\left(1 - \frac{4w_1}{3V}\right), \quad \beta = -\log\left(1 - \frac{4w_2}{3V}\right)
\end{equation}

To compare them, we define the dimensionless, small perturbation parameters $p_1 = \frac{4w_1}{3V}$ and $p_2 = \frac{4w_2}{3V}$, assuming $p_i \ll 1$. Since $w_2 > w_1$, we have $p_2 > p_1 > 0$.

We expand the logarithm: $-\log(1 - p) \approx p + \frac{1}{2}p^2$. Then
\begin{align}
    F_{\text{cost},{\text{disc}}}=  \frac{p_2\left(1 + \frac{1}{2}p_2\right) }{(p_2-p_1)\left[1 + \frac{1}{2}(p_2 + p_1)\right]} %
    \approx \left(\frac{w_2}{\Delta w}\right) \left[ 1 - \frac{1}{2}p_1\right]%
    =F_{\text{cost},{\text{cont}}}\left[ 1 - \frac{2w_1}{3V}\right].\label{eq:compare_cost_di_c}
\end{align}
This shows that for large $V$ the cost of the discrete noise injection based error mitigation is smaller than for the continuous noise amplification one. The benefit vanishes as $w_1/V\rightarrow 0$.

The advantage of discrete noise insertion is plausible considering the
effective noise rates applied to each component. For continuous noise amplification, both components are subjected to an identical added background noise rate ($\Delta \gamma$), while for discrete noise insertion, the concavity of the logarithm yields a different added noise rate for both components stretching them out.%

In particular from the $\log$ being concave it follows
\begin{align}
    \frac{\alpha}{w_1} < \frac{\beta}{w_2} \Leftrightarrow\frac{\beta}{\beta-\alpha} < \frac{w_2}{w_2-w_1}  ,
\end{align}
so that in terms of scaling $\mathcal{O}(F_{\text{cost},{\text{disc}}})<\mathcal{O}(F_{\text{cost},{\text{cont}}})$.

This also holds for multiple exponentials---the logarithm appearing in the discrete noise insertion causes the different $w_i$ to appear more distinct, providing an explanation for the efficiency gain through discrete noise insertion observed in the main text.

\subsubsection{Evaluating the 2-Norm Cost Overhead}
If $w_1$ and $w_2$ are unknown prior to sampling, the measurement scheme requires sampling $C(\gamma_1)$ and $C(\gamma_2)$ with uniform statistical accuracy. The total sampling variance overhead to predict the zero-noise estimate $\hat{C}(0) = a C(\gamma_1) + b C(\gamma_2)$ is then given by the 2-norm squared of the measurement coefficients: $\text{Cost} = 2(a^2 + b^2)$. Here, we outline the cost calculation similar to the one provided for the 1-norm, starting with continuous noise amplification and then comparing to the discrete one.

\emph{Continuous noise amplification}
We start with the previously derived solution
\begin{align}
    a &= \frac{e^{\gamma_1 w_2}\left(1 - e^{-\gamma_2 \Delta w}\right)}{1 - e^{-\Delta \gamma \Delta w}} \\
    b &= -\frac{e^{\gamma_1 w_2 + \Delta \gamma w_1}\left(1 - e^{-\gamma_1 \Delta w}\right)}{1 - e^{-\Delta \gamma \Delta w}}
\end{align}
and note that the results were the same for inserting $k$ noise events upon the replacement $\gamma_2 w_1\rightarrow k \alpha + \gamma_1w_1,\gamma_2 w_2\rightarrow k \beta + \gamma_1 w_2$.
Rewriting $a,b$, squaring and summing them yields
\begin{equation}
    a^2 + b^2 = e^{2\gamma_1 w_2} \left[ (1 + A)^2 + A^2 e^{2\Delta \gamma w_2} \right], \quad A=\frac{1 - e^{-\gamma_1 \Delta w}}{e^{\Delta \gamma \Delta w} - 1}.
\end{equation}

To find the optimal separation $\Delta \gamma$, we take the derivative of $a^2 + b^2$ with respect to $\Delta \gamma$ and set it to zero, yielding the transcendental equation
\begin{equation}
    w_1 - w_2 e^{-\Delta \gamma \Delta w} = \frac{\Delta w}{ 1 - e^{-\gamma_1 \Delta w}} e^{-2\Delta \gamma w_2} \left(e^{\Delta \gamma \Delta w}- e^{-\gamma_1 \Delta w}\right)
\end{equation}
Unlike the 1-norm case, the baseline noise $\gamma_1$ inherently couples into the exact optimal shift.%

We proceed and evaluate the expression in the limit $w_2 \gg \Delta w$. To study the perturbative limit, we define $\epsilon = \frac{\Delta w}{w_2} \ll 1$ and the scaled separation $y = \Delta \gamma w_2$, such that $\Delta \gamma \Delta w = y \epsilon$. Defining the dimensionless continuous baseline error parameter $\kappa = \gamma_1 w_2$, we approximate $1 - e^{-\gamma_1 \Delta w} \approx \gamma_1 \Delta w = \kappa \epsilon$ and arrive at

\begin{equation}
    y_0 - 1 = e^{-2y_0} \left(1 + \frac{y_0}{\kappa}\right) \label{eq:2norm_transcendental} %
\end{equation}
The solution $y_0$ now depends on $\gamma_1$ via $\kappa$. For large $\gamma_1$, $y_0$ approaches $1.1$, while for typical $\kappa \sim 0.5$--$2.0$, we find $y_0 \approx 1.16$--$1.28$, yielding the $3.2$--$3.5$ additional decay reported in the beginning of this section.

In this limit $A\approx \kappa / y_0$ so that the cost evaluates to
\begin{equation}\label{eq:final_scaling_result2}
    a^2 + b^2 \approx %
     e^{2\gamma_1 w_2} \left[ \left(1 + \frac{\kappa}{y_0}\right)^2 + \frac{\kappa (\kappa + y_0)}{y_0^2(y_0 - 1)} \right].
\end{equation}

\emph{Evaluation for discrete noise insertion and comparison to continuous noise amplification}
Like for the 1-norm, the mathematical structure between continuous noise amplification and discrete noise insertion is the same up to
replacing $\Delta \gamma \rightarrow k$, $w_2 \rightarrow \beta$, and $\Delta w \rightarrow (\beta - \alpha)$, yielding:
\begin{equation}
    a^2 + b^2 = e^{2\gamma_1 w_2} \left[ (1 + \tilde{A})^2 + \tilde{A}^2 e^{2\beta k} \right] \quad \text{where} \quad \tilde{A} = \frac{ 1 - e^{-\gamma_1 \Delta w}}{e^{(\beta - \alpha)k} - 1}.
\end{equation}
In the limit $w_2 \gg \Delta w, w_i \ll V$ we can proceed like before and find
 \begin{equation}
     \tilde{A} \approx A \left[ 1 - \frac{2w_1}{3V} \right].
     \label{eq:compare_cost_di_c2}
 \end{equation}
Hence, also in this case, the discrete noise insertion is more efficient compared to the continuous noise amplification.

\section{Tunable error cancellation}\label{app:tec}
In this appendix, we provide further details on tunable error cancellation (TEC).
First, we present additional details on the analytic properties of the TEC filter function, including derivations of several properties stated in the main text.
We then briefly discuss the numerical optimization of the filter function and compare its efficiency to the analytical expression for the filter function mostly used in the main text.
We then provide the details of our adaptive algorithm to select the threshold weight $w_*$ in a quantum experiment with an a priori unknown expectation value and reactivity function, with minimal additional sampling overhead.
We also provide details for a more general adaptive algorithm that can be applied to any tunable quantum error mitigation method.
Finally, we present our general framework incorporating Pauli-path zero-noise extrapolation and more general fitting-based quantum error mitigation methods into our filter function framework; beyond its conceptual appeal, this also enables a simple formula for the optimal allocation of shots in any fitting-based method.

\subsection{Further analytic details on the tunable error cancellation filter function}\label{app:tec_filter_details}

In this subsection, we provide further details on  the TEC filter function introduced in  the main text.
We abbreviate $a\equiv\gamma_0w$, $h_\tau(a) \equiv h_\tau(w/\gamma_0)$, and recall $\Delta\equiv\pi^2/(4\tau^2)$ and $r_j=1+\Delta j^2$.

Let us first derive the non-perturbative error of the filter function at small weights $w$.
Using the coefficients in Eq.~\eqref{eq:tec_main}, the damping of a Pauli path with weight $w$ after mitigation can be written as
\begin{equation}\label{eq:app_tec_resummed_start}
    e^{-a}h_\tau(a)
    =
    h_0\sum_{j=-\infty}^{\infty}
    \frac{(-1)^j e^{-(1+\Delta j^2)a}}{1+\Delta j^2},
    \qquad
    h_0=\frac{\sinh(2\tau)}{2\tau},
\end{equation}
where we extend the sum over $j$ to negative infinity, since the summand is even in $j$.
To proceed, we use the identities, $e^{-(1+\Delta j^2)a}/(1+\Delta j^2)=\int_a^\infty\!dt\,e^{-(1+\Delta j^2)t}$, and, $\sum_j(-1)^j e^{-A j^2}
    =
    \sqrt{\frac{\pi}{A}}
    \sum_\ell e^{-\pi^2(\ell+1/2)^2/A}$.
These give
\begin{equation}\label{eq:app_tec_small_weight_integral}
    1-e^{-a}h_\tau(a)
    =
    \frac{2\sinh(2\tau)}{\sqrt{\pi}}
    \int_0^a dt \, t^{-1/2}
    e^{-t-\tau^2/t}
    \left[1+\mathcal O\!\left(e^{-8\tau^2/t}\right)\right],
\end{equation}
where we retained only the two leading terms, $\ell=0,-1$.
For small $a$, the integral is dominated by its upper endpoint, yielding $1-e^{-\gamma_0w}h_\tau(w)
    \approx
    (2/\sqrt{\pi}\tau^2)
    \sinh(2\tau)
    (\gamma_0w)^{3/2}
    e^{-\tau^2/(\gamma_0w)-\gamma_0w}$, as quoted in the main text.

Let us now turn to the scaling of  $w_*$ with the tuning parameter $\tau$. %
This is most easily derived from a complementary probabilistic representation of the filter function.
Let $X_j$ be independent exponential random variables with rates $r_j$ and distribution $\Pr(X_j=x)=r_j e^{-r_jx}$ for each $j$.
We define the summed random variable, $S_{\tau}\equiv\sum_{j=0}^\infty X_j$.
Owing to the independence of  each $X_j$, $S_{\tau}$ has generating function $\mathbb E[e^{-sS_{\tau}}]=\prod_{j} r_j/(r_j+s)$.
Performing an inverse Laplace transform yields the survival function, $\Pr(S_{\tau}>a) = \sum_j h_j e^{-r_ja}$, where $h_j$ are precisely the TEC filter function coefficients.
Hence, we have
\begin{equation}\label{eq:app_tec_survival_probability}
    e^{-\gamma_0w}h_\tau(w)
    =
    \Pr(S_\tau>\gamma_0w).
\end{equation}
The residual error of the filter function is $1-\Pr(S_\tau>\gamma_0 w) = \Pr(S_\tau\leq\gamma_0w)$.
To derive the scaling of $w_*$ with $\tau$, we note that the mean and variance of  $S_\tau$ are
\begin{equation}\label{eq:app_tec_moments_short}
    \mu_\tau
    =
    \tau\coth(2\tau)+\frac12 \approx \tau + \mathcal{O}(1),
    \qquad
    \sigma_\tau^2
    =
    \frac{\tau^2}{\sinh^2(2\tau)}
    +\frac{\tau}{2}\coth(2\tau)+\frac12 \approx \tau/2 + \mathcal{O}(1),
\end{equation}
where the latter scalings hold at large $\tau$.
Thus, $S_\tau$ is centered at $\tau+\mathcal O(1)$ and has a width of order $\sqrt\tau$.
Moreover, for large $\tau$, $S_\tau$ receives contributions from many independent random variables, and hence approaches a Gaussian distribution.
Hence, to achieve error $|1-e^{-\gamma_0 w_*} h_\tau(w_*)| \leq \varepsilon$ at a given weight $w_*$, one needs $(\mu_\tau-\gamma_0 w_*)^2=\mathcal O(\sigma_\tau^2 \log(1/\varepsilon))$, and hence $\tau
    =
    \gamma_0w_*
    +
    \mathcal O(
        \sqrt{\gamma_0w_*\log(1/\varepsilon)}
        +\log(1/\varepsilon))$, as quoted in the main text.

Let us now address the sampling overhead of the filter function.
For amplified noise rates, the sampling overhead follows from the equality, $1+2\sum_{j\geq1}r_j^{-1}=2\tau\coth(2\tau)$.
This yields $\sum_j|h_j|=h_0(1+2\sum_{j\geq1}r_j^{-1})=\cosh(2\tau)$, and  $X(\tau)=\cosh^2(2\tau)$.
We next turn to discrete noise insertions.
For $w_*\ll V$, the Poisson distributions associated with different amplified noise rates have negligible overlap when translated to discrete noise events, and hence there is little cancellation between their discrete noise event coefficients.
This results in an equivalent sampling overhead to noise amplification.
This changes once the threshold weight $w_*$ is of order $V$, at which the coefficients and sampling overhead approach their PEC values (Fig.~\ref{fig:TEC}, main text).
To derive this approach analytically, we abbreviate $K\equiv3\gamma_0V/4$.
Applying the translation in Eq.~\eqref{eq:poisson_main} to each amplified noise rate gives the discrete noise event coefficients,
\begin{equation}\label{eq:app_tec_discrete_coefficients_short}
    \widetilde h_k(\tau)
    =
    \sum_{j=0}^{\infty}
    h_j e^{-K(r_j-1)}
    \frac{(K(r_j-1))^k}{k!}
    =
    \partial_K^k \left( \frac{\sinh(2\tau)}{2\tau}
    \sum_{j=-\infty}^{\infty}
    \frac{(-1)^j}{1+\pi^2 j^2/4\tau^2} e^{-K \pi^2 j^2/4\tau^2} \right)
    \frac{(-K)^k}{k!}.
\end{equation}
The term in large parentheses is equal to the filter function, $h_\tau(w)$, evaluated at weight $3V/4 = K/\gamma_0$.
For large $\tau$, the filter function approaches $h_\tau(3V/4) \rightarrow e^{\gamma_0 (3V/4)} = e^K$.
Hence, $\partial_K^k e^K \rightarrow e^K$, and we recover the PEC coefficients.
The sampling overhead in this limit is given by $(e^K\sum_kK^k/k!)^2=e^{4K}=e^{3\gamma_0V}$.

\subsection{Numerical optimization of filter functions} \label{ssec:filter_func_num}

\begin{figure}
    \centering
    \includegraphics[scale=1]{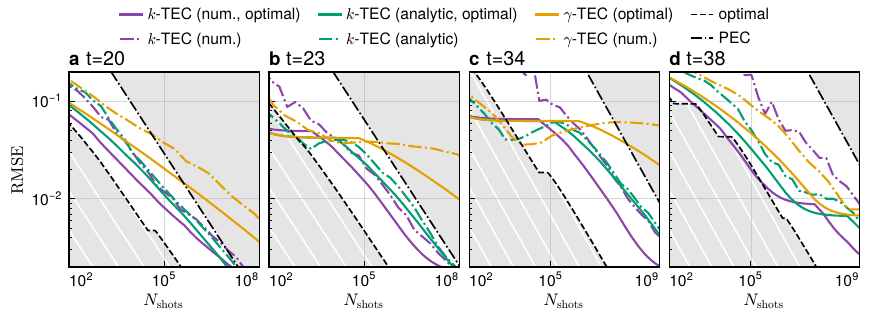}
    \caption{Performance of several variants of tunable error cancellation (TEC) in the 2D TFIM at four evolution times.
    Optimal and PEC curves are as described in the main text.
    Purple curves represent TEC with discrete noise insertions and numerically optimized filter functions, showing both the optimal performance that could be achieved with the method (solid) and the performance achieved by our adaptive algorithm (dashed).
    Here, we define the optimal performance as $B_f(w_*)^2 + \text{Var}(w_*)$, where $B_f(w_*) \equiv \max_{w_*' \geq w_*} B(w_*')$ and $B(w_*')$ is the bias at tuning parameter value $w_*'$; this is the best performance that one could hope to attain with any algorithm based on assessing convergence with increasing $w_*$.
    Green curves denote the same for TEC with discrete noise insertions and the analytic filter function $h_\tau(w)$ described in the main text.
    Our results are illustrative of a broader trend: the numerically optimized filter functions in general have slightly better optimal performance than the analytic filter function, yet suffer from difficulties in achieving this performance in practice (i.e.~using our adaptive algorithm).
    Yellow curves denote the optimal performance of TEC with noise amplification; in this scenario a simple adaptive algorithm performs well and is described in Sec.~\ref{app:opt_hyp_seq}.
    We remark that the 2D TFIM is a system where the reactivity comes surprisingly close to saturating the circuit volume; this explains the relative closeness of TEC with discrete noise insertions to PEC, as well as the comparatively poor performance of TEC with noise amplification.}
    \label{fig:num_optimized_filter_comparison}
\end{figure}

In this subsection, we provide the details for our numerical selection of filter functions for tunable error cancellation.
As summarized in Fig.~\ref{fig:num_optimized_filter_comparison}, we find that this selection leads to moderately more cost-efficient filter functions in practice, compared to our analytic filter function described in the main text and preceding subsection.
However, we also find that our numerical filter functions  occasionally lead to difficulties in selecting the best choice of the threshold weight $w_*$ in practice, using our adaptive algorithm described in the next subsection.
We anticipate that these difficulties could be lessened by a better design of an adaptive algorithm in future work.

We numerically select our filter function coefficients as follows. For discrete noise insertions, we assume $h(w) = \sum_k h_k (1-4w/3V)^k$ for $k = 0,\ldots, V$.
For amplified noise, we assume $h(w) = \sum_\gamma h_\gamma e^{-\gamma w}$ for some choice of inserted noise rates $\{ \gamma \}$.
Specializing to discrete noise insertions for brevity, we choose the coefficients $h_k$ such that they minimize the sampling overhead, $(\sum_k|h_{k}|)^2$, subject to $\sum_kh_{k}=1$ and
\[
 0\leq1-P(w)\leq0.02\frac{w}{w_*}
 \quad(1\leq w\leq w_*),\qquad
 P(w)=e^{-\gamma_0w}\sum_kh_{k}\left(1-\frac{4w}{3V}\right)^k,
 \qquad\gamma_0=0.0025,
\]
for a given choice of threshold weight $w_*$.
This is a convex optimization problem and can be easily solved numerically to give the optimal filter function coefficients.
Here, we select the tolerance, $1-P(w)\leq0.02\frac{w}{w_*}$, to increase linearly from zero at $w = 0$ to a finite small value at $w = w_*$, such that as $w_*$ is increased, the method converges to perfect mitigation.
Increasing or decreasing the small value, $0.02$, changes the relation between the threshold weight $w_*$ and the resulting filter functions, but in practice leads to only very minor changes in the final mitigation performance after $w_*$ is selected adaptively from experimental data.

\subsection{Adaptive algorithm for implementing TEC with discrete noise insertions}
\label{app:adaptive-discrete-tec}

In this subsection, we present our heuristic adaptive algorithm to select the parameter $\tau$ (for our analytic filter functions) and threshold weight $w_*$ (for our numerically-optimized filter functions) when applying tunable error cancellation to a quantum experiment with an a priori unknown noise response and reactivity function.
Our method assumes a fixed total budget of $M$ shots, and allocates these shots in stages.
At each stage, the chosen filter is updated, typically, increasing from smaller threshold weights with lower sampling overhead to larger threshold weights with higher sampling overhead, as the number of shots is increased.
In between each stage, the convergence of the estimate is assessed and the distribution of subsequent shots for the next stage is allocated.
This enables cost savings over performing each estimate individually without reusing shots.

Our algorithm is described in several steps as follows:

\vspace{3mm}
\noindent \emph{Filter coefficients and expectation values.} We index our filter function candidates by $j=0,\ldots,J$, with coefficients $h_j=(h_{jk})$ for measurements with $k$ inserted noise events. For the analytic filter functions, $h_{jk}=\widetilde h_k(\tau_j)$ where $\tau_j=0,0.02,\ldots,5$. For the numerical filter functions, $h_{jk}=h_k(w_j^*)$ where $h_k(w_j^*)$ are obtained as described in the previous subsection.
These filter functions and their coefficients are computed in advance and are fixed throughout our algorithm.
The sampling overhead is $X_j= (\sum_k |h_{jk}|)^2$.

We denote an estimate obtained with filter $j$ as $y_j$. Measurements with a given $k$ contribute to every $y_j$. This reuse avoids separate experiments for each filter.
It also correlates their statistical errors. Namely, if at some stage of the algorithm we have $n_k$ accumulated shots for $k$ insertions, this gives a covariance matrix $\Sigma_{ij}=\sum_{k:n_k>0}h_{ik}h_{jk}/n_k$, where we replace the standard deviation $\sqrt{4p_k(1-p_k)/n_k}$ of the binomial distribution with outcomes $\pm 1$ with its upper bound $\sqrt{1/n_k}$ for simplicity. The standard deviation of $y_a-y_b$ is therefore $s_{ab}=(\Sigma_{aa}+\Sigma_{bb}-2\Sigma_{ab})^{1/2}$.

\vspace{3mm}
\noindent \emph{Estimating the remaining bias.}
In the absence of knowledge of the ideal expectation value, our algorithm uses two proxies for the bias, $L_j$ and $G_j$, defined below. Their average, $B_j=(L_j+G_j)/2$, is used to select the choice of filter. To shorten notation, we define the common parameter $u_j$ as $u_j=\tau_j$ for analytic filter functions and $u_j=\gamma_0w_j^*$ for numerical filter functions. To reduce computation, in this comparison stage we only use every fifth filter function, i.e.~$u=0,0.1,0.2,\ldots$, including both endpoints.

Our first proxy is $L_a=\max_{b\in\mathcal I,\,b\geq a}[|y_a-y_b|-1.25s_{ab}]_+$ for $a\in\mathcal I$, with $[x]_+=\max(x,0)$, followed by linear interpolation to all $j$. Our second proxy is computed from a Gaussian-process fit with kernel $K(u,v)=0.04e^{-(u-v)^2/2}$ and matrix entries $K_{ij}=K(u_i,u_j)$. Here $y_{\mathcal I}$ contains the estimates on $\mathcal I$, $\mathbf1$ is a vector of ones, and $I$ is the identity matrix. The fitted constant mean $c$ and smoothed estimates $g_j$ are
\[
 A=K_{\mathcal I,\mathcal I}+\Sigma_{\mathcal I,\mathcal I}+\delta I,\qquad
 c=\frac{\mathbf1^TA^{-1}y_{\mathcal I}}{\mathbf1^TA^{-1}\mathbf1},\qquad
 g_j=c+K_{j,\mathcal I}A^{-1}(y_{\mathcal I}-c\mathbf1).
\]
The small diagonal term $\delta=10^{-10}\max\{\operatorname{tr}(\Sigma_{\mathcal I,\mathcal I})/|\mathcal I|,0.04,10^{-12}\}$ limits roundoff sensitivity in $A^{-1}$ when estimates are strongly correlated. The second proxy is $G_j=\max_{\ell\geq j}|g_\ell-g_J|$.

\vspace{3mm}
\noindent \emph{Choosing subsequent measurements.}
The first $m_0=1000$ shots are allocated before filter selection begins: 750 at $k=0$, and 250 across insertion counts according to the following rule. Each coefficient vector is normalized to $\bar h_j=h_j/\|h_j\|_1$. Its difference from $\bar h_0$ is then normalized to $d_j=(\bar h_j-\bar h_0)/\|\bar h_j-\bar h_0\|_1$, with $d_j=0$ when the difference vanishes. Of the 250 shots, the allocation to each $k$ is proportional to $\sqrt{\langle d_{jk}^2\rangle_j}$, with an equal-weight average over all filter functions satisfying $X_j\leq M$.

Subsequent measurements are grouped into $L=\max\{100,\lceil40\log_{10}(M/m_0)\rceil\}$ stages, with cumulative counts $m_r=m_0(M/m_0)^{r/L}$ growing exponentially at each stage $r$. At each intermediate stage, the preferred filter minimizes $R_j=B_j^2+X_j/M$, where $X_j/M$ is the minimum variance attainable by allocating all $M$ shots to that filter. The current choice, initially $j=0$, moves by at most one index, $j \rightarrow j-1,j,j+1$, toward this minimum at each stage.

Each stage allocates most of its shots to the current filter $q$, and a smaller subset of its shots to resolving the difference between a pair of filter functions, $a<b$ in $\mathcal I$. The selection of this pair uses an interval $[\ell_{ab},v_{ab}]$ for $|y_a-y_b|$ and the width $D_{ab}$ of $[\ell_{ab}^2,v_{ab}^2]$, defined by
\[
 \ell_{ab}=[|y_a-y_b|-1.25s_{ab}]_+,\qquad
 v_{ab}=|y_a-y_b|+1.25s_{ab},\qquad D_{ab}=v_{ab}^2-\ell_{ab}^2.
\]
The selected pair maximizes $D_{ab}/(\|h_a-h_b\|_1^2/M+0.05R_q)$. Here $\|h_a-h_b\|_1^2/M$ is the minimum variance for estimating the difference with $M$ shots. After $m=\sum_kn_k$ shots, the fraction devoted to this comparison is $\rho=0.1\sqrt{1-m/M}\,D_{ab}/(D_{ab}+R_q)$. The resulting allocation profile $p_k$ and updated counts $n'_k$ are
\[
 p_k=(1-\rho)\frac{|h_{qk}|}{\|h_q\|_1}
       +\rho\frac{|h_{ak}-h_{bk}|}{\|h_a-h_b\|_1},\qquad
 n'_k=\max(n_k,\lambda p_k),\qquad \sum_kn'_k=m_{r+1}.
\]
The value of $\lambda$ is chosen so that the total number of new shots, $\sum_k (n'_k-n_k)$, is equal to the  allocation, $m_{r+1}-m_r$.
We repeat this and the previous step $L-1$ times, for each stage $r = 1,\ldots, L-1$.

\vspace{3mm}
\noindent \emph{Final estimate and numerical implementation.} We perform a slightly modified filter function selection at the final stage of our algorithm, $r = L$.
In this last stage, each filter is scored using $B_j^2$ plus the variance attainable with all remaining shots and retained samples. The chosen index $j$ then moves at most five positions in either direction toward the minimum. The remaining shots are allocated to this filter. The final estimate  uses all accumulated shots.

\subsection{Adaptive algorithm for implementing general mitigation methods with a continuous tuning parameter}\label{app:opt_hyp_seq}

In this subsection, we present a general adaptive algorithm to select the tuning parameter for any tunable quantum error mitigation scheme with a continuous tuning parameter.
Unlike our adaptive algorithm in the previous section, to maintain generality, the algorithm we present here does not reuse shots between stages.
Hence, it can be less efficient than such specialized adaptive algorithms.

The starting point for the algorithm used in the previous section is the set of numerically optimized linear coefficients $h_r$, here for $\gamma=0.0025$ to $\gamma=0.01$ in $10$ steps, obtained as $h_r(\alpha,w_l) = \arg \min \sum_r|h_r|$ constrained to $0\leq 1-h(w) \leq \alpha$ for  $w<w_l$. The achievable target accuracy and optimal hyperparameters $\alpha,w_l$ may not be known in advance. We study this scenario by adapting $\alpha,w_l$ iteratively while increasing the number of shots. In particular, we start with $\alpha=1/2$ and $w_{l}=V/64$, where $V$ is the circuit volume. While increasing the number of shots, we evaluate the difference $\hat{C}_{\alpha,{w_l'}}(0)-\hat{C}_{\alpha,{w_{l}}}(0)$, where $w_l'$ is chosen so that the variance increases by a factor $\exp(1)$ for a given number of shots: $\mathrm{Var} \hat{C}_{\alpha,{w_l'}}(0)=\exp(1)\mathrm{Var}\hat{C}_{\alpha,{w_{l}}}(0)$. The factor $\exp(1)$ is motivated by being the optimal spacing if the needed number of shots for a target accuracy is unknown but reciprocally distributed, as we discuss in the following. Once the difference $\hat{C}_{\alpha,{w_l'}}(0)-\hat{C}_{\alpha,{w_{l}}}(0)$ is significant, i.e., larger than twice the error estimate on it, we set $w_l=w_l'$. The hyperparameter $\alpha$ is reduced by a factor of $2$ (chosen arbitrarily) whenever $\mathrm{Var}\hat{C}_{\alpha,{w_{l}}}(0)<\alpha^2$.

The factor $\exp(1)$ can be obtained from the following considerations. For simplicity, let $\upsilon$ denote the continuous tuning parameter.
We assume that the cost of the error mitigation method, $N_\upsilon$, increases monotonically in $\upsilon$.
We also assume that there exists an optimal tuning parameter value, $\upsilon_*$, with cost $N_{\upsilon_*}$, such that the mitigated expectation value, $\hat{C}_{\upsilon}(0)$, has converged to the ideal expectation value, $C(0)$, to within a desired precision for all $\upsilon \geq \upsilon_*$.
Our algorithm is optimal under the assumption that the a priori unknown cost, $N_{\upsilon_*}$, to evaluate $\hat{C}_{\upsilon_*}(0)$ with the target $\upsilon_*$ is reciprocally distributed, i.e.~$\log(N_{\upsilon_*})$ is uniformly distributed.

Our algorithm proceeds in $A+1$ stages.
At each stage $i$, one selects a tuning parameter value, $\upsilon_i$, and spends $N_{\upsilon_i}$ shots to measure the mitigated expectation value $\hat{C}_{\upsilon_i}(0)$.
This requires a total number of shots,
\begin{align}
    N_\mathrm{tot} = \sum_{i=0}^{A} N_{\upsilon_i}.
\end{align}
As an ansatz, we choose the sequence $\upsilon_i$ defined by increasing the \emph{cost} in every iteration by a factor of $q$,
\begin{align}
    N_{\upsilon_{i+1}} = q  N_{\upsilon_{i}}.
\end{align}
We imagine that this procedure is repeated until $\upsilon_i$ converges to near its ideal value, $\upsilon_A \approx \upsilon_*$.

We would like to determine the value of $q$ that minimizes the overall cost $N_\mathrm{tot}$.
Choosing $q$ too large could lead to overshooting, in which our achieved $N_{\upsilon_A}$ is far larger than the necessary $N_{\upsilon_*}$.
Meanwhile, choosing $q$ too small could lead to substantial shot overhead from many intermediate evaluations of $\hat{C}_{\upsilon_i}(0)$.
Given $q$, we find
\begin{align}
    A = \lceil \log(N_{\upsilon_*}/N_{\upsilon_0})/\log(q) \rceil,
\end{align}
and express $N_{\upsilon_*}$ as
\begin{align}
    N_{\upsilon_*} = N_{\upsilon_0} q^{A-1} r,
\end{align}
introducing the factor $r$. Since $\log(N_{\upsilon_*})$ is uniformly distributed, $\log(r)$ is also uniformly distributed on $(0,\log(q))$.
Then
\begin{align}
    N_\mathrm{tot} = \sum_{i=0}^{A} N_{\upsilon_0} q^i =N_{\upsilon_0}\frac{q^{A+1}-1}{q-1}.
\end{align}
Hence, in comparison to the target cost, the total cost is larger by a factor of
\begin{align}
    \frac{N_\mathrm{tot}}{N_{\upsilon_*}} = \frac{1}{rq^{A-1}}\frac{q^{A+1}-1}{q-1} \approx  \frac{1}{r}\frac{q^2}{q-1}.
\end{align}
Using that $\log(r)$ is uniformly distributed on $(0,\log(q))$ we find the expected overhead to be
\begin{align}
    \frac{1}{\log(q)}\int_0^{\log(q)}\!\mathrm{d} \log(r) \frac{1}{r}\frac{q^2}{q-1} = \frac{q}{\log(q)}.
\end{align}
This is minimal for $\partial_q (q/\log(q))=0 \Leftrightarrow q=e^1=e$. Hence, choosing $\upsilon_i$ so that $N_{\upsilon_{i+1}}=eN_{\upsilon_{i}}$ minimizes the expected relative cost overhead.

\subsection{Pauli-path zero-noise extrapolation as information filtering}\label{app:ppzne_as_filtering}

As mentioned in the main text, one can incorporate more general methods of quantum error mitigation, such as Pauli-path zero-noise extrapolation or other fitting-based methods, in our filter function framework by linearizing the cost function in the neighborhood of the optimal parameters.
In this subsection, we present the details of this mapping.

We consider fitting a model $c_{\kappa}(\gamma)$ to data $C(\gamma)$ by optimizing a loss function $L_{\kappa}=\sum_{\gamma}\mathcal{L}\big(c_{\kappa}(\gamma),C(\gamma)\big)$, with optimized parameters $\kappa^*$.
(For example, one can choose the mean squared error $\mathcal{L}\big(c_{\kappa}(\gamma),C(\gamma)\big)=\tfrac{M_{\gamma}}{\sigma_{\gamma}^2}\big(c_{\kappa}(\gamma)-C(\gamma)\big)^2$, where $M_{\gamma}$ is the number of shots taken for the estimate of $C(\gamma)$ and $\sigma^2_{\gamma}=(1-C(\gamma)^2)$ is the variance in a single shot.)
The parameter covariance matrix and variance in $c_{\kappa^*}(0)$ after optimization are
\begin{align}
    \Sigma^{-1}_{\kappa^*} = \sum_{\gamma}W_{\gamma}\bs{d}_{\gamma}\bs{d}_{\gamma}^T,\hspace{0.5cm}\sigma^2_0=\bs{d}_0^T\Sigma_{\kappa^*}\bs{d}_0,
\end{align}
where the derivative vectors $\bs{d}_{\gamma}=\nabla_{\kappa}c_{\kappa}(\gamma)$ capture the sensitivity of our model to the parameters we fit, and the weights $W_{\gamma}=\frac{\partial^2\mathcal{L}}{\partial c^2}\big(c_{\kappa}(\gamma),C(\gamma)\big)$ measure the relative contribution of the $\gamma$ datapoint to the cost function.
We linearize our fit by defining the coefficients $h_{\gamma}$
\begin{equation}\label{eq:parameterized_filter_coeff_def}
    h_{\gamma}=W_{\gamma}\bs{d}_{\gamma}^T\Sigma_{\kappa^*}\bs{d}_0, \hspace{1cm}\sum_{\gamma} h_{\gamma}\bs{d}_{\gamma}=\bs{d}_0.
\end{equation}
Under relatively general conditions, this propagates to the parametrized function itself, yielding
\begin{equation}
    \sum_{\gamma}h_{\gamma}C(\gamma)\approx\sum_{\gamma}h_{\gamma}c_{\kappa}(\gamma)=c_\kappa(0)\approx C(0),
\end{equation}
which is a linear combination of the form in Eq.~\eqref{eq:linear_combination}.

This result has two useful outcomes.
Firstly, for typical loss functions such as the MSE, $W_{\gamma}=M_{\gamma}\sigma^{-2}_\gamma$, and we can optimize the distribution of measurements $M_{\gamma}$
\begin{equation}\label{eq:shot_distribution_fitting}
    M_{\gamma}=M \frac{|h_{\gamma}|\sigma_{\gamma}}{\sum_{\gamma'}|h_{\gamma'}|\sigma_{\gamma'}},\hspace{1cm}\sigma^2_0=\frac{1}{M}\left(\sum_{\gamma}|h_{\gamma}|\sigma_{\gamma}\right)^2.
\end{equation}
It is challenging to use this result a priori in an experiment, as the coefficients $h_\gamma$ depend on the derivative vectors $\bs{d}_{\gamma}$ at the point $\kappa^*$ of optimization.
However, one could optimize the shot distribution by using a small fraction of a total shot budget to estimate Eq.~\eqref{eq:shot_distribution_fitting}, before proceeding with a larger experiment.
Secondly, we can obtain the filter function that one applies in Laplace space upon performing a parametrized fit, by substituting Eq.~\eqref{eq:parameterized_filter_coeff_def} into Eq.~\eqref{eq:filter_function_main}
\begin{equation}
    h_{\mathrm{fit}}(w)=\mathbf{d}_0^T\Sigma_{\kappa^*}\sum_{w'}\bs{s}(w')\Big(\sum_{\gamma}W_{\gamma}e^{-\gamma(w+w')}\Big).
\end{equation}
Here, $\bs{s}(w)=\nabla_{\kappa} r_{\kappa}(w)$, where $r_{\kappa}(w)$ is the reactivity of $c_{\kappa}(\gamma)$; $c_{\kappa}(\gamma)=\int dw e^{-\gamma w}r_{\kappa}(w)$.
One can observe that the term in the brackets acts as the kernel of the Fisher information matrix for the parameterized function $r_{\kappa}(w)$ relative to the obtained data $\{C(\gamma)\}$.

\section{Experimentally relevant methodologies}\label{sec:app_pp_zne_otoc}

Implementing the Pauli-path zero-noise extrapolation (ZNE) framework on physical quantum devices introduces significant challenges, particularly when mitigating deep circuits characterized by severe signal damping and low fidelities. In these regimes, the extrapolation process becomes highly sensitive to statistical noise. To achieve robust and accurate estimation of the zero-noise limit for an observable across an entire time series, e.g. an OTOC, a specialized data processing and parameter estimation pipeline was found to be highly beneficial. This section details the practical methodologies necessary to stabilize the mitigation protocol on real hardware.

\begin{figure}
    \centering
    \includegraphics[width=\columnwidth]{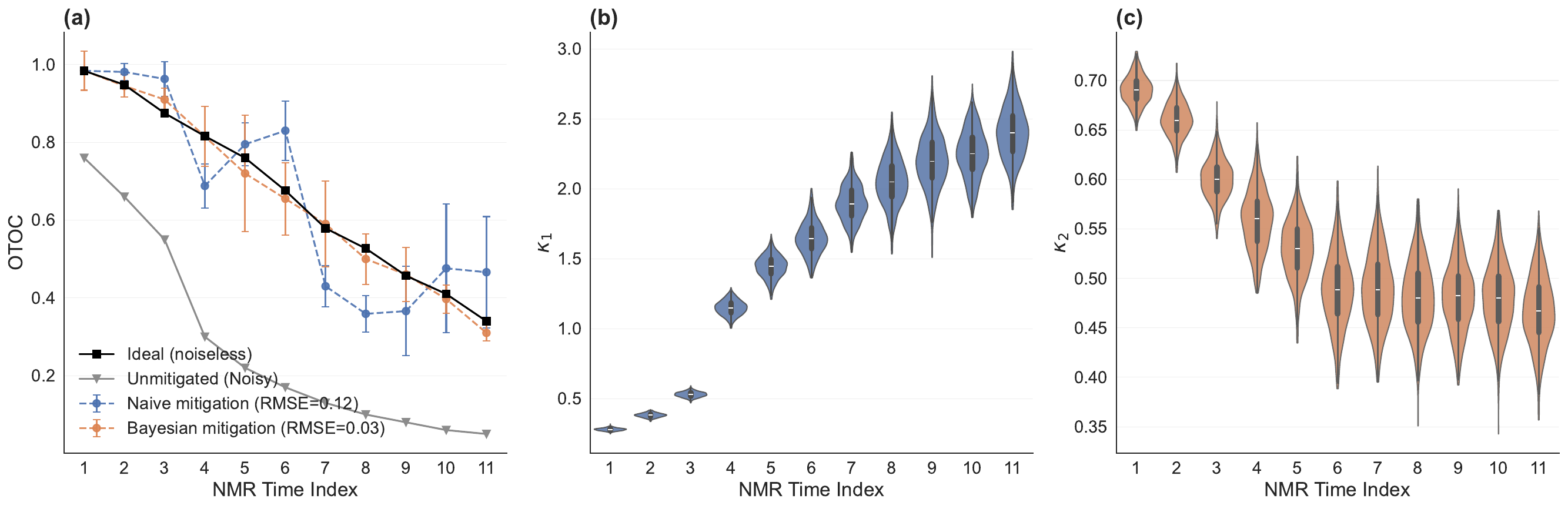}
    \caption{(a) Comparison of Out-of-Time-Order Correlator (OTOC) dynamics as a function of the NMR time index. The unmitigated data (gray triangles) exhibits severe signal damping at deep circuit depths. Standard, naive mitigation techniques (blue circles) struggle with the low signal-to-noise ratio, resulting in unstable estimates (RMSE = 0.12). By employing a time-correlated Bayesian update workflow (orange circles), the zero-noise limit is accurately and robustly extracted across the entire time series, closely mirroring the ideal, noiseless trajectory (black squares) with significantly improved accuracy (RMSE = 0.03).
(b, c) Parameter estimation distributions for the cumulant expansion variables (represented here by $\kappa_1$ and $\kappa_2$) across the time series, evaluated using bootstrap resampling. Because deep, noisy circuits are prone to non-physical local minima, the parameters are fitted sequentially. Reliable estimates from shallow, high-fidelity time steps are used to seed a Gaussian prior ($L_2$ penalty) for subsequent steps. This time-ordered regularization allows the parameters to smoothly diffuse, stabilizing the optimization protocol against severe hardware noise and transient fluctuations.}
    \label{fig:nmr_otoc_error_mitigation}
\end{figure}

For deep, noisy circuits, standard model-fitting techniques often fail to converge or get trapped in non-physical local minima due to poor signal-to-noise ratios. To overcome this, parameter estimation benefits strongly if structured as a time-correlated Bayesian update workflow, as shown in Fig.~\ref{fig:nmr_otoc_error_mitigation}.

The explicit advantage of this Bayesian framework is demonstrated in Fig.~\ref{fig:nmr_otoc_error_mitigation}. As seen in panel (a), the unmitigated OTOC signal decays rapidly as circuit depth increases, and naive mitigation attempts become highly unstable at later time indices. In contrast, the Bayesian protocol reliably reconstructs the ideal noiseless dynamics. This robustness stems directly from the regularized parameter distributions detailed in panels (b) and (c). By bounding the cumulant variables (in this case just $\kappa_2$) with time-ordered priors, the bootstrap-resampled parameter distributions exhibit a smooth, physical evolution across the time series, explicitly preventing the erratic, non-physical jumps characteristic of unconstrained fitting routines.

In practice, this is implemented by initially fitting the data from shallower, less-noisy time steps where circuit fidelity remains relatively high. The parameters extracted from these reliable, short-depth fits are then used to establish a Bayesian prior for higher-weight cumulants in the Pauli path expansion that are more susceptible to experimental noise. As the analysis progresses to deeper, noisier variants of the circuit, this prior iteratively seeds and constrains the subsequent model fit. By propagating the prior sequentially across the time series, the fitting algorithm retains memory of the physical system's behavior, thereby stabilizing the parameter estimation against severe signal damping.

We denote our time-ordered observable by $z(t)$, for example an NMR OTOC observable, produced from a series of quantum circuits that we seek to mitigate. Let $r$ index individual noise instances for a given experimentally realized noise strength $\gamma$; the associated uncertainty is given by $\sigma_i$. Lastly, $C^{(M)}$ denotes the truncated cumulant expansion as given by Eq. \ref{eq:app_cumulant_truncated}. Bayesian regularization is achieved by enforcing a prior on higher order cumulants using the previous, assumed higher-confidence estimate. We found good behavior of a Gaussian prior, effectively allowing higher order cumulants to diffuse slowly, i.e. penalizing discontinuous estimates across time-ordered circuit realizations. This effectively amounts to augmenting the log-likelihood to be optimized with an $L_2$ penalization, as written below
\begin{equation}
-\log L(z_r|\kappa) =
\displaystyle\sum_{r=1}^{n_r} \frac{1}{2\sigma_r^2}\big(z_r(t) - C^{(M)}\left(\gamma_r|\kappa\right)\big)^2 + \displaystyle\sum\limits_{m=2} \frac{1}{2\tau_m^2}\big(\kappa_m({t}) - \kappa_m({t-1})\big)^2.
\end{equation}

To ensure robustness during the Bayesian fitting process, it is recommended to utilize a global-parameter optimization algorithm; given the complex parameter space in the presence of experimental noise, standard gradient-descent methods are prone to premature convergence. We found an effective approach to be the combination of a local minimization routine with stochastic jumps across the parameter space, allowing the optimizer to escape local minima, i.e. basin hopping. Furthermore, this optimization routine must be paired with a rare-outlier rejection protocol to filter anomalous data points generated by transient hardware fluctuations or non-Markovian noise spikes.

Rigorous error bounding is critical for evaluating the performance of the ZNE protocol. Rather than relying on standard propagation of error—which may fail to capture the complex correlations inherent to the fitting and optimization routines—uncertainty should be evaluated using bootstrap resampling. By resampling over the distribution of executed mitigation circuit instances, one can empirically derive both the mitigated expectation values and their respective confidence intervals.

%
%
%

\clearpage
\section{Notation}\label{app:notation-clean}

\noindent Unless stated otherwise all systems are qubits, and indices run over the ranges given.

\onecolumngrid
\begingroup
\small
\tablehead{\hline\hline Symbol & Meaning \\ \hline}
\tabletail{\hline}
\tablelasttail{\hline\hline}
\begin{supertabular}{@{}p{0.28\textwidth}p{0.62\textwidth}@{}}
\multicolumn{2}{l}{\textbf{Circuit and system}}\\[0.2em]
$N$                     & number of qubits.\\
$t$, $T$                & layer (moment) index and total number of layers; $t=1,\dots,T$.\\
$s$, $S$                & shot-noise realization index and number of realizations.\\
$\rho_{\mathrm{init}}$  & initial state.\\
$O$                     & measured observable.\\
$U_t$, $\mathcal{U}_t$  & unitary of layer $t$, and the corresponding superoperator, $\mathcal{U}_t|\rho\rrangle\equiv|U_t\rho U_t^\dagger\rrangle$.\\
$\mathcal{N}_t(\gamma)$ & noise channel following layer $t$, with $\mathcal{N}_t[\bs 0]=\mathcal{I}$.\\[1em]
\hline
\multicolumn{2}{l}{\textbf{Double-bracket (Liouville) notation}}\\[0.2em]
$\mathcal{H}$           & the $N$-qubit Hilbert space, $\mathcal{H}=(\mathbb{C}^2)^{\otimes N}$, $\dim\mathcal{H}=2^N$.\\
$\mathcal{L}(\mathcal{H})$ & the $4^N$-dimensional operator space; superoperators act on it, $\mathcal{U}_t,\mathcal{N}_t\in\mathcal{L}(\mathcal{L}(\mathcal{H}))$.\\
$|\rho\rrangle$         & vectorized state, $|\rho\rrangle\equiv 2^N\rho$.\\
$\llangle O|$           & vectorized observable acting as $\llangle O|\cdot\equiv\mathrm{Tr}[O^\dagger\,\cdot\,]$.\\
$\llangle P|P'\rrangle$ & inner product, $\tfrac{1}{2^N}\mathrm{Tr}[PP']$, under which the Pauli operators are orthonormal.\\
$|P\rrangle\llangle Q|$ & outer product, $\tfrac{1}{2^N}P\,\mathrm{Tr}[Q\,\cdot\,]$.\\[1em]
\hline
\multicolumn{2}{l}{\textbf{Paulis and Pauli paths}}\\[0.2em]
$\mathbb{P}$            & single-qubit Pauli alphabet $\{I,X,Y,Z\}$; $\mathbb{P}^N$ the $N$-qubit Pauli operators.\\
$P$, $P_t$              & a Pauli operator; the Pauli at layer $t$ along a path.\\
$\vec{P}$               & a Pauli path, the tuple $(P_1,\dots,P_T)$ with one Pauli per layer; there are $4^{NT}$ of them.\\ 
$w_{P}$                 & single-layer weight of Pauli $P$, with $w_I=0$.\\
$w_{\vec{P}}$           & accumulated path weight, $w_{\vec P}=\sum_t w_{P_t}$.\\
$A_{\vec{P}}$           & path amplitude, $A_{\vec P}=\llangle O|\prod_t|P_t\rrangle\llangle P_t|\,\mathcal{U}_t|\rho_{\mathrm{init}}\rrangle$.\\[1em]
\hline
\multicolumn{2}{l}{\textbf{Noise rates and signal}}\\[0.2em]
$\gamma$                & common noise-rate parameter; all layer rates scale with it.\\
$\bs{\gamma}_t$         & the (possibly many) noise rates of layer $t$.\\
$\gamma_0$, $\gamma_{\max}$ & lowest and highest accessible noise rates; $\gamma_0$ is the device's own rate, the smallest achievable.\\
$\gamma_r$              & the $r$th amplified noise rate at which the signal is measured, $r=1,\dots,n_r$.\\
$n_r$                   & number of amplified noise rates used; $n_r=4$ and $n_r=10$ in the numerics.\\
$C(\gamma)$             & noisy signal, $C(\gamma)=\llangle O|\prod_t\mathcal{N}_t(\gamma)\,\mathcal{U}_t|\rho_{\mathrm{init}}\rrangle$.\\
$C(0)\equiv C_{\mathrm{exact}}$ & the noiseless target.\\
$\hat{C}(0)$            & an estimator of the noiseless target.\\[1em]
\hline
\multicolumn{2}{l}{\textbf{Reactivity}}\\[0.2em]
$w$                     & accumulated weight, the variable conjugate to $\gamma$.\\
$R(w)$                  & \emph{circuit reactivity} (with an argument $w$), $R(w)=\sum_{\vec P}\delta_{w_{\vec P},w}A_{\vec P}$; signed in general.\\
                        & The noisy signal is its Laplace transform: $C(\gamma)=\sum_w e^{-\gamma w}R(w)$.\\
$R_{\gamma_G}(w)$       & reactivity smoothed by a Gaussian kernel of width $1/\gamma_G$.\\
$\gamma_G$              & smoothing scale; features below $\Delta w\sim1/\gamma_G$ are removed.\\
$\bar{w}$, $\sigma_w^2$ & mean and variance of the normalized reactivity $R(w)/C(0)$.\\
$\Delta w$              & extent (width) of the reactivity about its peak.\\[1em]
\end{supertabular}
\newpage
\begin{supertabular}{@{}p{0.28\textwidth}p{0.62\textwidth}@{}}
\multicolumn{2}{l}{\textbf{Generalized framework (Appendix~\ref{app:react_Pauli paths})}}\\[0.2em]
$\lambda$               & abstract label for a circuit modification; $\lambda=0$ is the noise-free circuit. Becomes $\gamma_r$ for continuous amplification and $k$ for discrete insertion.\\
$F(\lambda,w)$          & modification function, $C(\lambda)=\sum_w F(\lambda,w)\,R(w)$; known, and simple by construction.\\
$\mathcal{S}_t$, $\vec{\mathcal{S}}$ & superoperator component of the noise channel at layer $t$, and a generalized path through them.\\
$A_{\vec{\mathcal{S}}}$ & generalized path amplitude, $A_{\vec{\mathcal S}}=2^N  \llangle O|\prod_t\mathcal{S}_t\mathcal{U}_t|\rho_{\mathrm{init}}\rrangle$.\\
$R(\vec{w})$            & multidimensional reactivity, before the uniform-amplification reduction to $R(w)$.\\
$|L_{\mathcal{S}}\rrangle$, $\llangle R_{\mathcal{S}}|$ & right and left eigenoperators (subscripted $\mathcal S$) of a diagonalizable noise channel, $\llangle R_{\mathcal S}|L_{\mathcal S'}\rrangle=\delta_{\mathcal S,\mathcal S'}$; here $w_{\mathcal S}$ may be complex or unbounded.\\
$w_{\mathcal{S}_t}$, $w_{\vec{\mathcal{S}}}$ & weight of mode $\mathcal S$ at layer $t$, and the accumulated weight of a generalized path, $w_{\vec{\mathcal S}}=\sum_t w_{\mathcal S_t}$. Real for stochastic Pauli noise; complex in general.\\
$\gamma_t$, $\vec{\gamma}$ & noise strength of layer $t$, and the vector $(\gamma_1,\dots,\gamma_T)$ of layer-resolved strengths. Homogeneous amplification is $\gamma_t=\gamma$ for all $t$.\\[1em]
\hline
\multicolumn{2}{l}{\textbf{Pauli generator and discrete insertion (App.~\ref{app:biorthogonal})}}\\[0.2em]
$\mathcal{V}_t$         & generator of the layer-$t$ Pauli channel, $\mathcal{N}_t=e^{-\gamma_t\mathcal{V}_t}$ with $\mathcal{V}_t(P)=\tfrac{3N}{4}\bigl(P-\sum_E q_t(E)EPE^\dagger\bigr)$. The prefactor $\tfrac{3N}{4}$ is fixed so that $w$ counts errors.\\
$E$, $q_t(E)$           & an inserted Pauli error, and the probability of error $E$ at layer $t$; $q_t\geq0$, normalized.\\
$s_{E,P}$               & commutation sign, $s_{E,P}=\tfrac{1}{2^N}\mathrm{Tr}(EPE^\dagger P)=\pm1$: $+1$ if $E$ and $P$ commute, $-1$ if they anticommute.\\
$m$                     & Hamming weight of a Pauli string, its number of non-identity sites. For uniform local depolarizing noise $w_{P_t}=\gamma_t m$.\\
$k$                     & number of deliberately inserted noise events.\\
$\bar{k}$               & mean number of native noise events, $\bar{k}=3N\gamma_0/4$. Sets the PEC coefficients $h_k^{\mathrm{PEC}}=e^{\bar{k}}(-\bar{k})^k/k!$ and the Poisson sampling distribution $p_k=e^{-\bar{k}}\bar{k}^k/k!$. Written $\bar{k}_r=3N\gamma_0(r-1)/4$ for amplification by a ratio $r$.\\
$\gamma_{\Sigma}$       & averaged native noise exposure of the circuit, $\gamma_{\Sigma}=T^{-1}\sum_t\gamma_t$; sets the insertion sampling weight $p(E,t)=\gamma_tq_t(E)/(T\gamma_{\Sigma})$.\\
$f_k(w)$                & discrete-insertion kernel, $f_k(w)=\bigl(1-4w/3N\gamma_{\Sigma}\bigr)^{k}=\bigl(1-w/w_0\bigr)^{k}$.\\
$w_0$                   & weight at which the kernel vanishes, $w_0=3N\gamma_{\Sigma}/4$; paths of this weight are erased from the signal.\\
$F_{\mathrm{disc}}(k,w)$ & discrete modification function, $F_{\mathrm{disc}}(k,w)=e^{-w}f_k(w)$ --- the $F(\lambda,w)$ of the response relation with $\lambda=k$.\\[1em]
\hline
\multicolumn{2}{l}{\textbf{Mitigation methods}}\\[0.2em]
$\kappa_m$              & $m$th cumulant of $e^{-\gamma_0 w}R(w)$; $\log C(\gamma)=\sum_m\frac{(\gamma_0-\gamma)^m}{m!}\kappa_m$.\\
$M$                     & truncation order of a cumulant or multi-exponential model.\\
$\eta$                  & rate parameter of the illustrative distribution under discussion in the convergence examples (Poisson, exponential); unrelated to the noise rates.\\
$h_r$, $h(w)$           & filter coefficients and the induced filter function, $h(w)=\sum_r h_r e^{-\gamma_r w}$.\\
$w_*$                   & threshold weight up to which the filter is constrained to reconstruct the signal.\\
$\alpha$                & filter tolerance, $0<1-h(w)<\alpha$ for $w<w_*$.\\
$N_{\mathrm{shots}}$    & total number of circuit executions.\\
$f$                     & signal fidelity, $f\equiv C(\gamma)/C(0)$, the coordinate in which mitigation cost is quoted.\\ 
\end{supertabular}
\endgroup

\medskip
\noindent Weights $w$ are dimensionless and accumulate over the circuit's spacetime volume; the
reactivity $R(w)$ collects the signal amplitude carried by all Pauli paths of a given accumulated
weight.

\medskip


\twocolumngrid

\end{document}